\documentclass[final,twocolumn,12pt]{elsarticle}
\usepackage{hyperref}
\usepackage[margin=1.5cm,includefoot]{geometry}
\usepackage{auto-paper}
\zexternaldocument*{supp}

\newcommand{\gprrmse}{0.0218}

\newcommand{\avgrmse}{0.1283}

\newcommand{\gprrmsePercReduction}{83}

\newcommand{\avgmae}{0.0955}

\newcommand{\nnomega}{2.8709 \pm 00.69112}

\newcommand{\symtime}{76}

\newcommand{\startpt}{red points}
\newcommand{\singlept}{magenta points}
\newcommand{\sympt}{dark blue points}
\newcommand{\singlesympt}{dark blue point}
\newcommand{\refpt}{white circle}
\newcommand{\vbordercolor}{black}
\newcommand{\vcellcolor}{light blue}
\newcommand{\inpt}{input}
\newcommand{\outpt}{prediction}
\newcommand{\vfzorepo}{\gls{vfzo} repository}
\newcommand{\mytitle}{Five Degree-of-Freedom Property Interpolation of Arbitrary Grain Boundaries via \glsentrytitlecase{vfzo}{long} Framework}
\makeglossaries
\GlsXtrEnableEntryCounting{abbreviation}{3}
\glssetcategoryattribute{abbreviation}{nohyper}{true}

\newabbreviation[longplural=five degrees of freedom]{5dof}{5DOF}{five degree-of-freedom}
\newabbreviation[longplural=three degrees of freedom]{3dof}{3DOF}{three degree-of-freedom}
\newabbreviation[longplural=degrees of freedom]{dof}{DOF}{degree of freedom}
\newabbreviation{ebsd}{EBSD}{electron backscatter diffraction}
\newabbreviation[longplural={grain boundaries}]{gb}{GB}{grain boundary}
\newabbreviation{fcc}{FCC}{face-centered cubic}
\newabbreviation{sem}{SEM}{scanning electron microscope}
\newabbreviation{fea}{FEA}{finite element analysis}
\newabbreviation{bcs}{BCs}{boundary conditions}
\newabbreviation[longplural={triple junctions}]{tj}{TJ}{triple junction}
\newabbreviation{gpr}{GPR}{Gaussian process regression}
\newabbreviation{gprm}{GPRM}{Gaussian process regression mixture}
\newabbreviation{ann}{ANN}{artificial neural network}
\newabbreviation{nn}{NN}{nearest neighbor}
\newabbreviation{rmse}{RMSE}{root mean square error}
\newabbreviation{mae}{MAE}{mean absolute error}
\newabbreviation{brk}{BRK}{Bulatov Reed Kumar}
\newabbreviation{gbed}{GBED}{grain boundary energy distribution}
\newabbreviation{gbcd}{GBCD}{grain boundary character distribution}
\newabbreviation{mfz}{MFZ}{misorientation fundamental zone}
\newabbreviation{bp}{BP}{boundary plane}
\newabbreviation{bpfz}{BPFZ}{boundary plane fundamental zone}
\newabbreviation{knn}{kNN}{k-nearest neighbor}
\newabbreviation{gbe}{GBE}{grain boundary energy}
\newabbreviation{gbo}{GBO}{grain boundary octonion}
\newabbreviation{nbo}{NBO}{no-boundary octonion}
\newabbreviation{oslerp}{oSLERP}{octonion Spherical Linear Interpolation}
\newabbreviation{loocv}{LOOCV}{leave-one-out cross validation}
\newabbreviation{kfcv}{kFCV}{k-fold cross validation}
\newabbreviation{seo}{SEO}{symmetrically equivalent octonion}
\newabbreviation{fex}{FEX}{file exchange}
\newabbreviation{idw}{IDW}{inverse-distance weighting}
\newabbreviation{fic}{FIC}{fully independent conditional}
\newabbreviation{svd}{SVD}{singular value decomposition}
\newabbreviation{gbc}{GBC}{grain boundary character}
\newabbreviation{fz}{FZ}{fundamental zone}
\newabbreviation{vfz}{VFZ}{Voronoi fundamental zone}
\newabbreviation{vfzo}{VFZO}{Voronoi fundamental zone octonion}
\newabbreviation{lobpcg}{LOBPCG}{locally optimal block preconditioned conjugate gradient}
\newabbreviation{lkr}{LKR}{Laplacian kernel regression}
\newabbreviation{ms}{MS}{molecular statics}

\begin{document}

	\sloppy 

	\begin{frontmatter}

		\title{\mytitle{}}

		\author[myu]{Sterling G. Baird\corref{cor1}}
\ead{ster.g.baird@gmail.com}
\author[myu]{Eric R. Homer}
\author[myu]{David T. Fullwood}
\author[myu]{Oliver K. Johnson}

\address[myu]{Department of Mechanical Engineering, Brigham Young University, Provo, UT 84602, USA}

\cortext[cor1]{Corresponding author.}

\date{February 2020}

		\begin{abstract}
			In this work we introduce the \gls{vfzo} interpolation framework for \gls{gb} structure-property models and surrogates. The \gls{vfzo} framework offers an advantage over other \gls{5dof} based property interpolation methods because it is constructed as a point set in a manifold. This means that directly computed Euclidean distances approximate the original octonion distance with significantly reduced computation runtime ($\sim$7 CPU minutes vs. 153 CPU days for a $\num{50000}\times\num{50000}$ pairwise-distance matrix). This increased efficiency facilitates lower interpolation error through the use of significantly more input data.
			We demonstrate \gls{gbe} interpolation results for a non-smooth validation function and simulated bi-crystal datasets for Fe and Ni using four interpolation methods: barycentric interpolation, \gls{gpr} or Kriging, \gls{idw}, and \gls{nn} interpolation. These are evaluated for \num{50000} random \inpt{} \glspl{gb} and \num{10000} random \outpt{} \glspl{gb}. The best performance was achieved with \gls{gpr}, which resulted in a reduction of the \gls{rmse} by \calcnum{(\avgrmse-\gprrmse)/\avgrmse*100}\% relative to \gls{rmse} of a constant, average model. Likewise, interpolation on a large, noisy, \gls{ms} Fe simulation dataset improves performance by \SI{34.4}{\percent} compared to \SI{21.2}{\percent} in prior work. Interpolation on a small, low-noise \gls{ms} Ni simulation dataset is similar to interpolation results for the original octonion metric (\SI{57.6}{\percent} vs. \SI{56.4}{\percent}). A vectorized, parallelized, MATLAB interpolation function (\matlab{interp5DOF.m}) and related routines are available in our \vfzorepo{} (\url{github.com/sgbaird-5dof/interp}) which can be applied to other crystallographic point groups. The \gls{vfzo} framework offers advantages for computing distances between \glspl{gb}, estimating property values for arbitrary \glspl{gb}, and modeling surrogates of computationally expensive \gls{5dof} functions and simulations.
		\end{abstract}


	\end{frontmatter}


	\section{Introduction} \label{sec:intro}

	\subsection{Motivation}
	\label{sec:intro:motivation}
	High fidelity \gls{gb} structure-property models can accelerate the design and understanding of materials for \gls{gb} engineering applications such as grain growth (\gls{gbe} \cite{jinColossalGrainGrowth2018}, mobility \cite{brandenburgMigrationFacetingLowangle2014}, and grain rotation \cite{huangGrainRotationLattice2015,trauttCapillarydrivenGrainBoundary2014,sharmaObservationChangingCrystal2012,wareGrainBoundaryPlane2018}), stress-corrosion cracking (diffusivity \cite{liAnisotropyHydrogenDiffusion2017,oudrissGrainSizeGrainboundary2012}, solubility \cite{metsueHydrogenSolubilityVacancy2016}, and segregation \cite{huangHydrogenEmbrittlementGrain2017}) \cite{xiaApplingGrainBoundary2011,demkowiczThresholdDensityHelium2020,hansonCrystallographicCharacterGrain2018,jothiInvestigationMicromechanismsHydrogen2016,zhouChemomechanicalOriginHydrogen2016}, strength \cite{huangNaturalImpactresistantBicontinuous2020,wangAdditivelyManufacturedHierarchical2018,linMeasuringNonlinearStresses2016}, ceramics \cite{yinCeramicPhasesOnedimensional2019,guanAnalysisThreedimensionalMicrostructure2011}, electronics \cite{vlassioukEvolutionarySelectionGrowth2018,hanSubnanometreChannelsEmbedded2018}, and thermoelectrics \cite{sunEnhancingPowerFactor2020}. With the increased use of nanomaterials \cite{hanSubnanometreChannelsEmbedded2018,huangNaturalImpactresistantBicontinuous2020}, \glspl{gb} take on increasingly larger roles as the \gls{gb} volume fraction becomes significant; this is complicated by the fact that properties of \glspl{gb} can span orders of magnitude depending on the five macroscopic \glspl{dof} \cite{johnsonInferringGrainBoundary2015,yangMeasuringRelativeGrain2001,zhangGrainBoundaryMobilities2020} as well as the three microscopic \glspl{dof} \cite{hanGrainboundaryMetastabilityIts2016,weiDirectImagingAtomistic2021}. However, the mentioned studies generally only consider a binary classification of \glspl{gb} or variation of a few \glspl{dof} which represents a small "slice" of the full \gls{gbc} space. Recent advances in high-throughput simulation \cite{bostanabadStochasticMicrostructureCharacterization2016,Homer2019c,Jothi2015h,pirgaziAlignment3DEBSD2019,pirgaziThreedimensionalCharacterizationGrain2015,speidelCrystallographicTextureCan2018,zhangGrainBoundaryMobilities2020,zhengGrainBoundaryProperties2020}, experimental characterization \cite{keinanIntegratedImagingThree2018,Seita2016,speidelCrystallographicTextureCan2018,winiarskiBroadIonBeam2017,zhangGrainBoundaryMobilities2020}, and availability of rich \gls{gb} datasets \cite{kimIdentificationSchemeGrain2011,liAtomisticSimulationsEnergies2019,liRelativeGrainBoundary2009,olmstedSurveyComputedGrain2009,olmstedSurveyComputedGrain2009a,pirgaziThreedimensionalCharacterizationGrain2015,randleFiveparameterGrainBoundary2008,saylorMisorientationDependenceGrain2000,saylorRelativeFreeEnergies2003,yangAtomisticSimulationsEnergies2019,zhengGrainBoundaryProperties2020} warrant high-fidelity structure-property models capable of handling large amounts of input data to aid in the aforementioned applications.

	\subsection{Prior Work}
	\label{sec:intro:prior}
	In prior work, a number of strategies have been developed for predicting\footnote{We use the term "predict" throughout this work to refer to interpolation, inference, and/or extrapolation as some approaches can individually involve multiple prediction types. } \gls{5dof} \gls{gb} properties from experimental or simulated data. Because different works use different validation functions and data, it is difficult to objectively compare their performance. To facilitate meaningful comparisons, in addition to quoting absolute performance in terms of \gls{rmse} or \gls{mae}, we will also report the percent reduction in error compared to a constant-valued control model whose value is chosen as the mean of the respective \inpt{} data.

	Several researchers have taken the approach of discretizing unsymmetrized \gls{5dof} \gls{gbc} space, and then using a least squares objective function and gradient descent to fit a piecewise-constant function, resulting in \gls{5dof} \glspl{gbed} for nickel \cite{liRelativeGrainBoundary2009}, yttria \cite{dillonCharacterizationGrainboundaryCharacter2009}, and copper \cite{randleFiveparameterGrainBoundary2008} based on experimentally characterized 3D microstructures.

	\citet{restrepoUsingArtificialNeural2014} used an \gls{ann} and approximately \num{17000} and \num{51000} Fe bicrystal simulations from \citet{kimIdentificationSchemeGrain2011} as training and validation data, respectively, to achieve \glspl{mae} of \SI{0.0486}{\J\per\square\meter} and approximately \SI{0.09}{\J\per\square\meter} in the best fitted \glspl{ann} for randomly selected and special \glspl{gb}, respectively. If a constant, average value (i.e. average of the \inpt{} \glspl{gbe}) was chosen as the model, the \gls{mae} would be \SI{0.0617}{\J\per\square\meter}, implying that predictions of randomly selected \glspl{gb} were improved by \calcnum{-(0.0486-0.0617)/0.0617*100}\% relative to this simple, control model. Others have combined machine learning approaches with large lists of macroscopic and microscopic descriptors \cite{guziewskiMicroscopicMacroscopicCharacterization2021,huGeneticAlgorithmguidedDeep2020}.

	Recently, a new \gls{gb} representation, \glspl{gbo}, was reported \cite{francisGeodesicOctonionMetric2019} and tested \cite{chesserLearningGrainBoundary2020}. The \gls{gbo} representation is valuable for a number of applications. Most relevant to the present work is the resulting distance metric. The \gls{gbo} distance metric offers an advantage over other metrics in that it "correctly determines the angular distances between \glspl{gb} with a common normal or misorientation" and "closely approximates the geodesic metric on $SO(3) \times SO(3)$ \textit{for all grain boundary pairs} while maintaining the ability to be analytically minimized with respect to the $U(1)$ symmetry" \cite{francisGeodesicOctonionMetric2019}. In this context, \citet{francisGeodesicOctonionMetric2019} derived \gls{oslerp} and provided examples showing that \gls{oslerp} produces smooth, minimum distance paths through \gls{gb} character space between two arbitrary \glspl{gb}.

	\Gls{lkr} (similar to \gls{idw}) involving scaled pairwise distance matrices was later used with \glspl{gbo} to predict properties of arbitrary \glspl{gb} from a set of known values \cite{chesserLearningGrainBoundary2020}. Using \gls{kfcv} with $k=10$ for \num{388} Ni \gls{gbe} simulations \cite{olmstedSurveyComputedGrain2009a} and an optimized scaling parameter, a \gls{rmse} of \SI{0.0977}{\J\per\square\meter} was obtained compared to a constant, average model \gls{rmse} of \SI{0.2243}{\J\per\square\m} (\SI{56.4}{\percent} improvement). Due to computation time of pairwise distance matrices, this approach is currently "limited to datasets with several thousand or fewer" \glspl{gb} \cite{chesserLearningGrainBoundary2020}.

	\subsection{\glsentrytitlecase{vfzo}{long} Framework}
	\label{sec:intro:vfzo}
	We present a new method for interpolating and predicting \gls{gb} properties from a set of measured/calculated values (e.g. \gls{gbe} from \gls{ms} simulations). We term our approach the \gls{vfzo} framework. It is highly efficient and facilitates the use of large data sets to enhance prediction accuracy. We discuss motivation for (\cref{sec:intro:motivation}) and prior implementations of \gls{5dof} property prediction (\cref{sec:intro:prior}) and then highlight unique properties of the \gls{vfzo} framework that offer advantages over other methods (\cref{sec:intro:vfzo}).

	The \gls{vfzo} interpolation framework introduced in this work offers an advantage over other methods because it is defined as a \gls{vfz} point set in a manifold\footnote{"In mathematics, a manifold is a topological space that locally resembles Euclidean space" \cite{morawiecDistancesGrainInterfaces2019}. By removing the Euclidean approximation in the \gls{vfzo} framework, the metric becomes intrinsic \cite{morawiecDistancesGrainInterfaces2019}. } for which directly computed, scaled Euclidean distances approximate the original octonion distance given by \citet{francisGeodesicOctonionMetric2019}. This advantage is manifest in the ability to triangulate a mesh using standard routines (e.g. quickhull \cite{barberQuickhullAlgorithmConvex1996}) and interpolate using barycentric coordinates or machine learning methods such as \gls{gpr}. Building on prior work on \glspl{gbo} \cite{francisGeodesicOctonionMetric2019,chesserLearningGrainBoundary2020}, we create a \gls{vfz} point set by obtaining a set of octonions minimized with respect to Euclidean distance and an arbitrary reference octonion after considering all \glspl{seo}. Because \glspl{gbo} are guaranteed to reside on the surface of a hypersphere \cite{francisGeodesicOctonionMetric2019} (a type of Riemannian manifold) a point set which locally resembles Euclidean space is the result (\cref{sec:methods:framework:vfz-dist}). Below we provide the detailed description of the method, followed by numerical test results (\cref{sec:results}).

	We also provide a vectorized, parallelized implementation of the \gls{vfzo} framework and related functions. These are contained in what we will refer to as the \vfzorepo{}, which is available at \url{github.com/sgbaird-5dof/interp}. In what follows, when we refer to built-in MATLAB functions, we refer to them with parentheses as in \matlab{interp1()}. When we refer to functions in the \vfzorepo{}, we do so with the \matlab{.m} extension as in \matlab{interp5DOF.m} unless specifying the usage with arguments as in \matlab{interp5DOF(qm,nA,qm2,nA2,y)}.

	\section{Methods} \label{sec:methods}

	We describe methods related to the \gls{vfzo} framework  (\cref{sec:methods:framework:vfz}), generation of random \glspl{gb} (\cref{sec:methods:rand}), and four different \gls{gb} property interpolation schemes (\cref{sec:methods:interp}). We also describe details regarding two simulated literature datasets that we use (\cref{sec:methods:litdata}).

	\subsection{The \glsentrytitlecase{vfzo}{long} Framework}
	\label{sec:methods:framework}

	The core operations of the \gls{vfzo} framework are:
	\begin{enumerate}
		\item generating \glspl{gbo} (\cref{sec:methods:framework:vfz})
		\item mapping \glspl{gbo} into a \gls{vfz} (\cref{sec:methods:framework:proj})
		\item calculating distances within the \gls{vfz} (\cref{sec:methods:framework:vfz-dist})
	\end{enumerate}

	\subsubsection{Defining the \glsentrytitlecase{vfz}{long}}
	\label{sec:methods:framework:vfz}

\begin{figure*}
		\centering
		\includegraphics[scale=1]{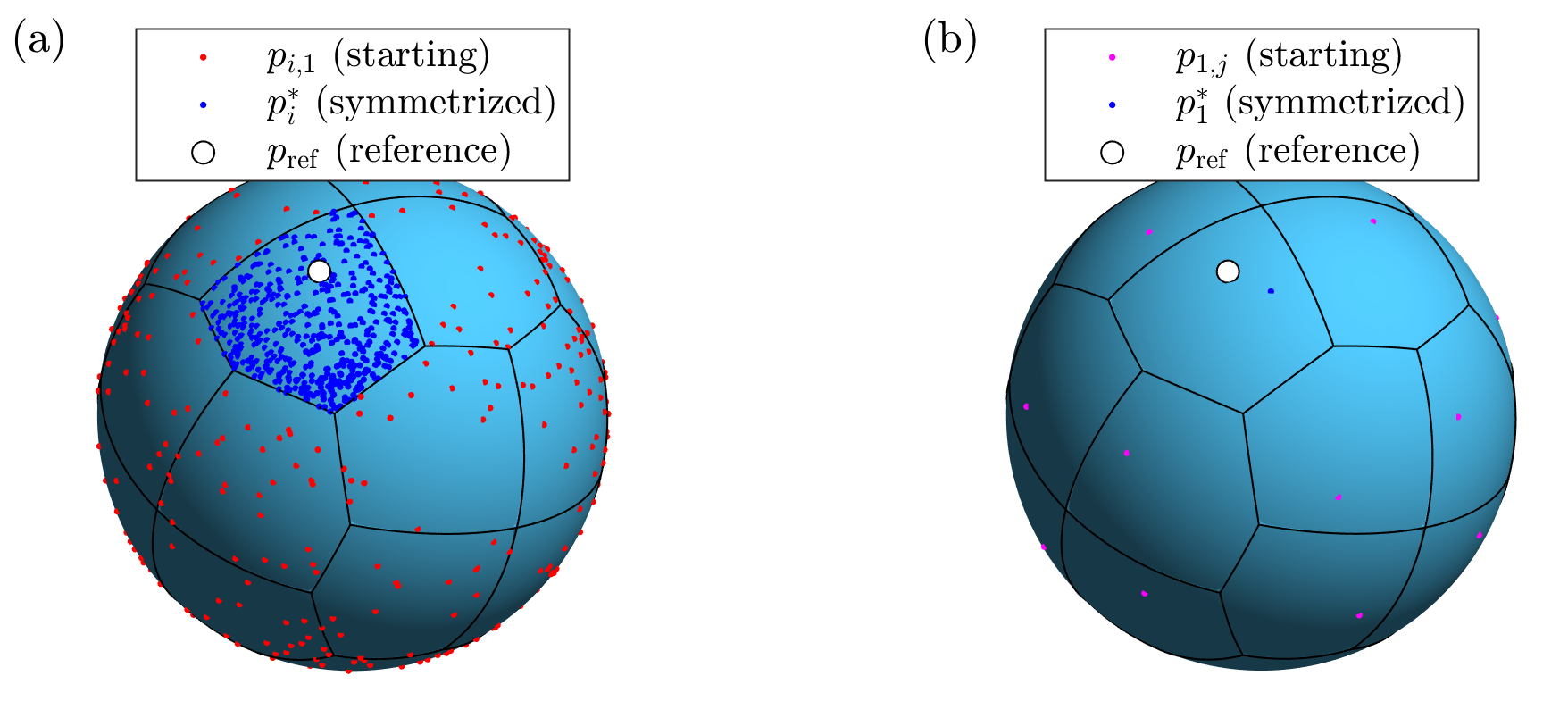}
		\caption{(a) 3D Cartesian analogue to a non-degenerate 7D Cartesian representation of U(1)-symmetrized \glspl{gbo} and \glspl{vfzo} (\glspl{vfzo} are inherently U(1)-symmetrized) which demonstrates the symmetrization of many points relative to a fixed reference point (\refpt{}). This produces a 3D Cartesian \gls{vfz} point set (\sympt{}). (b) To further illustrate, a single input point (\singlept{}) is symmetrized (\singlesympt{}) relative to a fixed reference point (\refpt{}), demonstrating that only one symmetrized point is found within the borders (\vbordercolor{}) of each of the Voronoi cells (\vcellcolor{}). The Voronoi tessellation is defined by the symmetric images of the reference point, and the spherical Voronoi diagram for this illustration is constructed using a modified version of \cite{luongVoronoiSphere2020}.}
		\label{fig:voronoi}
	\end{figure*}

	\Gls{3dof} \glspl{fz} have typically been defined using linear inequalities (e.g. the orientation \cite{heinzRepresentationOrientationDisorientation1991} and misorientation \cite{grimmerUniqueDescriptionRelative1980,heinzRepresentationOrientationDisorientation1991} \glspl{fz}). Instead of using linear inequalities\footnote{If desired, linear inequalities can be obtained for a \gls{vfz} by determining a Voronoi tessellation's junction points (similar to what is shown in \cref{fig:voronoi} by e.g. \matlab{voronoin()}), transforming to 6D Cartesian coordinates via a \gls{svd} transformation (\cref{sec:app:bary}) and defining the bounded region by e.g. MATLAB FEX function \matlab{vert2lcon.m}.}, we take a numerical approach to define what we will call a \gls{vfz}.

	To define a \gls{vfz}, an arbitrary, fixed, low-symmetry reference \gls{gbo} is chosen ($o_{\text{ref}}$) and the \gls{vfz} is formally defined as the region of $\mathbb{S}^7$ (the unit 7-sphere in 8 dimensions) closer to $o_{\text{ref}}$ than any of its symmetric images\footnote{We also refer to lower-dimensional representations of the 8D Cartesian \gls{vfz} as \glspl{vfz} (described in \cref{sec:methods:framework:proj}) and describe which dimensionality we are referring to as appropriate. }. However, use of the \gls{vfz} does not require its explicit construction. Rather, practical calculations require only the selection of the single point $o_{\text{ref}}$ (which completes the definition of the \gls{vfz}), followed by mapping of query points into the \gls{vfz} by comparison of their \glspl{seo} with $o_{\text{ref}}$.

	To illustrate the process of mapping points into the \gls{vfz}, we describe a 3D Cartesian analogue (\cref{fig:voronoi}) to a 7D Cartesian non-degenerate (i.e. U(1) degeneracy removed) representation of a \gls{vfz}. A set of \num{500} points ($p_i, i\in[1,500]$) randomly scattered on the surface of the 2-sphere comprise the data (\startpt{} in \cref{fig:voronoi}a). A random point, $p_{\text{ref}}$, also on the surface of the 2-sphere, is chosen as the reference point (\refpt{}). In this illustration, $O_h$ or $m\bar{3}m$ point group rotations are used as symmetry operators, $S_j,\ j\in[1,N_p]$, where $N_p$ is the number of proper rotations as before and $N_p = 24$ for the $O_h$  point group. For each data point, \num{24} symmetrically equivalent representations ($p^{\text{sym}}_{i,j} = S_j(p_i),\ j\in[1,24]$) are produced by applying each of the relevant symmetry operators. After calculating the Euclidean distance between $p_{\text{ref}}$ and $p^{\text{sym}}_{i,j}$, the point ($p^{*}_i$) closest to $p_{\text{ref}}$ is chosen and retained as the unique representative of $p^{\text{sym}}_{i,j}$. As illustrated in \cref{fig:voronoi}a, the projected points $p^{*}_i$ (dark blue points) all fall in the \gls{vfz} without ever having to construct or define it explicitly, we call this group of projected points a \textit{\gls{vfz} point set}. Note also that there is only one $p^{*}_i$ in the \gls{vfz} for each $p^{\text{sym}}_{i,j}$ (see \cref{fig:voronoi}b).

	To calculate the distance between a given octonion, and the reference octonion, we employ the standard 8D Euclidean distance
	\begin{equation}
		\label{eq:8Deuclidean_dist}
		d_{\text{E}}\!\left(o_{A},o_{B}\right) = {\left(\sum_{k=1}^{8} {\left(o_{A,k} - o_{B,k}\right)}^2 \right)}^{1/2}
	\end{equation}
	where $o_{A,k}$ and $o_{B,k}$ represent the $k$-the element of normalized octonions $o_A$, and $o_B$, respectively.

	Euclidean distance is an approximation to the true geodesic arc length on $\mathbb{S}^7$, which is given by
	\begin{equation}
		\label{eq:7sphere_arc_length}
		d_{\text{S}}\!\left(o_{A},o_{B}\right)=\cos ^{-1}\left(o_A\cdot o_B\right)
	\end{equation}
	where $\cdot$ is the dot product, $\cos ^{-1}$ is the inverse cosine operator, and $o_A$ and $o_B$ are each normalized and $d_{\text{S}}\simeq d_{\text{E}}$ (\cref{fig:dist-parity}). In \cite{francisGeodesicOctonionMetric2019}, the original octonion distance metric was defined by
	\begin{equation}
		\label{eq:omega}
		d_\Omega\!\left(o_{A},o_{B}\right) = 2\cos ^{-1}\left(o_A\cdot o_B\right)
	\end{equation}
	where $o_A$ and $o_B$ are each normalized and $d_\Omega$ can be seen to be simply twice the geodesic arc length: $d_\Omega = 2 d_{\text{S}}$. Thus,
	$d_\text{E}\simeq \frac{1}{2}d_\Omega$.

	The definition of $d_\Omega$ has certain aesthetic benefits in that it mirrors the definition of a misorientation angle, $\omega_{AB}$, between two crystal orientations in the quaternion parameterization: $\omega_{AB} = 2 \cos^{-1}{\left(q_A \cdot q_B\right)}$.

	Our choice to use $d_{\text{E}}$ instead of $d_{\text{S}}$ or $d_\Omega$ is motivated by the fact that it enables the use of standard algorithms, for a variety of operations, that require or assume Euclidean distances. In addition to enabling us to leverage the machinery of efficient and established algorithms, this choice can be justified by the following observations:
	\begin{itemize}
		\item The minimum Euclidean distance \gls{seo} will be the same as the minimum arc length distance \gls{seo} because $d_{\text{S}}$ is a monotonically increasing function of $d_{\text{E}}$, for $d_{\text{S}}\!\left(d_{\text{E}}\right)\in[0,\pi]$ (\cref{fig:dist-parity}).
		\item For the FCC point group symmetry ($m\bar{3}m$) the portion of $\mathbb{S}^7$ subtended by the \gls{vfz} is sufficiently small that the approximation $d_{\text{E}} \simeq d_{\text{S}}$ holds to very high accuracy\footnote{This is true for a specific pair of octonions within a \gls{vfz}. When calculating the \emph{minimum} distance between \glspl{seo} of two points, there are additional considerations that must be attended to as discussed in detail in \cref{sec:methods:framework:vfz-dist}.} as shown in \cref{fig:dist-parity}.
		\item Calculation of $d_{\text{E}}$ does not require the use of any inverse trigonometric functions and is about \SI{23}{\percent} faster than calculation of $d_{\text{S}}$ or $d_\Omega$.
	\end{itemize}

	For applications other than interpolation which require precise quantification of high-dimensional volume, a mapping between Euclidean-approximated volumes and true volumes may be necessary\footnote{We have not tested to what extent a Euclidean-approximated volume will differ from the true volume; however, Euclidean-approximated volumes can be obtained by using the triangulation methods discussed in \cref{sec:app:bary:tri} (i.e. \matlab{convhulln()}.} or the Euclidean approximation may be removed altogether\footnote{i.e. by setting the matlab{dtype} argument of \matlab{GBdist4.m} to \matlab{'omega'} rather than \matlab{'norm'}.}. The latter allows for the (non-ensembled) \gls{vfzo} metric to be intrinsic (see \citet{morawiecDistancesGrainInterfaces2019} for an in-depth treatment of intrinsicality).

	The expectation that a single, unique \gls{seo} will be found (within numerical tolerance and given a low-symmetry reference \gls{gbo}\footnote{The probability that a randomly generated \gls{gbo} will fall exactly on a high-symmetry boundary vanishes in the limit of infinite precision. }) is verified by several manual tests and internally within the symmetrization sub-routine \matlab{get\_octpairs.m} \cite{bairdFiveDegreeofFreedom5DOF2020} that is part of the \matlab{interp5DOF.m} package. Similar numerical tests reveal that inappropriately selecting a high-symmetry reference \gls{gbo} to (attempt to) define a \gls{vfz} results in many degenerate minimum distance \glspl{seo}, with the identity octonion ($\{1,0,0,0,0,0,0,0\}\in\mathbb{R}^8$) \cite{francisGeodesicOctonionMetric2019} giving the highest degeneracy.

	\subsubsection{Mapping \glsfmtshortpl{gb} to the \glsentrytitlecase{vfz}{long}}
	\label{sec:methods:framework:proj}

	As described above in the 3D analogy, with a reference \gls{gbo} chosen ($o_{\text{ref}}$), and consequently the \gls{vfz} defined (\cref{sec:methods:framework:vfz}), a \gls{gbo} is mapped into the \gls{vfz} by finding among its \glspl{seo} the one that is closest to $o_{\text{ref}}$ according to $d_{\text{E}}$ (\cref{eq:8Deuclidean_dist}). This is performed for all \inpt{} and \outpt{} points with respect to $o_{\text{ref}}$, and the result is a \gls{vfz} point set.

	\begin{figure*}
		\centering
		\includegraphics[scale=1]{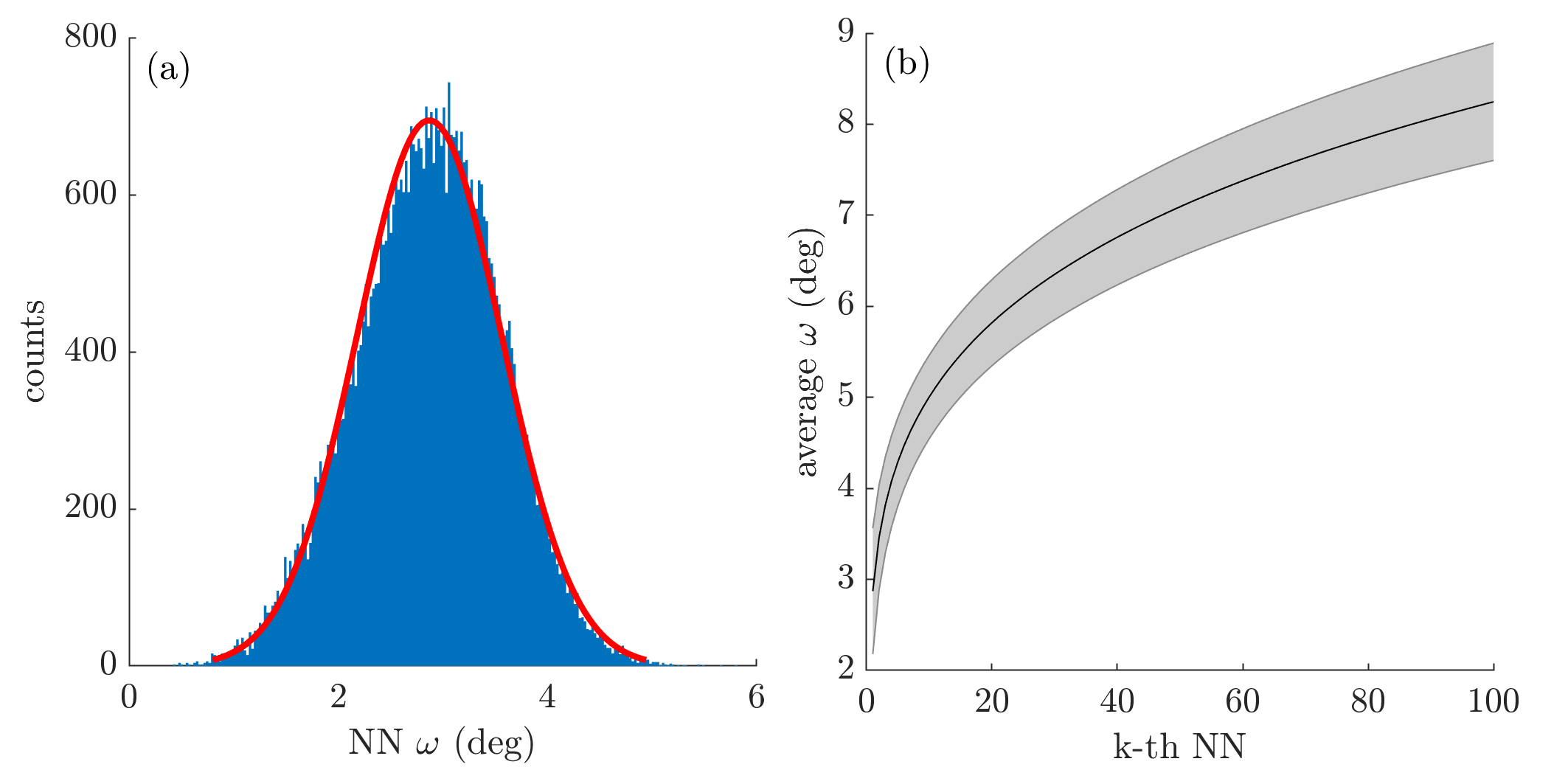}
		\caption{(a) Histogram of \gls{nn} octonion distances ($\omega$) in a \gls{vfzo} set of \num{50000} points. The average \gls{nn} distance was \SI{\nnomega}{\degree}. (b) The average k-th nearest neighbor distances demonstrate that many nearest neighbors fall within a tight tolerance (less then \SI{10}{\degree}) out of approximately 10 trial runs.}
		\label{fig:nnhist-knn-50000}
	\end{figure*}

	\subsubsection{Distance Calculations in the \glsentrytitlecase{vfz}{long}}
	\label{sec:methods:framework:vfz-dist}

\begin{figure*}[!ht]
		\centering
		\includegraphics[scale=1]{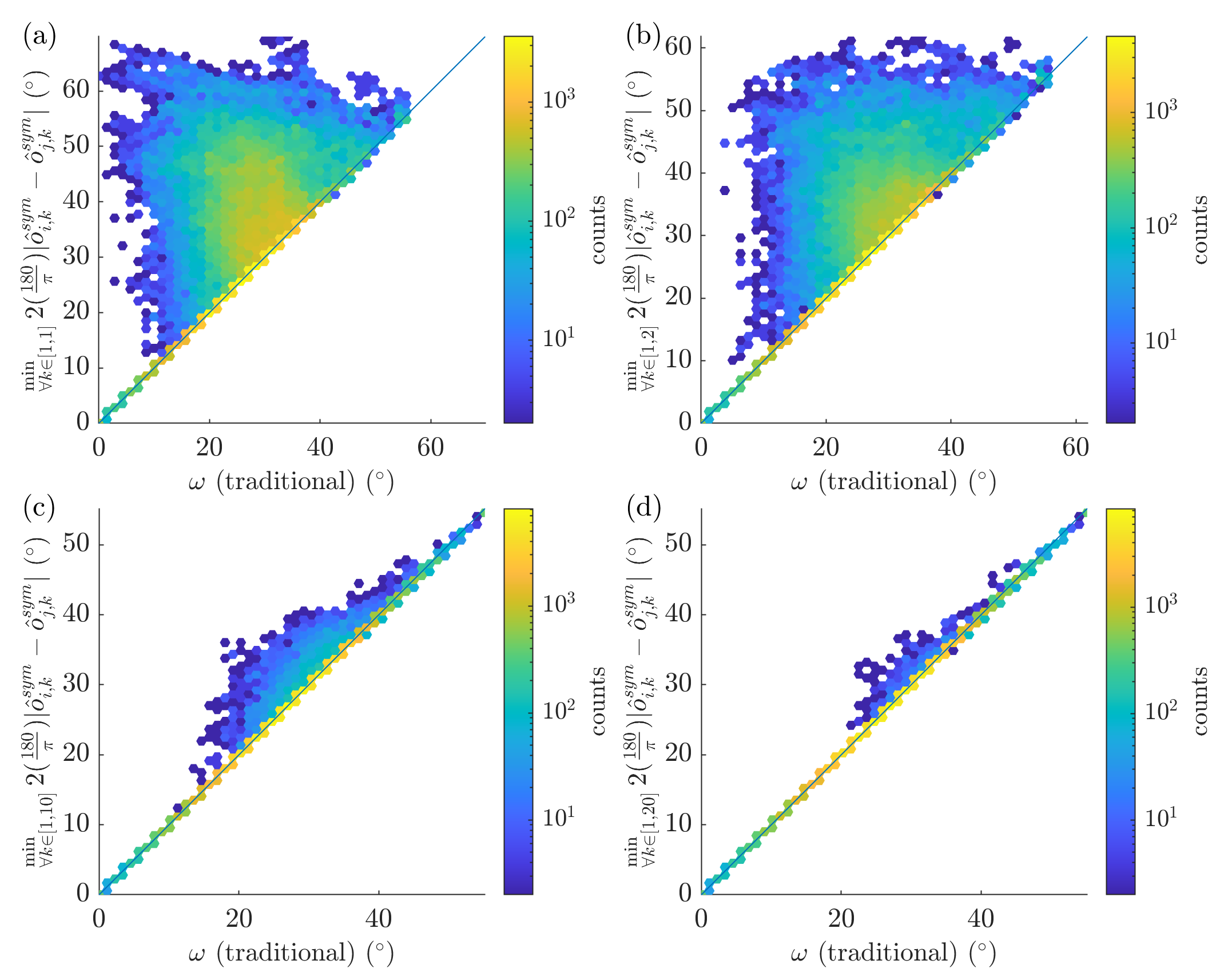}
		\caption{Hexagonally binned parity plots of pairwise distances of 388 Ni bicrystals \cite{olmstedSurveyComputedGrain2009a}. Euclidean distance approximation is converted to octonions ($x_{i,j,k}=2\left(\frac{180}{\pi}\right)|\hat{o}_{i,k}^{\text{sym}}-\hat{o}_{j,k}^{\text{sym}}|$) for comparison with the traditional octonion metric \cite{chesserLearningGrainBoundary2020}. The minimum distance among an ensemble of \gls{vfzo} sets ($\min_{\forall k \in [1,k_{max}]}x_{i,j,k}$) is used for (a) 1, (b) 2, (c) 10, and (d) 20 \gls{vfzo} sets. As the number of \gls{vfzo} sets increases, the correlation between the Euclidean distance and the traditional octonion distance improves.}
		\label{fig:dist-ensemble-k1-2-10-20}
	\end{figure*}

	Euclidean distances are an accurate approximation of arc length distances in a \gls{vfz} because the difference between the two metrics for the maximum pairwise distance ($pd_{max} \simeq \SI{60}{\degree}$) in a \gls{vfz} is small as shown in \cref{fig:dist-parity}. However, when compared with the traditional octonion distance \cite{francisGeodesicOctonionMetric2019}, due to the presence of low-symmetry \glspl{gb} near the exterior of a \gls{vfz}, some \gls{gb} pairs will exhibit larger Euclidean or arc length distances than is truly representative (see e.g. \cref{fig:dist-ensemble-k1-2-10-20}a). In other words, moving "past" the low-symmetry border of a \gls{vfz} will result in an instantaneous relocation to a possibly distant point in the \gls{vfz} that in reality is highly correlated.

	This is a limitation of the \gls{vfzo} framework, which generates a \gls{vfz} with low-symmetry \glspl{gb} at the borders in contrast to typical \glspl{fz} \cite{patalaSymmetriesRepresentationGrain2013,homerGrainBoundaryPlane2015}. While defining a \gls{fz} with high-symmetry \glspl{gb} at the borders (especially mirror-symmetry \glspl{gb}) will certainly increase interpolation accuracy, the favorable interpolation results presented in this work are obtained because overestimation is infrequent within a small correlation length (e.g. \SI{10}{\degree} \cite{olmstedSurveyComputedGrain2009}, which many \glspl{nn} fall within for a \num{50000} \gls{vfzo} set, see \cref{fig:nnhist-knn-50000}b), and underestimation is non-existent within numerical precision. Naturally, smaller dataset pairwise distance matrices will exhibit more frequent distance overestimation.

	Overestimation imposes a "sparseness" of data within a local region of influence common to the interpolation methods in this work, whereas underestimation would give erroneous high correlations between uncorrelated \glspl{gb}. Because only overestimation relative to traditional octonion distances exist in this work (as shown in \cref{fig:dist-ensemble-k1-2-10-20}), we expect that large errors will occur infrequently (\cref{sec:results:accuracy}).

	While distance calculations are subject to these infrequent overestimates, they are largely immaterial for interpolation. This is because all interpolation methods in this work involve a region of influence that is small, so that if the distance to a \gls{nn} is overestimated it simply does not contribute to the interpolation (the "sparseness" referred to earlier). Consequently the accuracy of the interpolation is not significantly impacted by infrequent distance overestimates, and excellent results can be achieved without addressing this limitation. However, if even greater accuracy is desired it can be obtained for a relatively minor cost by considering multiple \glspl{vfz}.

\begin{figure}[ht]
		\centering
		\includegraphics[scale=1]{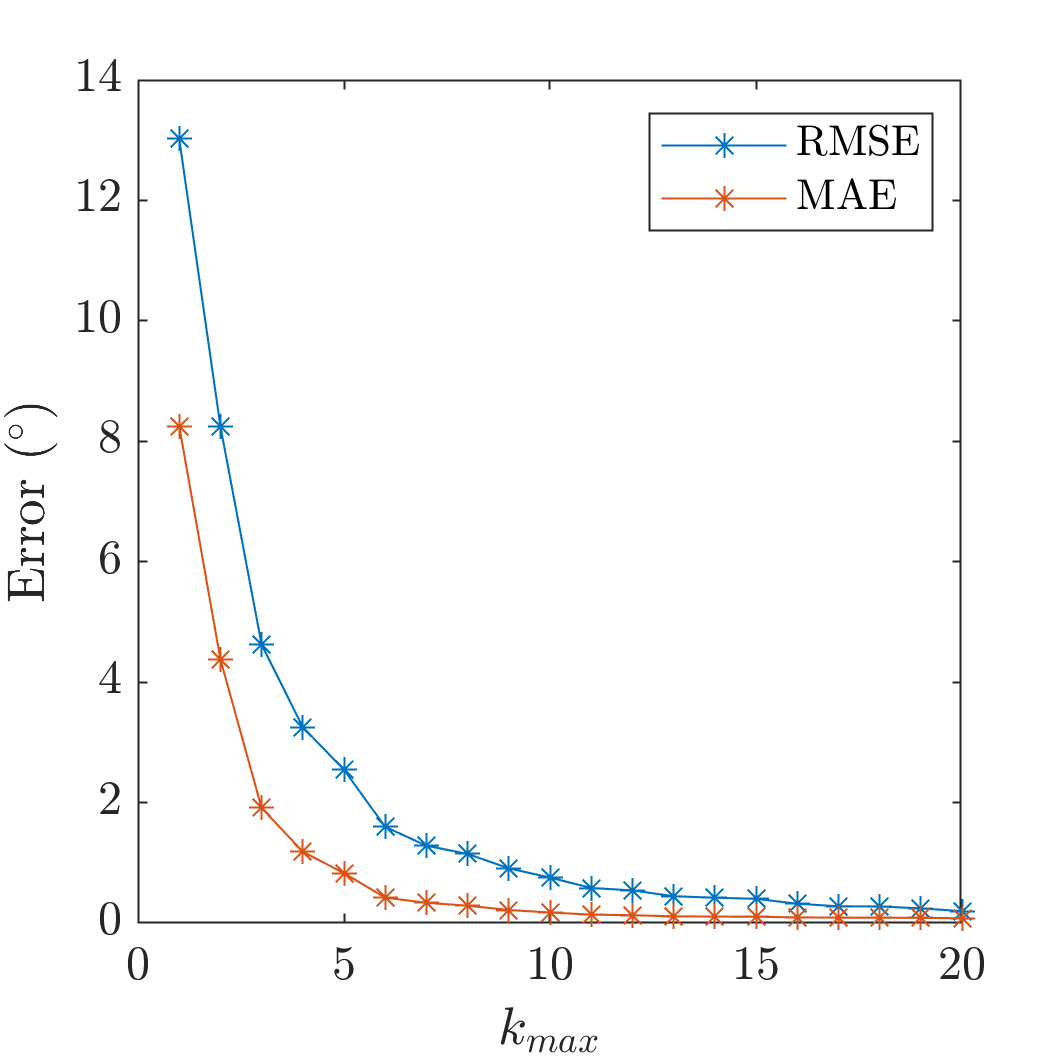}
		\caption{\Gls{rmse} and \gls{mae} of pairwise distance errors for 388 Ni bicrystals \cite{olmstedSurveyComputedGrain2009} of scaled Euclidean distance approximation relative to the traditional octonion metric \cite{chesserLearningGrainBoundary2020} (compare with \cref{fig:dist-ensemble-k1-2-10-20}). The minimum distance among an ensemble of \gls{vfzo} sets ($\min_{\forall k \in [1,k_{max}]}x_{i,j,k}$, where $x_{i,j,k}$ is the scaled Euclidean distance) is taken, iteratively adding consecutive sets up to $k_{max} = 20$. As the number of \gls{vfzo} sets increases, \gls{rmse} and \gls{mae} between the scaled Euclidean distance approximation and the traditional octonion distance decreases.}
		\label{fig:dist-ensemble-rmse-mae}
	\end{figure}

	We find that taking the minimum distance among several \gls{vfzo} sets defined by separate reference octonions leads to better correlation between the Euclidean approximation and the traditional octonion metric as shown in \cref{fig:dist-ensemble-k1-2-10-20}. Additionally, \cref{fig:dist-ensemble-rmse-mae} shows that the error between scaled Euclidean distance and the traditional octonion metric decreases rapidly as the number of ensemble \gls{vfzo} components increases. This confirms that employing a small ensemble of \gls{vfzo} sets results in significant improvement to the Euclidean distance approximation (\cref{fig:dist-ensemble-k1-2-10-20,fig:dist-ensemble-rmse-mae}) of the traditional octonion metric. However, as already mentioned, improvements to interpolation results are expected to be less significant since they are already robust to occasional distance overestimates. In terms of computational runtime, use of an ensemble of 10 \glspl{vfz} will increase runtime by a factor of $\sim$10 via a loop-based implementation. For a symmetrized $\num{50000}\times\num{50000}$ pairwise distance matrix, this results in a runtime of approximately 1 CPU hour instead of $\sim$7 CPU minutes for a single \gls{vfz}. However, this is still much faster than the original octonion approach used in \cite{chesserLearningGrainBoundary2020}, which would take an estimated 6.6 CPU years using the original implementation (or 153 CPU days if one \gls{gb} in the \gls{gb} pair is fixed according to the assumption in \citet{morawiecDistancesGrainInterfaces2019}). Additionally, it may be worthwhile to make the distance calculations GPU-compatible for further speed-up. 

	\Gls{vfzo} Euclidean, hyperspherical arc length, and octonion distances are computed via \vfzorepo{} function \matlab{GBdist4.m} which is used in the symmetrization function \matlab{get\_octpairs.m} and an example of ensemble \gls{vfzo} distance calculations is given in \matlab{plotting.m}.

	In addition to their use for distance calculations alone, ensembles of \gls{vfzo} sets can be employed with interpolation methods to increase overall interpolation accuracy, but there is a computational cost (e.g. approximately 10$\times$ using an ensemble of 10 \gls{vfzo} sets). For \num{50000} \inpt{} points, use of an ensemble with 10 \gls{vfzo} sets decreases \gls{rmse} and \gls{mae} from \SIlist{0.0241;0.0160}{\J\per\square\m} to \SIlist{0.0187;0.0116}{\J\per\square\m}, respectively (single trial run). We expect these overall accuracy improvements occur because \gls{gbe} predictions near the exterior of the \gls{vfz} where data may be sparse are improved. Ensemble interpolation results as a function of ensemble size and parity plots for mean, median, minimum, and maximum functions applied to the ensemble are shown in \cref{fig:ensemble-interp-rmse-mae} and \cref{fig:ensemble-interp}, respectively. Further details of ensemble interpolation are given in \cref{sec:ensemble-interp}.

	\subsubsection{Comparison with Traditional Octonion Framework}

	We compare the \gls{vfzo} framework with the traditional octonion metric (\cref{tab:closed-mesh-comparison}) and give examples that illustrate the computational complexity of each approach.

	\begin{table*}
		\caption{Comparison between \glsxtrlong{vfzo} and traditional octonion frameworks. *6D Cartesian representation used only for mesh triangulation efficiency in barycentric interpolation and *7D Cartesian representation only required for barycentric interpolation. 7D Cartesian representation is also implemented (though not required) for \gls{gpr}, \gls{nn}, and \gls{idw}. For pairwise distance complexity, $N_p$ is the number of proper rotations ($N_p=24$ for $m\Bar{3}m$ \gls{fcc} point group) and $L$ is the number of \glspl{gb}.}
		\centering
		\begin{tabular}{ccc}
			\toprule
			Property & Traditional & This Work \\
			\midrule
			Symmetrizing Distance & \gls{gbo} & \gls{vfz} Euclidean \\
			Dimensionality & 8D Cartesian & 6*/7*/8D Cartesian \\
			Bounded by \gls{fz} & No & Yes \\
			Pairwise Distance Complexity & $O(N_p^2L^2)$ & $O(N_p^2L)$ \\
			Rotation Convention & Passive & Active \\
			\bottomrule
		\end{tabular}
		\label{tab:closed-mesh-comparison}
	\end{table*}

	The construction of the \gls{vfz} dramatically reduces the computational burden of pairwise distance calculations. The mechanism by which this reduction is achieved can be illustrated with an example. Let $o_1$ and $o_2$ denote two \glspl{gb} represented in \gls{gbo} coordinates.
	To perform a traditional symmetrized \gls{gbo} distance calculation according to \citet{francisGeodesicOctonionMetric2019}, we compare all \glspl{seo} of $o_1$ to all of the \glspl{seo} of $o_2$ and take the smallest distance. If $N_p$ is the number of proper rotations of the crystallographic point group, this single minimum distance calculation requires a total of $4N_p^4$ \glspl{seo} to be considered (Sections 4.3 and 4.5 of \citet{francisGeodesicOctonionMetric2019}). The total number of \gls{seo} computations will be $4N_p^4L^2$. However, it is possible to fix a single \gls{gb} in the \gls{gb} pair and still obtain accurate\footnote{Compared with the pairwise distance matrix of the 388 Olmsted \glspl{gb}, we obtained a \gls{rmse} of \SI{1.6566E-7}{\degree} for this computation which completed in \SI{133}{\s} using 6 cores (see \matlab{get_pd_fix.m})} due to isometry equivalence (see Section 7 of \cite{morawiecDistancesGrainInterfaces2019} and \cref{fig:pd-fix}).

	In contrast, for a single distance calculation using the \gls{vfzo} framework, $o_1$ and $o_2$ are first mapped into the \gls{vfz}, and then only a single distance calculation is required between them. Mapping $o_1$ into the \gls{vfz} requires comparison of $8N_p^2$ \glspl{seo}\footnote{This is 8 instead of 4 because the simplifying assumption that only two of the four double cover cases need to be considered \cite{francisGeodesicOctonionMetric2019} does not apply in the \gls{vfzo} framework. This is confirmed by applying \matlab{uniquetol()} on a set of $4608$ octonions which has a final set size of $4608$, where $4608=8\times N_p^2$ and $N_p=32$ (see \matlab{osymset.m}).} of $o_1$ with a fixed reference \gls{gb} in the interior of the \gls{vfz}; and likewise for $o_2$. Consequently, a single distance calculation between $o_1$ and $o_2$ under the \gls{vfzo} framework requires $O(N_p^2)$ \gls{seo} computations. If one desires to compute a pairwise distance matrix between $L$ \glspl{gb}, the total computational cost\footnote{See \cref{sec:results:efficiency:symruntime} for a detailed explanation of why this is \emph{not} $O(N_p^2L^2)$.} will be $O(N_p^2L)$, which represents a dramatic reduction compared to the traditional approach. A summary of the differences between the two approaches is provided in \cref{tab:closed-mesh-comparison}.

	\subsection{Generating Random \glsentrytitlecase{vfzo}{long}s}
	\label{sec:methods:rand}
	In addition to the 3 core operations of the \gls{vfzo} framework described in \cref{sec:methods:framework}, it will be necessary for our tests, and useful for other applications, to be able to generate random \glspl{gbo} from \gls{5dof} representations. We briefly explain here our process for accomplishing this.

	First, random \glspl{gbo} are formed by taking random misorientation quaternion (\matlab{qm}) and \gls{bp} normal (\matlab{nA}) pairs. Random misorientation quaternions are obtained via cubochoric sampling \cite{singhOrientationSamplingDictionarybased2016} (\matlab{get\_cubo.m}) and random \gls{bp} vectors are sampled from a multivariate Gaussian distribution ($\mu=0$, $\sigma=1$) in $\mathbb{R}^3$ and normalized\footnote{Several methods for uniform sampling of points on a sphere, including the one mentioned here, are described in \url{https://mathworld.wolfram.com/SpherePointPicking.html}.}. After this, they are converted to \glspl{gbo} via \vfzorepo{} function \matlab{five2oct.m}. The \vfzorepo{} function \matlab{get\_five.m} returns the result of these several operations. These (\matlab{qm},\matlab{nA}) pairs are then converted to an octonion representation, \matlab{o}, using \vfzorepo{} function \matlab{o=five2oct(qm,nA)} (see also \vfzorepo{} function \matlab{get\_ocubo.m} for generating random \glspl{gbo} directly).

	The \glspl{gbo} are then symmetrized (i.e. they become \glspl{vfzo}) via \matlab{osym=get\_octpairs(o)}. A default reference octonion\footnote{This is generated by \matlab{get\_ocubo.m} using a random number generator seed of 10. We expect that \matlab{five2oct.m} combined with \matlab{get\_five.m} will generate near identical statistical properties to \matlab{get\_ocubo.m} which is supported by a visual comparison of pairwise distance histograms (not shown in this work), and indirectly by an assertion in Section 5.3 of \citet{morawiecDistancesGrainInterfaces2019}. } is used for these calculations, unless specified by the user. We use the active convention for \matlab{qm}, \matlab{nA}, and \matlab{o} (see \cref{sec:app:convention} for further details of conventions).

	For the present work we use this procedure to randomly generate \gls{vfzo} sets containing between \num{100} to \num{50000} \glspl{vfzo} where each trial run has its own unique set of \glspl{gb}. We use these to perform the validation and performance evaluation tests described later. For reference, we note that the average \gls{nn} distance (over approximately 70 trials) of such sets ranges between \SI{10.7175 \pm 0.3684}{\degree} and \SI{2.6479 \pm 0.2254}{\degree}, respectively.
	\begin{figure}
		\centering
		\includegraphics[scale=1]{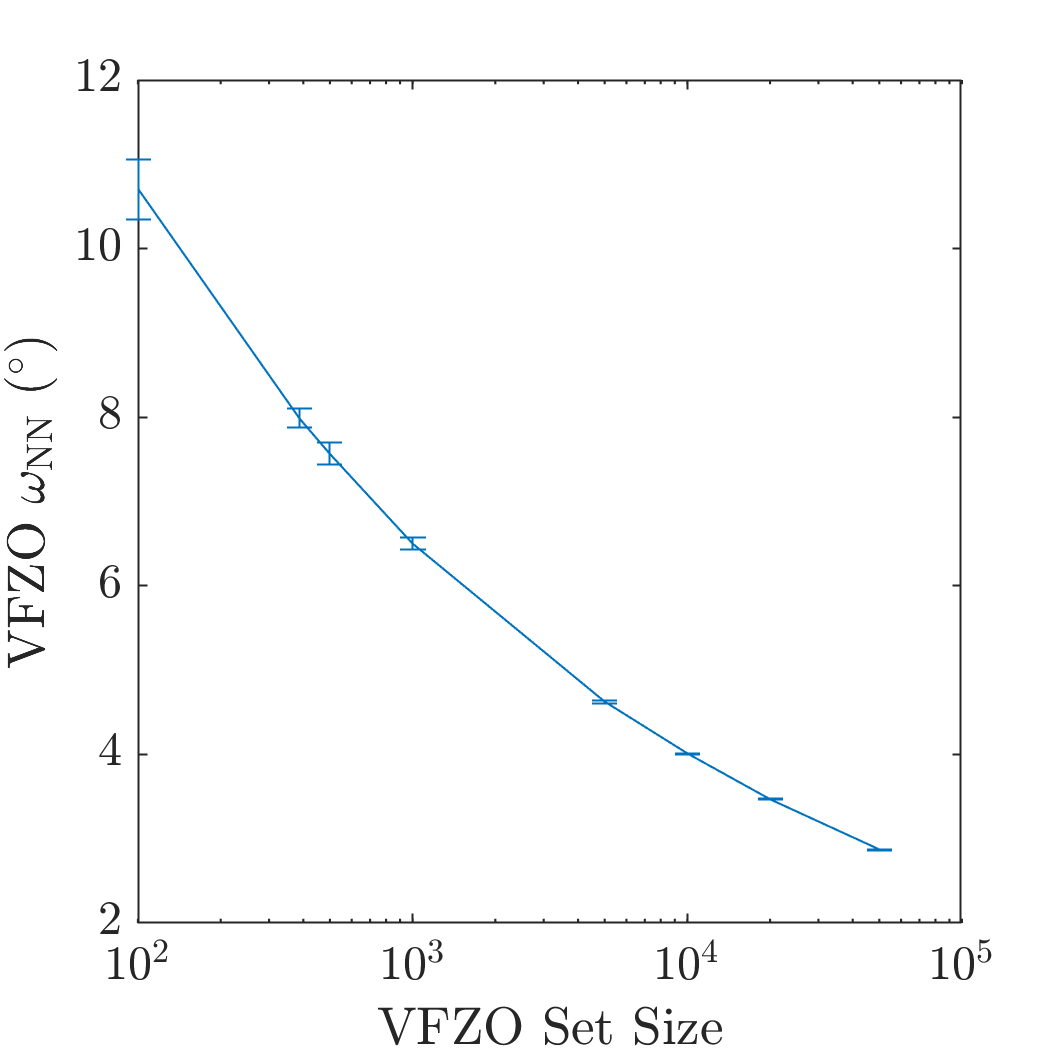}
		\caption{\Gls{nn} \gls{vfzo} ($\omega_{\text{NN}}$) distances ($^{\circ}$) versus \gls{vfzo} set size out of 70-80 random \gls{vfzo} sets per set size.}
		\label{fig:nndist-vs-setsize}
	\end{figure}

	\Cref{fig:nndist-vs-setsize} illustrates how the \gls{vfzo} average \gls{nn} distance varies with the cardinality of the set (i.e. number of random \glspl{vfzo} in the set). For a specific \num{50000} \gls{vfzo} set, the \gls{nn} octonion distance is \SI{\nnomega{}}{\degree} (\cref{fig:nnhist-knn-50000}a) while the average 100-th \gls{nn} distance is within \SI{10}{\degree} (\cref{fig:nnhist-knn-50000}b). This indicates that, on average, \outpt{} \glspl{vfzo} fall within a typical \gls{gb} correlation length (\SI{10}{\degree} \cite{olmstedSurveyComputedGrain2009}) of \inpt{} \glspl{vfzo} in large set sizes.

	\subsection{Interpolation in the \glsentrytitlecase{vfzo}{long} Framework}
	\label{sec:methods:interp}

	With the \gls{vfzo} framework established, it is possible to define interpolation schemes over the \gls{vfz} to predict the properties of new \glspl{gb} from the known properties of other \glspl{gb}. For one application of interest to us, it is necessary to evaluate multiple different functions over a fixed set of \inpt{} and \outpt{} \glspl{gbo}. In this section we first present a barycentric interpolation method that we have developed to efficiently accomplish this specialized task by pre-computing the interpolation weights (which remain fixed when only the function being evaluated changes). We then present adaptations of three other interpolation methods---\gls{gpr} (\cref{sec:methods:interp:gpr}), \gls{idw} (\cref{sec:methods:interp:idw}), \gls{nn} (\cref{sec:methods:interp:nn})---that are useful for general applications (an additional interpolation method---\gls{gprm}---which we developed specifically for a non-uniformly distributed, noisy, simulation dataset is described in \cref{sec:methods:gprmix}). 
	Usage instructions for the \gls{vfzo} repository can be found at the GitHub page (\url{github.com/sgbaird-5dof/interp}) and in \cref{sec:methods:repofn}.

	\subsubsection{Barycentric Interpolation}
	\label{sec:methods:interp:bary}

	Barycentric coordinates are a type of homogeneous coordinate system that reference a \outpt{} point within a simplex \cite{langerSphericalBarycentricCoordinates2006} or convex polytope \cite{floaterGeneralizedBarycentricCoordinates2015,meyerGeneralizedBarycentricCoordinates2002,langerSphericalBarycentricCoordinates2006} based on "masses" or weights at the vertices, which can be negative. The \outpt{} point is assumed to be the barycenter (center of mass) of the simplex or convex polytope, and weights at the vertices necessary to make this assumption true are determined. We utilize rigid \gls{svd} transformations and a standard triangulation algorithm (quickhull \cite{barberQuickhullAlgorithmConvex1996} via \matlab{delaunayn()} in \vfzorepo{} function \matlab{sphconvhulln.m}) to define a simplicial mesh (\cref{sec:app:bary:tri}). We then use barycentric weights (i.e. coordinates) for computing intersections of a point within a simplicial facet (\cref{sec:app:bary:int}) and for interpolation (\cref{sec:app:bary-interp}) \cite{langerSphericalBarycentricCoordinates2006}. A detailed explanation of the process is provided in \cref{sec:app:bary}. 

	\subsubsection{\glsentrytitlecase{gpr}{long}}
	\label{sec:methods:interp:gpr}

	\Gls{gpr} or Kriging uses the notion of similarity between points to fit Gaussian processes (random variables) to data based on prior information and provides uncertainty information in addition to interpolated or inferred values. For a general treatment of \gls{gpr}, see \citet{rasmussenGaussianProcessesMachine2006}. We use MATLAB's built-in function, \matlab{fitrgp()}, with all default parameters\footnote{MATLAB R2020b was used for the Fe simulation dataset, all other results employed MATLAB R2019b, the latest installed version on our computing cluster.} except that a \gls{fic} approximation is used (\matlab{PredictMethod = 'fic'}) regardless of the number of \inpt{} points. We assume a Euclidean approximation of the \gls{vfz} (see \cref{sec:methods:framework:vfz-dist} and \cref{fig:dist-parity}). A slower, more accurate, and more memory-intensive prediction method that doesn't use sparse approximation (\matlab{PredictMethod = 'exact'}) is also available (\cref{sec:results:efficiency}). 

	\subsubsection{\glsentrytitlecase{idw}{long} Interpolation}
	\label{sec:methods:interp:idw}

	\Gls{idw} interpolation applies a weighted average to points within a neighborhood of a query point to obtain an interpolated value. \matlab{interp5DOF.m} implements a simple \gls{idw} approach based on \cite{tovarInverseDistanceWeight2020}. A default radius of influence of $r=\sqrt{2} \mu$ is used, where $\mu$ represents the mean \gls{nn} distance, and where octonion distance is approximated by the Euclidean distance or 2-norm (see \cref{sec:methods:framework:vfz-dist}, and \cref{fig:dist-parity}). \gls{nn} interpolation (\cref{sec:methods:interp:nn}) is used for a given query point when there are no \inpt{} points in the radius of influence. 

	\subsubsection{\glsentrytitlecase{nn}{long} Interpolation}
	\label{sec:methods:interp:nn}

	\Gls{nn} interpolation takes the nearest \inpt{} point relative to a query point and assigns the value of the \gls{nn} \inpt{} point to the query point. This is implemented via the built-in MATLAB function \matlab{dsearchn()} using a Euclidean approximation of octonion distance (see \cref{sec:methods:framework:vfz-dist}, and \cref{fig:dist-parity}). 

	\subsection{Literature Datasets}
	\label{sec:methods:litdata}
	In addition to performing validation tests of the \gls{vfzo} framework, we also describe results in which we apply it to actual \gls{gb} property data available from literature sources. Here we briefly mention details related to the retrieval and processing of two \gls{ms} simulation datasets from the literature. We describe \gls{gpr} applied to Fe (\cref{sec:methods:gprsim}) and Ni (\cref{sec:methods:gprsim-Ni}) simulations, as well as a specialized \gls{gprm} model applied to Fe to address non-uniformity and noise concerns (\cref{sec:methods:gprmix}).

	\subsubsection{\glsentrytitlecase{gpr}{long} for Fe Simulation Dataset}
	\label{sec:methods:gprsim}
	The Fe simulation data is obtained from \cite{kimPhasefieldModeling3D2014} rather than \cite{kimIdentificationSchemeGrain2011} due to a mistake in the earlier dataset file\footnote{We were informed of the error during an email discussion with the corresponding author of \cite{kimPhasefieldModeling3D2014}.}. \Glspl{gb} with a \gls{gbe} less than \SI{0.01}{\joule\per\square\meter} are removed to get rid of "no-boundary" \glspl{gb}. Repeated \glspl{gb} are then identified and removed by converting all \glspl{gb} into a \gls{vfzo} set (see \matlab{Kim2oct.m}) and sorting the repeated \glspl{gb} into "degenerate sets"\footnote{A degenerate "set" is distinct from a \glspl{vfzo} "set", the former of which is discussed in greater detail in Supplementary Information \cref{sec:supp:kim-interp:quality}. This sorting occurs via \matlab{avgrepeats.m} with \matlab{avgfn='min'}.}, and only the average \gls{gbe} (and a single \gls{gb}) within each degenerate set was retained. We estimate the intrinsic \gls{rmse} and \gls{mae} of the Fe simulation dataset to be \SIlist{0.06529;0.06190}{\joule\per\square\meter}, respectively. Minimum and maximum error was \SIlist{-0.2625;0.2625}{\joule\per\square\meter}, respectively. See \cref{sec:supp:kim-interp:quality} for further details on methods used to estimate intrinsic error of the Fe simulation dataset.

	\subsubsection{\glsentrytitlecase{gpr}{long} for Ni Simulation Dataset}
	\label{sec:methods:gprsim-Ni}

	We use the \glspl{gbo} representations \cite{chesserLearningGrainBoundary2020} of \glspl{gb} from \cite{olmstedSurveyComputedGrain2009} (\matlab{'olm_octonion_list.txt'} \cite{chesserGBOctonionCode2019}), importing and converting them to the active sense by taking the quaternion inverse of each of the octonions' quaternions. We take \gls{gbe} values (first column of \matlab{'olm_properties.txt'}, \cite{chesserGBOctonionCode2019}), and use a \gls{gpr} model (\cref{sec:methods:interp:gpr}).

	\subsubsection{\glsentrytitlecase{gprm}{long} for Fe Simulation Dataset}
	\label{sec:methods:gprmix}
	Separate from the four main methods analyzed in this work, a \gls{gprm} model is developed to better predict low \gls{gbe} using the non-uniformly distributed, noisy, Fe simulation dataset described in \cref{sec:methods:gprsim}. An exponential rather than a squared exponential kernel was used for the subset \gls{gpr} model (\cref{sec:supp:kim-interp}) to accommodate sharper transitions to better approximate low \glspl{gbe}.  Further details of the \gls{gprm} model are given in Supplementary Information (\cref{sec:supp:kim-interp}).

	\section{Results and Discussion} \label{sec:results}

	To illustrate the utility of the \gls{vfzo} framework for one application, namely interpolation, we compare the (i) accuracy (\cref{sec:results:accuracy}), and (ii) efficiency (\cref{sec:results:efficiency}) of the four previously described interpolation methods implemented over the \gls{vfz} with each other and with existing methods from the literature (see \cref{sec:intro}). For these tests, we use the \gls{5dof} \gls{gb} energy function by \citet{bulatovGrainBoundaryEnergy2014} (trained on Ni bicrystal simulation data \cite{olmstedSurveyComputedGrain2009}) as a validation function which we refer to as the \gls{brk} function.

	Following this validation study, we also demonstrate \gls{vfzo} \gls{gpr} interpolation applied to a large, noisy, \gls{ms} Fe bicrystal simulation dataset \cite{kimPhasefieldModeling3D2014} and a small, low-noise, \gls{ms} Ni bicrystal simulation dataset \cite{olmstedSurveyComputedGrain2009} (\cref{sec:results:simulation:compare}), to evaluate performance on real \gls{gb} property data.

	\subsection{Interpolation Accuracy}
	\label{sec:results:accuracy}

	Accuracy of \gls{gpr}, barycentric, \gls{nn}, and \gls{idw} interpolation methods are given w.r.t. the \gls{brk} validation function (\cref{sec:results:accuracy:interp}). Context is given to these error metrics through comparison with a constant-valued control model (\cref{sec:results:accuracy:control}) and the uncertainty associated with experimental and simulated datasets (\cref{sec:results:accuracy:exp-sim}).

	\subsubsection{Accuracy of Four Interpolation Methods}
	\label{sec:results:accuracy:interp}
\begin{figure*}[!ht]
		\centering
		\includegraphics[scale=1]{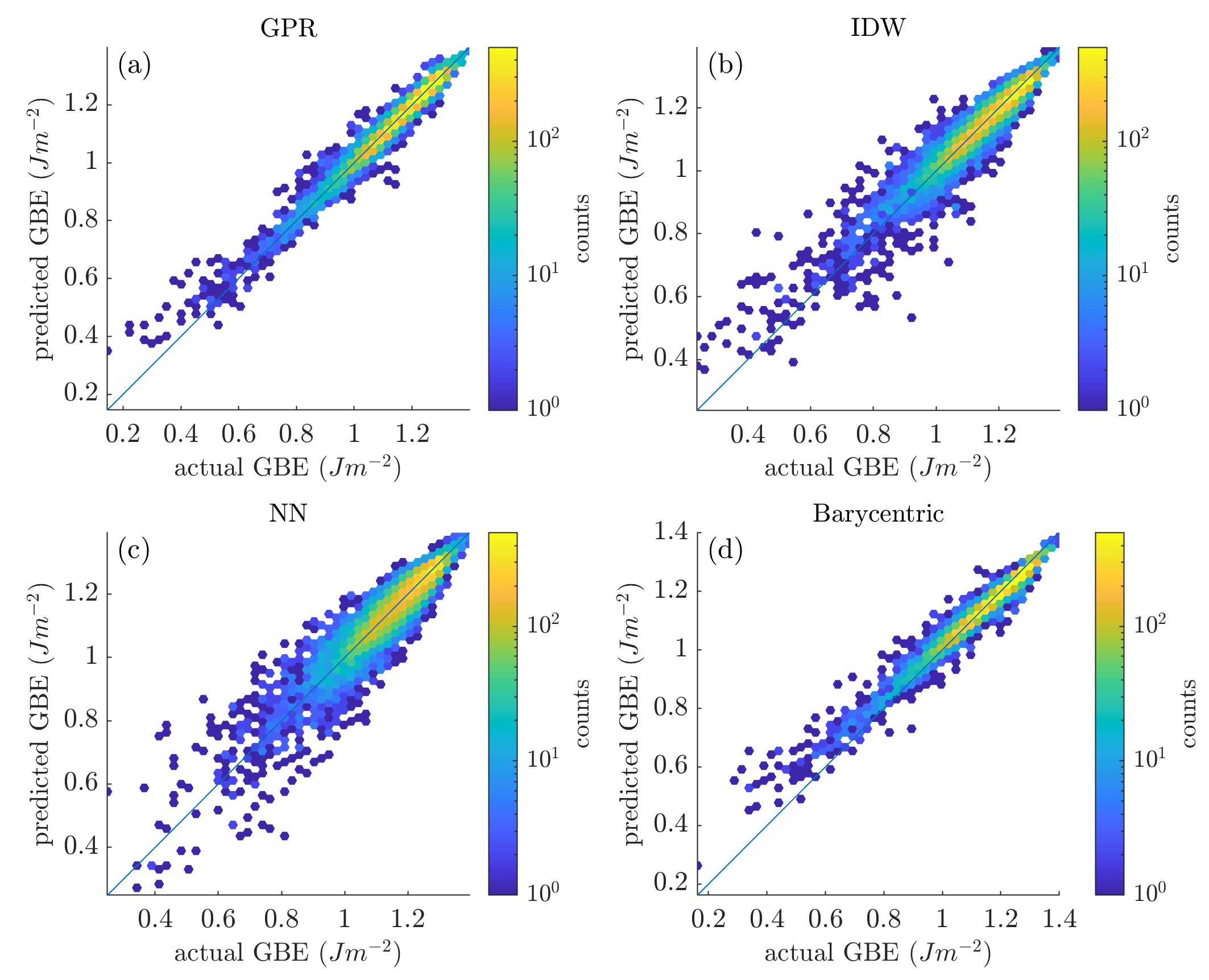}
		\caption{Hexagonally binned parity plots for \num{50000} \inpt{} and \num{10000} \outpt{} octonions formed via pairs of a random cubochorically sampled quaternion and a spherically sampled random boundary plane normal. Interpolation via (a) \gls{gpr}, (b) \gls{idw}, (c) \gls{nn}, and (d) barycentric coordinates.  \gls{brk} \gls{gbe} function for \gls{fcc} Ni \cite{bulatovGrainBoundaryEnergy2014} was used as the test function.}
		\label{fig:brkparity50000}
	\end{figure*}

	\Cref{fig:brkparity50000} provides hexagonally binned parity plots (\matlab{parityplot.m} via modified version of \cite{beanHexscatter2020}) for each of the four interpolation methods using \num{50000} \inpt{} \glspl{gb}. Results for \num{388} and \num{10000} \glspl{gb} are given in \cref{fig:brkparity388} and \cref{fig:brkparity10000}, respectively.

	All of the methods permit successful interpolation, and the highest density region in all cases falls squarely on the parity line. The \gls{gpr} and barycentric results show a slight asymmetry such that low energy values are overpredicted more often than they are underpredicted. The width of the point clouds provides a qualitative indication of the dispersion in the prediction errors, and the logarithmically scaled color indicates the frequency of errors of a given magnitude. As can be seen, the vast majority of errors are very small (the highest density---yellow region---is concentrated on the line of parity). Quantitative measures of the overall accuracy are presented for \gls{rmse} (\cref{tab:rmse-error-comparison}) and \gls{mae} (\cref{tab:mae-error-comparison}), and will be discussed in detail below (see \matlab{get\_errmetrics.m}).

	\begin{table*}[!ht]
		\centering
		\caption{Comparison of average interpolation \gls{rmse} (approximately 10 trial runs) for each interpolation method in the present work, using \num{50000} points in the definition of the \gls{vfz} and \glspl{gbe} obtained by evaluating the \gls{brk} validation function (\cite{bulatovGrainBoundaryEnergy2014}) at these points. A constant model (Cst, Avg \gls{rmse}), whose value was chosen to be the mean of the \inpt{} \gls{gbe} was used as a control. The last two columns represent the reduction ($\downarrow$) in \gls{rmse} in absolute units of \SI{}{\J\per\square\meter} and \% relative to the control model, respectively.}
		\label{tab:rmse-error-comparison}
		\begin{tabular}{@{}llllllll@{}}
			\toprule
			Method &
			Distance &
			Dataset &
			\thead{\# \glspl{gb}} &
			\thead{\gls{rmse} \\   (\SI{}{\J\per\square\meter})} &
			\thead{Cst, Avg \gls{rmse} \\   (\SI{}{\J\per\square\meter})} &
			\thead{\gls{rmse} $\downarrow$ \\   (\SI{}{\J\per\square\meter})} &
			\thead{\gls{rmse}   $\downarrow$ \\ (\%)} \\ \midrule
			\gls{gpr}   & \glsxtrshort{vfz} & \glsxtrshort{brk} & \num{50000} & \num{0.0218} & \num{0.1283} & \num{0.1065} & \num{83}   \\
			Barycentric & \glsxtrshort{vfz} & \glsxtrshort{brk} & \num{50000} & \num{0.0238} & \num{0.1283} & \num{0.1045} & \num{81.4} \\
			\gls{idw}   & \glsxtrshort{vfz} & \glsxtrshort{brk} & \num{50000} & \num{0.0356} & \num{0.1283} & \num{0.0927} & \num{72.3} \\
			\gls{nn}    & \glsxtrshort{vfz} & \glsxtrshort{brk} & \num{50000} & \num{0.0445} & \num{0.1283} & \num{0.0838} & \num{65.3} \\ \bottomrule
		\end{tabular}
	\end{table*}

	\begin{table*}
		\centering
		\caption{Comparison of average interpolation \gls{mae} (approximately 10 trial runs) for each interpolation method in the present work, using \num{50000} points in the definition of the \gls{vfz} and \glspl{gbe} obtained by evaluating the \gls{brk} validation function (\cite{bulatovGrainBoundaryEnergy2014}) at these points. A constant model (Cst, Avg \gls{mae}), whose value was chosen to be the mean of the \inpt{} \gls{gbe} was used as a control. The last two columns represent the reduction ($\downarrow$) in \gls{mae} in absolute units of \SI{}{\J\per\square\meter} and \% relative to the control model, respectively.}
		\label{tab:mae-error-comparison}
		\begin{tabular}{@{}llllllll@{}}
			\toprule
			Method &
			Distance &
			Dataset &
			\# \glspl{gb} &
			\thead{\gls{mae} \\   (\SI{}{\J\per\square\meter})} &
			\thead{Cst, Avg \gls{mae} \\   (\SI{}{\J\per\square\meter})} &
			\thead{\gls{mae} $\downarrow$ \\   (\SI{}{\J\per\square\meter})} &
			\thead{\gls{mae}   $\downarrow$ \\ (\%)} \\ \midrule
			\gls{gpr}   & \glsxtrshort{vfz} & \glsxtrshort{brk} & \num{50000} & \num{0.0145} & \num{0.0955} & \num{0.081}  & \num{84.8} \\
			Barycentric & \glsxtrshort{vfz} & \glsxtrshort{brk} & \num{50000} & \num{0.0145} & \num{0.0955} & \num{0.081}  & \num{84.8} \\
			\gls{idw}   & \glsxtrshort{vfz} & \glsxtrshort{brk} & \num{50000} & \num{0.0225} & \num{0.0955} & \num{0.073}  & \num{76.4} \\
			\gls{nn}    & \glsxtrshort{vfz} & \glsxtrshort{brk} & \num{50000} & \num{0.0307} & \num{0.0955} & \num{0.0648} & \num{67.9} \\ \bottomrule
		\end{tabular}
	\end{table*}

	As shown in \cref{tab:mae-error-comparison,tab:rmse-error-comparison}, of the four interpolation methods from this work, \Gls{gpr} has the lowest error, both in terms of \gls{rmse} and \gls{mae}, while \gls{nn} has the highest error. Compared to a constant valued control model, \gls{gpr} interpolation reduced the prediction \gls{rmse} by \SI{\gprrmsePercReduction}{\percent}, which outperforms all of the interpolation methods in this work with respect to accuracy, as well as those considered from the literature. After \gls{gpr} the next most accurate methods are barycentric,
	\gls{idw}, and \gls{nn}
	We also note that the \gls{rmse} interpolation error for the \gls{gpr} and barycentric methods is comparable to the minimum achievable noise-free experimental interpolation error which is the estimated error in experimental data (\cref{sec:results:accuracy:exp-sim}). 

\begin{figure*}
		\centering
		\includegraphics[scale=1]{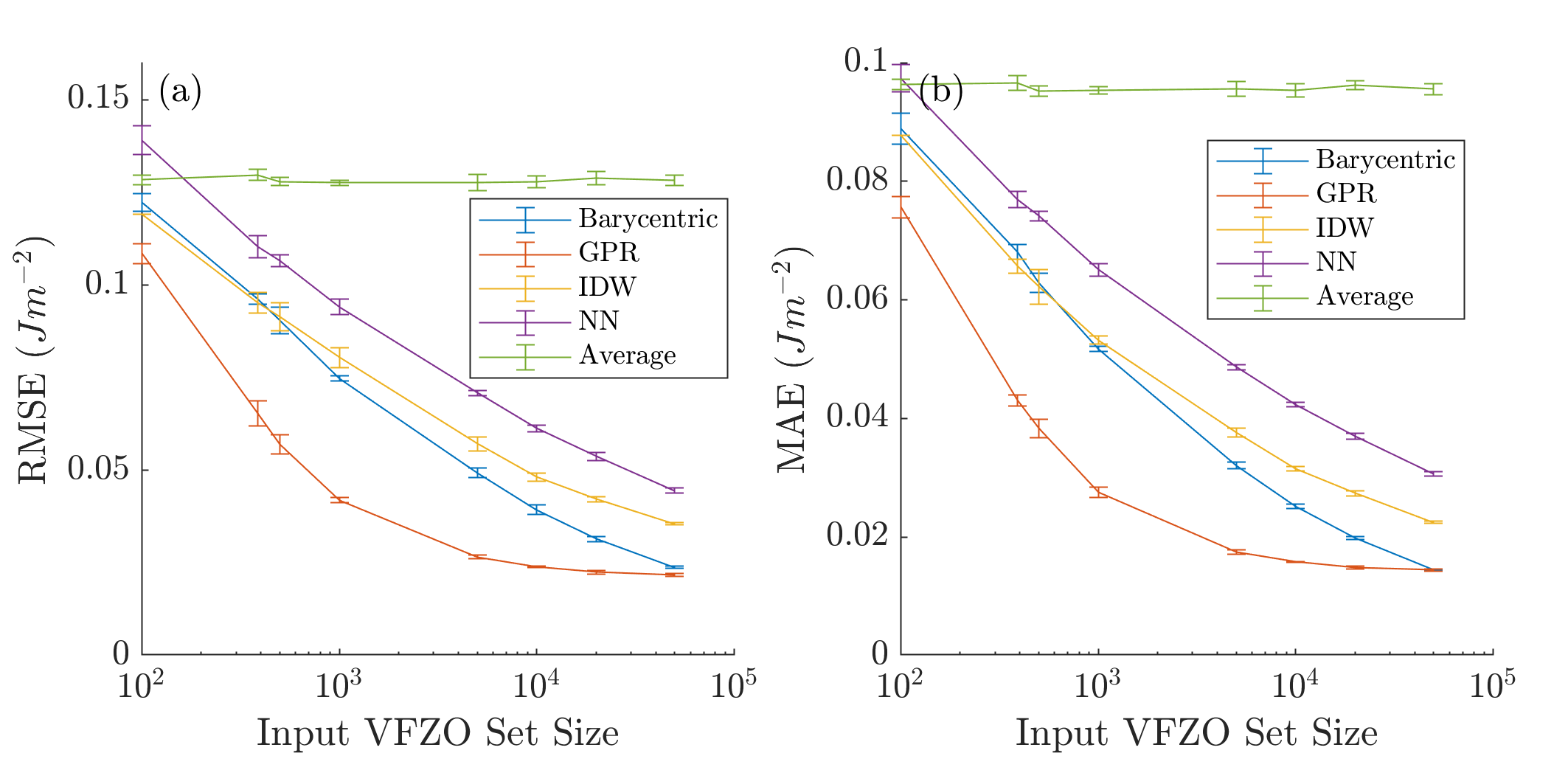}
		\caption{(a) Average \gls{rmse} and (b) average \gls{mae} vs. number of \inpt{} points for (planar) barycentric (blue), \gls{gpr} (orange), \gls{idw} (yellow), and \gls{nn} (purple) interpolation for approximately 10 random runs with different \inpt{} and \outpt{} points. Standard deviations of approximately 10 runs are also included. Compare with approximately \SI{\avgrmse{}}{\J\per\square\meter} and \SI{\avgmae{}}{\J\per\square\meter} \gls{rmse} and \gls{mae}, respectively, for a constant, average model (green) using the average of the \inpt{} properties (approximately \SI{1.16}{\J\per\square\meter}).}
		\label{fig:brkerror}
	\end{figure*}

	The accuracy of the predictions made using the \gls{vfz} methods depends on the \gls{vfzo} set size and distribution. 
	\Cref{fig:brkerror} compares the prediction accuracy for each of the 4 methods to the constant valued control model, as a function of the number of \inpt{} \glspl{vfzo} (\matlab{ninputpts}). As expected, higher density \gls{vfzo} sets result in lower error, but eventually give diminishing returns. Moreover, the standard deviations produced via multiple runs are tightly constrained and generally shrink as the \gls{vfzo} set size increases.

	\Gls{gpr} consistently gives lower error than the other three interpolation methods for all \gls{vfzo} set sizes.
	\Gls{nn} interpolation produces the worst error of the four methods, but is better than a constant valued control model (i.e. average of the \inpt{} \glspl{gbe}) so long as \matlab{ninputpts} exceeds a few hundred \inpt{} points.

	It is worthwhile to note that both \gls{gpr} and \gls{idw} are kernel-based in that a model parameter controls the size of the region that can influence the interpolation results. In the \gls{gpr} case, this is automatically calculated via an internal fitting routine of \matlab{fitrgp()}. \gls{nn} distance distributions (\cref{fig:nnhist-knn-50000}) can lead to insight about correlation lengths in a given \gls{vfzo} set and are used in the \gls{idw} implementation. For \gls{idw}, the radius of influence is set to $r=\sqrt{2} \mu$, where $\mu$ is the mean \gls{nn} distance. It is likely that better tuning of the kernel parameters in these two methods (such as use of built-in hyperparameter optimization in the case of \matlab{fitrgp()}) could further decrease their interpolation errors. Additionally, for \gls{gpr}, use of the \matlab{'exact'} \matlab{predictMethod} or a larger \matlab{'fic'} set size will also likely reduce interpolation error.

	By contrast, barycentric interpolation automatically adjusts its effective region of influence because the size of the simplices in the mesh decreases as the number of vertices increases. More uniformly distributed meshes (such as obtained via constrained optimization \cite{dolanBenchmarkingOptimizationSoftware2004,ConstrainedElectrostaticNonlinear2020}) will likely result in lower, more uniform interpolation error, especially for this simplex-based approach which can exhibit high-aspect ratio facets and non-intersections outside the bounds of the mesh (\cref{fig:high-aspect-non-int}). While the barycentric interpolation error is always higher than \gls{gpr} for the considered set sizes, at \num{50000} \glspl{vfzo}, the errors of \gls{gpr} and barycentric interpolation are nearly identical.
	\begin{table*}[]
		\centering
		\caption{Approximate coordinates of \glspl{vfzo} A and B used for the interpolation in \cref{fig:tunnel-50000}. Individual quaternions of each octonion are given in the active sense and in the laboratory reference frame with an assumed \gls{gb} normal pointing in the +z direction, also in the laboratory reference frame.}
		\label{tab:tunnel-AB}
		\begin{tabular}{@{}lllllllll@{}}
			\toprule
			Octonion & o(1)   & o(2)    & o(3)    & o(4)    & o(5)    & o(6)   & o(7)    & o(8)   \\ \midrule
			A        & 0.8658 & -0.4269 & -0.1270 & 0.2280  & 0.2810  & 0.8390 & -0.3852 & 0.2622 \\
			B        & 0.4684 & -0.7657 & -0.4100 & -0.1617 & -0.1483 & 0.8204 & -0.3588 & 0.4198 \\ \bottomrule
		\end{tabular}
	\end{table*}
	\begin{figure}[!ht]
		\centering
		\includegraphics{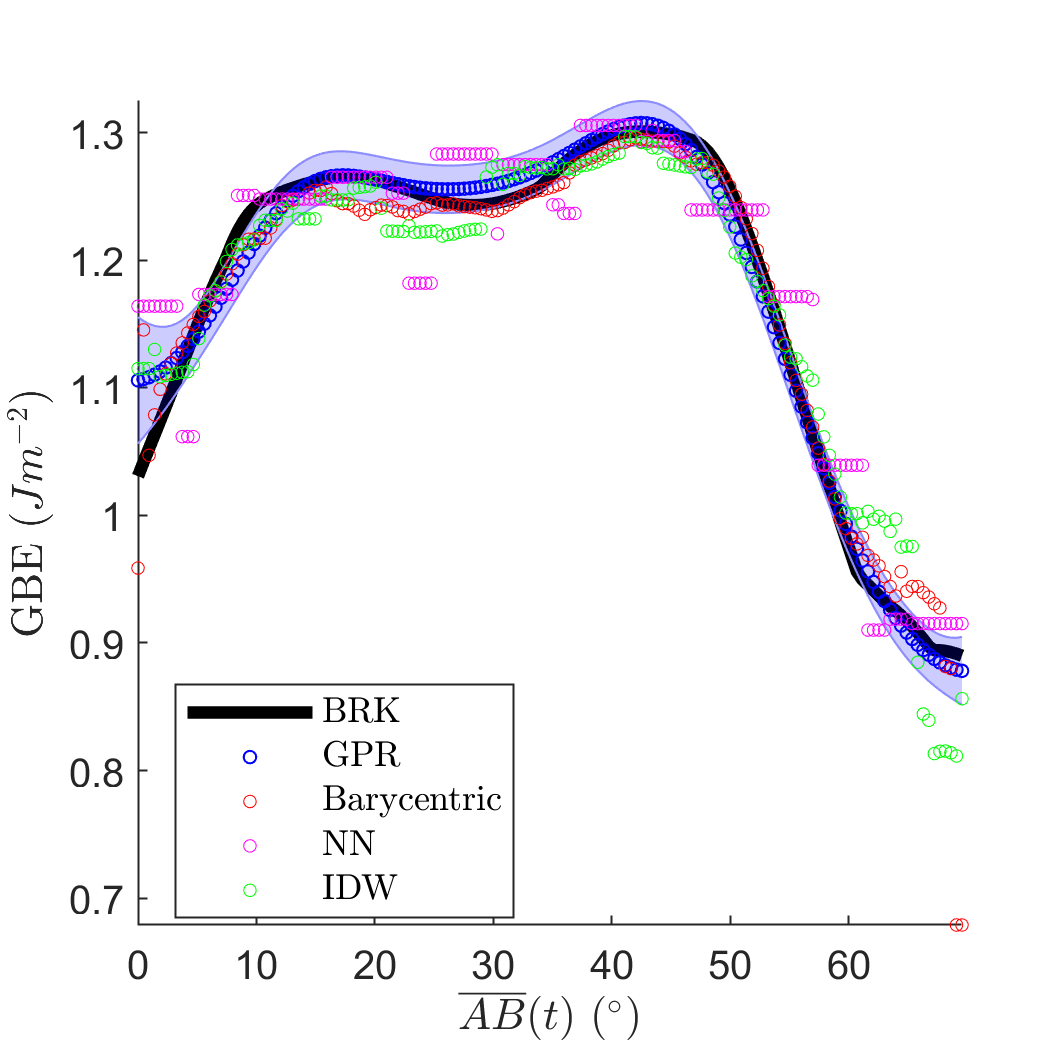}
		\caption{Predictions of \gls{gpr} (blue circles), barycentric (red circles), \gls{nn} (magenta circles), and \gls{idw} (green circles) as a function of distance along a 1D arc ($\overline{AB}$) between two \glspl{vfzo} ($A$ and $B$). The true, underlying \gls{brk} function is also shown (black line). \num{50000} random \inpt{} \glspl{vfzo} were generated and used for each of the models. \num{150} equally spaced points between $A$ and $B$ obtained via \gls{oslerp} \cite{francisGeodesicOctonionMetric2019} were used as \outpt{} points. \gls{gpr} uncertainty standard deviation is plotted as shaded error band.}
		\label{fig:tunnel-50000}
	\end{figure}

	\subsubsection{Constant-Valued Control Models}
	\label{sec:results:accuracy:control}
	To aid in objective interpretation of the error metrics, comparison is made to a constant valued control model, whose value is chosen to be the average of \matlab{y} (approximately \SI{1.16}{\J\per\square\meter} in the limit of $\matlab{ninputpts} \rightarrow \infty$) resulting in \gls{rmse} and \gls{mae} values of approximately \num{\avgrmse{}} and \SI{\avgmae{}}{\J\per\square\meter}. This comparison with the relevant constant-valued function gives a sense of the complexity and variability of the validation function and allows for a more objective comparison between differing works. For example, the \gls{rmse} for the relevant constant function compared to the validation function employed for the \gls{ann} interpolation method in \cite{restrepoUsingArtificialNeural2014} is \SI{0.0854}{\J\per\square\meter}; in contrast, the \gls{rmse} for the relevant constant function compared to the BRK validation function used in this work is \SI{0.1302}{\J\per\square\meter} (see \cref{tab:rmse-error-comparison}). This suggests that the BRK validation function is more complex and therefore less well approximated by a constant than the validation function used to test the \gls{ann} interpolation method in \cite{restrepoUsingArtificialNeural2014}. Consequently, the improved performance of the present methods (see \cref{sec:results:simulation:compare} and \cref{tab:mae-error-simulation,tab:rmse-error-simulation}) is even more notable in that the validation function employed here is more difficult to interpolate.

	\subsubsection{Experimental and Simulation Error}
	\label{sec:results:accuracy:exp-sim}
	To give further context to the results of this and prior works, it is useful to consider what the intrinsic error is for typical GB property data. This provides an idea of the minimum possible interpolation error, since one cannot reliably \emph{detect} lower error in the interpolation than already exists in the observed data itself.

	One such estimate for error is furnished by the work of \citet{shenDeterminingGrainBoundary2019}, who introduced a non-discretizing approach to extract relative \gls{gb} energies from polycrystalline samples using the \gls{lobpcg} method. Their approach utilizes regularization imposed on \gls{tj} equilibrium equations and \gls{knn} distances. Using \num{60000} \glspl{tj} (\num{180000} \glspl{gb}) and a custom, non-smooth validation function they obtained \gls{gbe} \gls{rmse} values of \SI{0.0076}{\J\per\square\meter} and \SI{0.0277}{\J\per\square\meter} for \gls{gbe} values greater than \SI{0.9}{\J\per\square\meter} and less than \SI{0.9}{\J\per\square\meter}, respectively. This suggests that an optimistic estimate for the error in noise-free\footnote{These errors are based on Figure 8 from \citet{shenDeterminingGrainBoundary2019}, which employed synthetic triple junctions with a custom validation function, rather than experimental data. While the authors did also consider the addition of noise, we use the noise-free results as an estimate of the best-case scenario.} \emph{experimental} \gls{gbe} data obtained using such a method is on the order of \SIrange{0.0076}{0.0277}{\J\per\square\meter}, which also serves as an estimate of the minimum achievable noise-free experimental interpolation error for any of the interpolation methods described here. Similar analysis for noisy \SI{0}{\kelvin} \gls{ms} \emph{simulation} data is provided in \cref{sec:results:simulation} and \cref{sec:supp:kim-interp:quality} giving a \gls{rmse} and \gls{mae} of \SIlist{0.06529;0.06190}{\joule\per\square\meter}, respectively.

	Again comparing the relevant constant-valued control model\footnote{We use the mean of the true \glspl{gbe} from their validation function to define the constant-valued control model instead of the mean of the \inpt{} \glspl{gbe} because the latter does not exist for polycrystalline data.} to the validation function employed by \citet{shenDeterminingGrainBoundary2019}, we calculate a \gls{rmse} and \gls{mae} of \SIlist{0.0976;0.0466}{\joule\per\square\meter}, respectively. This implies that the validation model used by \citet{shenDeterminingGrainBoundary2019} is also simpler\footnote{\citet{shenDeterminingGrainBoundary2019} used 8 cusps of varying depths and widths based on the Read-Shockley model and unity \gls{gbe} everywhere else.} than the \gls{brk} validation model employed in the present work.

	\subsection{Interpolation Efficiency}
	\label{sec:results:efficiency}

	Below, we present interpolation efficiency results in terms of computational runtime and memory for the four interpolation schemes used in this work (\cref{sec:results:efficiency:methods}). Additionally, an in-depth treatment of the improved symmetrization runtime (separate from interpolation runtime) relative to the original octonion metric is given (\cref{sec:results:efficiency:symruntime}).

	\subsubsection{Efficiency of Four Interpolation Methods}
	\label{sec:results:efficiency:methods}

	We discuss runtime and memory requirements for barycentric, \gls{gpr}, \gls{idw}, and \gls{nn} interpolation methods. Computational runtimes of the various methods are shown in \cref{tab:runtime}.

	\begin{table*}
		\centering
		\caption{Comparison of average~runtime~(\SI{}{\second}) for \num{10} trials for barycentric, \gls{gpr}, \gls{idw}, and \gls{nn} interpolation methods for various \inpt{} \gls{vfzo} set sizes using 12 cores and evaluated on \num{10000} \outpt{} \glspl{vfzo}. Because \gls{gpr}, \gls{idw}, and \gls{nn} method defaults do not use \matlab{parfor} loops but may have internal multi-core vectorization, it is unclear to what extent the number of cores affects the runtime of methods other than barycentric interpolation. \Gls{vfzo} symmetrization runtime was not included; however, symmetrization of \num{50000} \glspl{gbo} takes approximately \SI{\symtime}{seconds} on \SI{6}{cores} (Intel i7-10750H, 2.6 GHz) and is a common step in every interpolation method (i.e. it is fundamental to the \gls{vfzo} framework). We used the \gls{brk} validation function for \gls{gbe} \cite{bulatovGrainBoundaryEnergy2014}. }
		\label{tab:runtime}
		\begin{tabular}{lllll}
			\cline{2-5}
			& \multicolumn{4}{c}{Runtime (s)}                                                     \\ \hline
			\gls{vfzo} Set Size & Barycentric       & GPR                 & IDW                 & NN                  \\ \hline
			\num{100}           & $191.8 \pm 19.57$ & $0.4187 \pm 0.4342$ & $0.034 \pm 0$       & $0.0367 \pm 0.0041$ \\
			\num{388}           & $388.4 \pm 18.84$ & $0.943 \pm 0.3481$  & $0.0904 \pm 0.0224$ & $0.0705 \pm 0.0129$ \\
			\num{500}           & $455.7 \pm 55.28$ & $0.6104 \pm 0.3138$ & $0.1352 \pm 0.0364$ & $0.0724 \pm 0.0051$ \\
			\num{1000}          & $536.5 \pm 35.26$ & $1.743 \pm 0.9464$  & $0.1948 \pm 0.0395$ & $0.1203 \pm 0.0184$ \\
			\num{5000}          & $998.9 \pm 54.48$ & $5.216 \pm 0.4816$  & $0.8726 \pm 0.1529$ & $0.9277 \pm 0.2418$ \\
			\num{10000}         & $1516 \pm 56.59$  & $5.609 \pm 0.8756$  & $1.631 \pm 0.3915$  & $0.8938 \pm 0.1717$ \\
			\num{20000}         & $2526 \pm 119.5$  & $11.45 \pm 3.29$    & $3.191 \pm 0.4752$  & $1.275 \pm 0.3423$  \\
			\num{50000}         & $5743 \pm 361.3$  & $13.69 \pm 4.05$    & $7.635 \pm 1.872$   & $3.817 \pm 0.5884$  \\ \hline
		\end{tabular}
	\end{table*}

	Barycentric interpolation takes the longest, in spite of the fact that it is the only parallelized method by default (not accounted for in \cref{tab:runtime}). In other words, since 12 cores were used to obtain these runtime results, the total runtime across all cores is much higher compared with the other methods; however, it is possible that other methods used multi-threading via built-in vectorized functions. The long computation times of barycentric interpolation result primarily from the large number of facets present in a high-dimensional mesh triangulation and the interconnectedness of facets with respect to each other.

	\Gls{gpr} is fast compared to barycentric interpolation; however, the entire process has to be reevaluated (in the current implementation) if the \inpt{} points (i.e. \glspl{vfzo}) or \inpt{} property values (i.e. \glspl{gbe}) change
	(typically referred to as predictors/features and responses, respectively, in the machine learning community).
	On the other hand, barycentric interpolation is fast if the triangulation and intersections are pre-computed and only input property values change (\matlab{interp\_bary\_fast.m}), but slow if the \inpt{} or \outpt{} points change, which requires recomputing the triangulation and intersections. Additionally, \gls{gpr} is the second-longest in terms of of runtime. 

	\Gls{nn} and \gls{idw} interpolation have vectorized implementations and are much simpler than the barycentric and \gls{gpr} methods. Consistent with expectations, \gls{nn} and \gls{idw} exhibit almost negligible runtimes. 
	It should also be noted that barycentric interpolation has much higher memory requirements than \gls{gpr}, \gls{nn}, and \gls{idw} due to the need to store large matrices. If \matlab{PredictMethod = 'exact'} in \matlab{fitrgp()}, then \gls{gpr} also has high memory requirements for large \gls{vfzo} sets. For \num{50000} input points with sufficient RAM (e.g. $\sim$32 GB) and 12 cores available, the \matlab{'exact'} method runtime is \SI{535.1 \pm 392.6}{seconds}. However, because the \matlab{'fic'} approximation is always used in this work, memory requirements are similar to \gls{nn} and \gls{idw}.

	Because the default implementation of \gls{idw} uses a radius cut-off, the distance and weight matrices can be stored as sparse objects, dramatically reducing both the final memory storage requirements and computational complexity of this method. We expect that a \gls{knn} approach would produce similar results both in terms of runtime and error when a relatively uniform sampling of \gls{gbc} is obtained.

	\subsubsection{Symmetrization Runtime Comparison with Traditional Octonion Metric}
	\label{sec:results:efficiency:symruntime}
	In addition to the interpolation runtime of the methods just presented, it is valuable to consider the runtime of the \gls{vfz} symmetrization step (not included in \cref{tab:runtime}). The symmetrization step is at the core of the \gls{vfzo} framework and is a key to its overall performance. It is a common step for both (i) distance calculations and (ii) all of the interpolation methods presented here.

	Directly computed, scaled Euclidean and arc length distances in the \gls{vfzo} framework approximate the original octonion distance by \citet{francisGeodesicOctonionMetric2019}, and the calculation speed is even higher than explicit \gls{gbo} distance calculations using the original octonion distance. For example, \num{50000} \glspl{gbo} can by symmetrized into \glspl{vfzo} in approximately \SI{\symtime}{seconds} using \SI{6}{cores} (\matlab{get\_octpairs.m}), and the corresponding \num{50000} $\times$ \num{50000} pairwise-distance matrix can be computed in approximately \SI{10}{seconds} (\matlab{pdist()}), giving a total runtime of approximately \SI{86}{seconds} (\num{466} total CPU seconds). Compared to the original octonion metric distance calculations \cite{chesserLearningGrainBoundary2020} in the Fortran-based EMSoft package \cite{degraefEMSoft2020}, this represents an improvement in computational speed by $\sim$\num{5} orders of magnitude using our MATLAB implementation in the \vfzorepo{} \cite{bairdFiveDegreeofFreedom5DOF2020}.

	Improvement per distance calculation per core of the \vfzorepo{} is about \num{4e5} relative to the EMSoft \cite{degraefEMSoft2020} metric of 26 minutes using 8 cores for a $\num{388}\times\num{388}$ pairwise distance matrix. This EMSoft timing information is directly reported in \citet{chesserLearningGrainBoundary2020}. In other words, computation of a $\num{50000}\times\num{50000}$ using the traditional octonion metric and EMSoft implementation would take approximately 6.6 CPU years (or 153 CPU days by applying the isometry equation in Section 7 of \citet{morawiecDistancesGrainInterfaces2019}). Since most interpolation methods will depend on computing new distances, probing the model at new \glspl{gb} will also be expensive. For example, it would take at minimum $\sim$30 CPU days (after isometry equivalence has been applied) to perform property interpolation for \num{10000} \outpt{} \glspl{gb} assuming the pairwise-distance matrix relative to \num{50000} \inpt{} \glspl{gbo} needs to be computed. This presents an issue for iterative simulations (e.g. mesoscale grain growth) in which \num{1000}'s of new \gls{gb} segments would need to be sampled at each time step. By contrast, property values for \num{10000} new \glspl{gb} would be sampled in our approach in $\sim$\num{90} CPU seconds. For perspective, a phase-field simulation might have \num{10000} or more time steps with thousands of \glspl{gb} \citet{kimPhasefieldModeling3D2014,dimokratiSPFMModelIdeal2020}.  Recently, \citet{miyoshiLargescalePhasefieldStudy2021} presented Reed-Shockley anisotropic 3D phase-field grain growth results for initially \num{3125000} grains with as many as \num{125000} time steps to reach $\sim$\num{10000} final grains. Performing such a simulation with even the efficient \gls{vfzo} framework would require ~56 CPU years for the property sampling alone \footnote{For such an application, a GPU implementation of the \gls{vfzo} framework, batch implementation of the \gls{seo} considerations, directly tracking \glspl{gb} movement within a \gls{vfz}, and/or other approaches would likely be necessary to make the problem more tractable.}. 

	This significant speed up stems from the fact that in the \gls{vfzo} framework \glspl{seo} only need to be considered once per \gls{gb}, $O(L)$, rather than once per distance calculation, $O(L^2)$,
	and that \glspl{seo} only need to be considered once in a \gls{gb} pair, $O(N_p^2)$, rather than for every combination between the two \glspl{gb}, $O(N_p^4)$. The \gls{seo} computation complexity is thus $O(N_p^2L)$, a significant improvement compared with the original \gls{seo} complexity of $O(N_p^4L^2)$ \cite{chesserLearningGrainBoundary2020}, where $N_p$ is the number of proper rotations of the crystallographic point group ($N_p=24$ for $m\Bar{3}m$ \gls{fcc} point group) and $L$ is the number of \glspl{gb}.

	Empirically, to compute a pairwise-distance matrix for $L$ = \num{50000} \glspl{gb} using the \vfzorepo{} \cite{bairdFiveDegreeofFreedom5DOF2020}, the full $O(N_p^2L)$ symmetrization operations take about \SI{\symtime{}}{seconds} $\times\ 6$ cores $= \SI{456}{seconds}$ of CPU time, whereas the subsequent pairwise-distance computation is $O_{\text{pd}}(L^2)$ and takes approximately \SI{10}{seconds} for a \num{50000} $\times$ \num{50000} matrix. Even though $O(N_p^2L) \ll O_{\text{pd}}(L^2)$, the symmetrization step takes far more time than the pairwise distance calculation (even for large $L$) because of the cost of generating \glspl{seo}. Because Euclidean distances---which can be computed faster than trigonometric inverse functions---are employed, and built-in, vectorized MATLAB functions are utilized, there is a further speed enhancement in the \gls{vfzo} approach. 

	\subsection{Interpolation Visualization}
	We present interpolation results plotted in a 1D arc in the full \gls{5dof} \gls{gb} space (\cref{sec:results:vis:arc}) followed by discussion of potential to use numerical derivatives and identify local minima (\cref{sec:results:vis:apps}).

	\subsubsection{Interpolation Along a 1D Arc}
	\label{sec:results:vis:arc}

	To provide a visual illustration of the property predictions, \cref{fig:tunnel-50000} shows the predicted \gls{gbe} for each of the four interpolation methods as a function of distance along a 1D arc ($\overline{AB}$) between two \glspl{vfzo}, $A$ and $B$. Approximate coordinates for $A$ and $B$ are given in \cref{tab:tunnel-AB}, and each intermediate point between $A$ and $B$ resides on the surface of a hypersphere. The \num{150} intermediate points were obtained using \gls{oslerp} \cite{francisGeodesicOctonionMetric2019}. Each model used its own set of \num{50000} random \inpt{} \glspl{vfzo} with \gls{gbe} sampled via the \gls{brk} validation function. The two \glspl{vfzo} were chosen by taking the furthest apart pair out of \num{20000} \glspl{vfzo} which thus approximates the largest dimension of the \gls{vfz} where each endpoint is close to the true \gls{vfz} exterior.

	Comparison of the predictions from the four interpolation methods with the true values of the \gls{brk} validation function along this 1D path shows that all methods yield reasonable agreement with the true model. The \gls{gpr} and barycentric methods appear to agree most with the true model, followed by \gls{idw} and \gls{nn}. The \gls{nn} method shows the piecewise-constant (stair-step) artifact typical of \gls{nn} methods. We also note that while the fidelity of the predictions is quite good for all methods in the interior of the \gls{vfz}, the performance does degrade at the extreme limits of the \gls{vfz} (note the deviations at the left and right limits of \cref{fig:tunnel-50000}). This effect seems to be particularly pronounced for the barycentric method, and much less so for the \gls{gpr} method.

	We believe this is the first\footnote{\Gls{oslerp} results from \citet{francisGeodesicOctonionMetric2019} plots \gls{gb} \emph{structure} continuously between two \glspl{gb}, \citet{chesserLearningGrainBoundary2020} performs cross-validation on the simulated Olmsted Ni \glspl{gb}, and \cite{morawiecDistancesGrainInterfaces2019} plots distances between \glspl{gb} on a geodesic with another \gls{gb}. The results in these works are distinct from what is presented here: a plot of continuously interpolated \glspl{gbe} between two arbitrary \glspl{gb}.} plot of a \gls{gb} property continuously interpolated between two arbitrary \glspl{gb} (i.e. neither residing entirely in a single \gls{mfz} nor a single \gls{bpfz}). Such visualizations can naturally be extended to 2D and 3D by plotting colored points in a triangle or tetrahedron, respectively, all of which (1D, 2D, and 3D) represent small "slices" of the \gls{gbc} space.

	\subsubsection{Potential for Numerical Derivatives}
	\label{sec:results:vis:apps}

	Additionally, such visualizations suggest the ability to estimate numerical derivatives or gradients of \gls{gb} properties without being restricted to a \gls{gb} subspace (e.g. \gls{mfz} or \gls{bpfz}) which can be a useful mathematical construct for the \gls{gb} community. For example, steepest descent paths and all local \gls{gbe} minima can be estimated and used in grain growth simulations. In such contexts, use of ensembled \gls{vfzo} interpolation may be necessary to mitigate discontinuity artifacts when crossing the exterior of a \gls{vfz} as discussed in \cref{sec:methods:framework:vfz-dist} which we plan to explore in future work.

	\subsection{Literature Datasets}
	\label{sec:results:simulation}

	In addition to validation results (\cref{sec:results:accuracy}), we also apply the \gls{vfzo} framework to real \gls{gb} property data from two sources in the literature. This allows more direct comparison to previous methods as well as demonstration of the performance of the the \gls{vfzo} framework for typical \gls{ms} data. Specifically, we present \gls{gpr} interpolation results for \gls{ms} Fe and Ni simulation datasets and compare them with prior work (\cref{sec:results:simulation:compare}). Finally, because \gls{gpr} overestimates the low \gls{gbe} for the non-uniformly distributed, noisy Fe simulation dataset, we also provide results for an adaptation called the \gls{gprm} model that compensates for this effect (\cref{sec:results:simulation:gprm}).

	\subsubsection{Comparison with Prior Work}
	\label{sec:results:simulation:compare}

	The \gls{gpr} interpolation method of the present work was used with the same number of \inpt{} \glspl{gb} as was supplied in \citet{restrepoUsingArtificialNeural2014} for Fe (\num{17176}) and \citet{chesserLearningGrainBoundary2020} for Ni (\num{388}) to provide a more consistent comparison with prior work. For Fe, the remainder of the simulation data was used for testing, consistent with \citet{restrepoUsingArtificialNeural2014}, except that zero-energy \glspl{gb} and degenerate \glspl{gb} were treated differently as described in \cref{sec:methods:gprsim}. For Ni, a \gls{loocv} scheme was used, consistent with \citet{chesserLearningGrainBoundary2020}.

	Hexagonally binned parity plots for the Fe and Ni simulation datasets are shown in \cref{fig:kim-interp-teach}d and \cref{fig:olmsted-Ni-loocv}, respectively. \Gls{rmse} and \gls{mae} comparisons along with improvement relative to a constant, average model are given in \cref{tab:mae-error-simulation} and \cref{tab:rmse-error-simulation}, respectively.

	\begin{table*}
		\centering
		\caption{Comparison of interpolation \gls{mae} (1 trial run) for \SI{0}{\kelvin} \glsxtrfull{ms} datasets. A constant model (Cst, Avg \gls{mae}), whose value was chosen to be the mean of the \inpt{} \gls{gbe} was used as a control. The last two columns, \gls{mae} $\downarrow$ (\SI{}{\J\per\square\meter}) and \gls{mae} $\downarrow$ (\%)), represent the reduction in \gls{mae} in units of \SI{}{\J\per\square\meter} and \% relative to the control model, respectively. Non-sym refers to distances calculated in \citet{restrepoUsingArtificialNeural2014} without regard for crystal symmetries. }
		\label{tab:mae-error-simulation}
		\begin{tabular}{@{}llllllll@{}}
			\toprule
			Method &
			Distance &
			Dataset &
			\# \glspl{gb} &
			\thead{\gls{mae} \\   (\SI{}{\J\per\square\meter})} &
			\thead{Cst, Avg \gls{mae} \\   (\SI{}{\J\per\square\meter})} &
			\thead{\gls{mae} $\downarrow$ \\   (\SI{}{\J\per\square\meter})} &
			\thead{\gls{mae}   $\downarrow$ \\ (\%)} \\ \midrule
			\gls{gpr}                                            & \glsxtrshort{vfz} & \glsxtrshort{ms} Fe & \num{17176} & \num{0.0405} & \num{0.0617} & \num{0.0212} & \num{34.4} \\
			\gls{ann}   \cite{restrepoUsingArtificialNeural2014} & Non-sym        & \glsxtrshort{ms} Fe & \num{17176} & \num{0.0486} & \num{0.0617} & \num{0.0131} & \num{21.2} \\
			\gls{lkr}   \cite{chesserLearningGrainBoundary2020}  & \glsxtrshort{gbo} & \glsxtrshort{ms} Ni & \num{388}   & \NA          & \num{0.1752} & \NA          & \NA        \\ \bottomrule
		\end{tabular}
	\end{table*}

	\begin{table*}
		\centering
		\caption{Comparison of interpolation \gls{rmse} (1 trial run) for \SI{0}{\kelvin} \glsxtrfull{ms} datasets. A constant model (Cst, Avg \gls{rmse}), whose value was chosen to be the mean of the \inpt{} \gls{gbe} was used as a control. The last two columns, \gls{rmse} $\downarrow$ (\SI{}{\J\per\square\meter}) and \gls{rmse} $\downarrow$ (\%)), represent the reduction in \gls{rmse} in units of \SI{}{\J\per\square\meter} and \% relative to the control model, respectively. Non-sym refers to distances calculated in \citet{restrepoUsingArtificialNeural2014} without regard for crystal symmetries. }
		\label{tab:rmse-error-simulation}
		\begin{tabular}{@{}llllllll@{}}
			\toprule
			Method &
			Distance &
			Dataset &
			\thead{\# \glspl{gb}} &
			\thead{\gls{rmse} \\   (\SI{}{\J\per\square\meter})} &
			\thead{Cst, Avg \gls{rmse} \\   (\SI{}{\J\per\square\meter})} &
			\thead{\gls{rmse} $\downarrow$ \\   (\SI{}{\J\per\square\meter})} &
			\thead{\gls{rmse}   $\downarrow$ \\ (\%)} \\ \midrule
			\gls{ann}   \cite{restrepoUsingArtificialNeural2014} & Non-sym        & \glsxtrshort{ms} Fe & \num{17176} & \NA          & \num{0.0854} & \NA          & \NA        \\
			\gls{gpr}                                            & \glsxtrshort{vfz} & \glsxtrshort{ms} Ni & \num{388}   & \num{0.0951} & \num{0.2243} & \num{0.1292} & \num{57.6} \\
			\gls{lkr}   \cite{chesserLearningGrainBoundary2020}  & \glsxtrshort{gbo} & \glsxtrshort{ms} Ni & \num{388}   & \num{0.0977} & \num{0.2243} & \num{0.1266} & \num{56.4} \\ \bottomrule
		\end{tabular}
	\end{table*}

	For the Fe case, we see a larger improvement than prior work likely due to our incorporation of \gls{gb} symmetry, which was not considered in \citet{restrepoUsingArtificialNeural2014}. For the Ni case, there is a slight improvement relative to prior work, indicating that accuracy is similar to the original octonion metric while maintaining the significant computational benefits of the \gls{vfzo} framework.

\begin{figure*}
		\centering
		\includegraphics[scale=1]{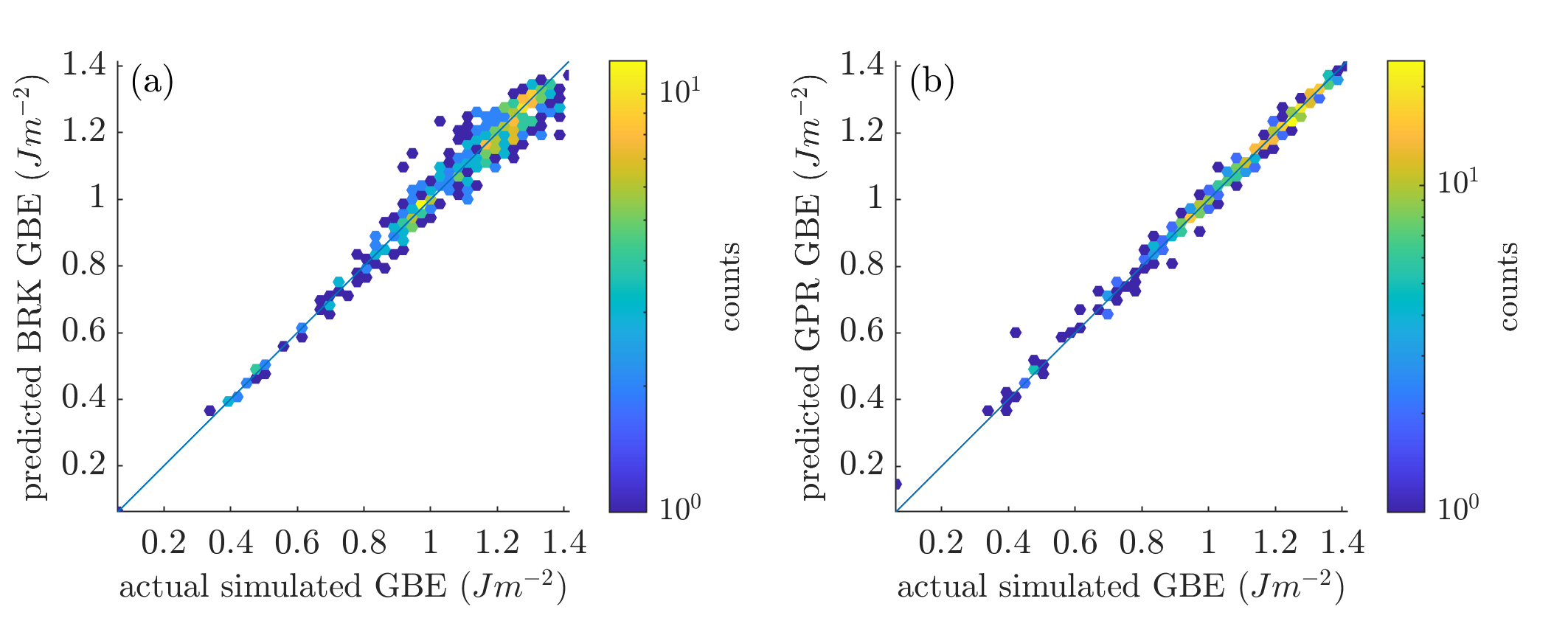}
		\caption{Hexagonally binned parity plots of (a) \gls{brk} and (b) \gls{gpr} model \glspl{gbe} fitted using Olmsted Ni simulation data vs. Olmsted Ni simulation \glspl{gbe}. \Gls{mae} is \SIlist{0.00975;0.03626}{\J\per\square\m} for (a) and (b), respectively. Likewise, \gls{rmse} is \SIlist{0.01727;0.04972}{\J\per\square\m}, respectively.}
		\label{fig:resubloss-ni}
	\end{figure*}

	Since the \gls{brk} validation function is also an interpolation function on the Ni simulation data, \gls{gpr} within the \gls{vfzo} framework and the \gls{brk} function results are directly compared via parity plot in \cref{fig:resubloss-ni}.

	For the \gls{brk} and \gls{gpr} interpolations, \gls{mae} is \SIlist{0.00975;0.03626}{\J\per\square\m}, respectively. Likewise, \gls{rmse} is \SIlist{0.01727;0.04972}{\J\per\square\m}, respectively. From \cref{fig:resubloss-ni}a, we see that low \gls{gbe} is predicted more accurately and high \gls{gbe} less accurately with \gls{brk} interpolation vs. \gls{gpr} in the \gls{vfzo} framework. Without access to the original fitting routines used to produce the \gls{brk} function, we have not performed \gls{loocv} which would allow for a safer model evaluation (i.e. one in which fair results are less likely due to overfitting). \Gls{loocv} results for the \gls{gpr} case are, however, shown in \cref{fig:olmsted-Ni-loocv}, indicating that the model performs much worse in such a data-limited regime at points the model has never seen before.

	\subsubsection{\glsentrytitlecase{gprm}{long} Applied to Metastable Fe Simulation Data}
	\label{sec:results:simulation:gprm}
\begin{figure*}
		\centering
		\includegraphics{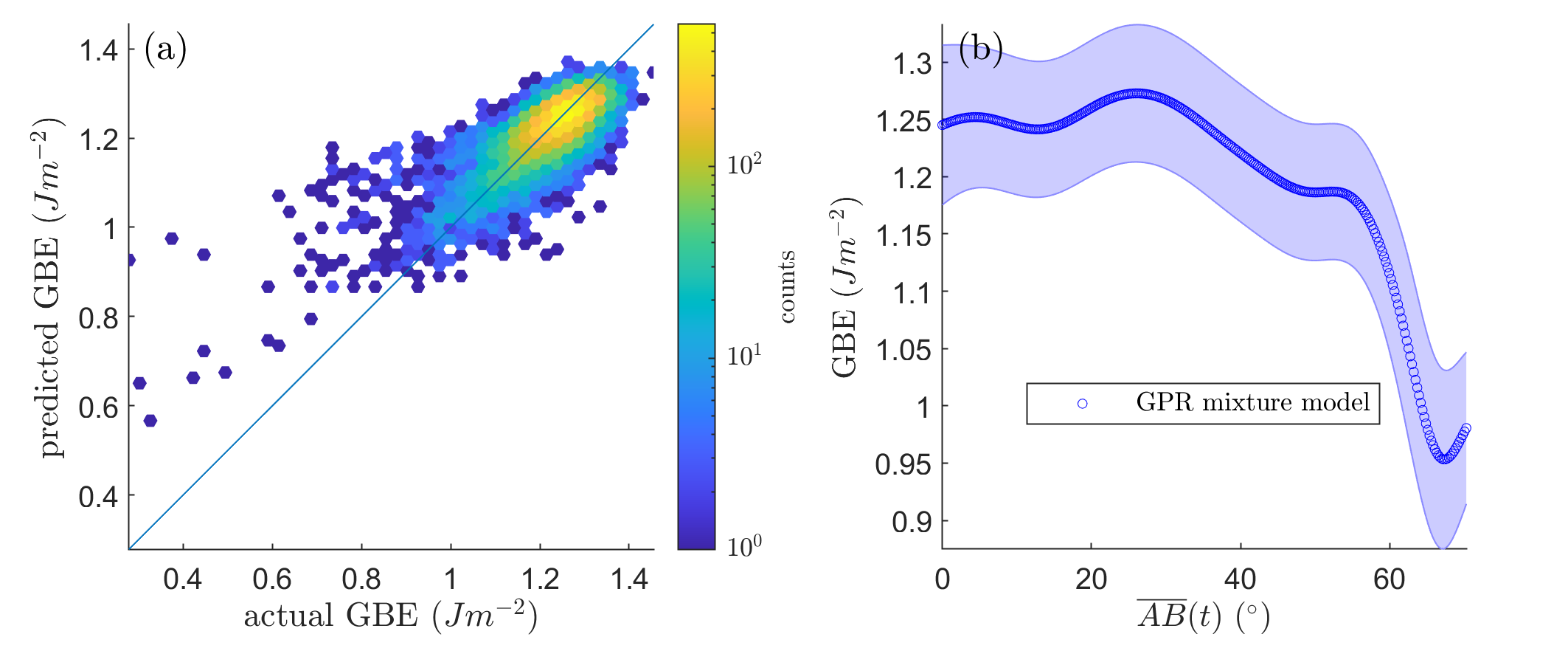}
		\caption{Interpolation results for a large Fe simulation database \cite{kimPhasefieldModeling3D2014} using \num{46883} \inpt{} \glspl{gb} and \num{11721} \outpt{} \glspl{gb} in an 80\%/20\% split and a \gls{gprm} model to better approximate low \glspl{gbe}. Use of a \gls{gprm} model predicts low \gls{gbe} better than the standard \gls{gpr} model (compare with \cref{fig:kim-interp-teach}d). (a) Hexagonally binned parity plot of the \gls{gpr} mixing model with \gls{rmse} and \gls{mae} of \SIlist{0.055035;0.039185}{\J\per\square\meter}, respectively, relative to typical, constant average models of \SIlist{0.0854;0.0617}{\joule\per\square\meter}, respectively. (b) Predictions of \gls{gprm} model (blue circles) as a function of distance along a 1D arc ($\overline{AB}$) between two \glspl{vfzo} ($A$ and $B$). }
		\label{fig:kim-interp}
	\end{figure*}

	In addition to \gls{gpr}, a \gls{gprm} model (\cref{fig:kim-interp-teach}) based on a sigmoid mixing function (\cref{fig:gprmix-sigmoid}) is used to better predict low \gls{gbe} values of the non-uniformly distributed, noisy Fe dataset (\cref{sec:methods:gprmix})\footnote{Alternatively, including \glspl{nbo} may likewise improve low \gls{gbe} performance, but possibly at the expense of high \gls{gbe} predictive accuracy.}. \Gls{gprm} interpolation results for the Fe \gls{gbe} simulations \cite{kimPhasefieldModeling3D2014} are shown in \cref{fig:kim-interp}, where approximate coordinates for the octonions $A$ and $B$ in \cref{fig:kim-interp}b are given in \cref{tab:tunnel-AB2}.

	We find that:
	\begin{itemize}
		\item the model error is on par with the intrinsic error of the data
		\item the predictions likely exhibit overprediction bias relative to the true minimum for a given \gls{gb}
		\item future availability of multiple metastable state \glspl{gbe} is anticipated to greatly improve the model performance
	\end{itemize}
	We now elaborate each of these points.

	\begin{table*}
		\centering
		\caption{Approximate coordinates of \glspl{vfzo} $A$ and $B$ used for the \gls{ms} Fe simulation dataset interpolation in \cref{fig:kim-interp}. Individual quaternions of each octonion are given in the laboratory reference frame with an assumed \gls{gb} normal pointing in the +z direction, also in the laboratory reference frame.}
		\label{tab:tunnel-AB2}
		\begin{tabular}{lllllllll}
			\hline
			Octonion & o(1)   & o(2)    & o(3)    & o(4)    & o(5)    & o(6)   & o(7)    & o(8)   \\ \hline
			A        & 0.8716 & -0.4124 & -0.1857 & 0.1893  & 0.3146  & 0.8359 & -0.3815 & 0.2382 \\
			B        & 0.4391 & -0.7856 & -0.4142 & -0.1360 & -0.1376 & 0.8082 & -0.3705 & 0.4366 \\ \hline
		\end{tabular}
	\end{table*}

	First, because only a single metastable state was used for each \gls{gbe} simulation, both the training and validation data are subject to noise, consistent with a wide lateral spread of predictions in both \cref{fig:kim-interp} and the intrinsic error estimation (\cref{fig:kim-interp-degeneracy-results}). The Fe simulation dataset \gls{gprm} model gives lower \gls{rmse} (\SI{0.055035}{\joule\per\square\meter}) and \gls{mae} (\SI{0.039185}{\joule\per\square\meter}) than the intrinsic error estimates. This indicates that the intrinsic error itself is somewhat overestimated\footnote{The \outpt{} error of a model typically cannot be less than the noise of the \outpt{} data of a model even if the model is estimating the true \outpt{} values with better accuracy than the noise (which is very possible and even expected with \gls{gpr} models when the noise in the \inpt{} data is approximately Gaussian).}. The fact that both model and intrinsic error metrics are relatively close and the prediction and intrinsic error parity plots (\cref{fig:kim-interp} and \cref{fig:kim-interp-degeneracy-results}b, respectively) are similar suggests that the model is performing well. It also suggests that further improvements in the model relative to the "true" values will be "hidden", i.e. they will probably not manifest as lower \gls{rmse} or \gls{mae} nor as more tightly distributed parity plots, etc.

	Next, given the theoretical existence of a true minimum \gls{gbe} for a given \gls{gb}, the predictions which were based on metastable \glspl{gbe} can be assumed to have an overprediction bias relative to the true minimum. On average, we expect this overprediction bias relative to the true minimum \gls{gbe} (rather than the most likely metastable state) may be on the order of a few hundred \SI{}{\milli\J\per\square\meter} and may vary as a function of true minimum \gls{gbe}. In other words, the model obtained is probably an estimate of the most likely metastable \gls{gbe} rather than the true minimum \gls{gbe}. This is akin to saying that we obtain from this data a model that approximates the non-equilibrium, Stillinger quenched red curve of Figure 4(c1) in \cite{hanGrainboundaryMetastabilityIts2016}, not the minimum \gls{gbe} blue curve of the same chart. See \cite{hanGrainboundaryMetastabilityIts2016} for an in-depth treatment of equilibrium and metastable \gls{gbe}.

	Finally, datasets where multiple metastable \glspl{gbe} (e.g. 3-10 repeats) are provided for each \gls{gb} will likely greatly improve the performance of the \gls{gpr} model in predicting either the most likely metastable \gls{gbe} (when all \glspl{gbe} are considered) or the true minimum \gls{gbe} (when only the minimum \gls{gbe} is considered for each \gls{gb}) and may even negate the need for a \gls{gprm} approach. Thus, it is suggested that, where feasible, future large-scale \gls{gb} bicrystal simulation studies will report all property data for repeated trial runs rather than a single trial run or a single value from a set of trial runs. Ideally, data for the three additional microscopic \glspl{dof} for \glspl{gb} (which falls into the category of epistemic uncertainty in this work) would also be included. We believe it is likely that minimum energy paths (i.e. paths of steepest descent) in the \gls{gbe} landscape depend on both macroscopic and microscopic \glspl{dof} (in total, 8DOF) and could offer a more holistic view of \gls{gb} behavior that better mimics and explains experimental grain growth observations. Indeed, it has been experimentally observed that at least some \gls{gb} migration mechanisms involve structural transformations between equilibrium \glspl{gb} via metastable states \cite{weiDirectImagingAtomistic2021}.

	\section{Conclusion} \label{sec:conclusion}
	In this work, we presented the \gls{vfzo} framework for (i) computing distances between GBs and (ii) predicting the properties of GBs from existing measurements. We found that distance calculations in the \gls{vfzo} framework are dramatically more computationally efficient 
	than traditional methods 
	at the expense of infrequent, large distance overestimation which can be addressed through ensemble techniques at a small computational cost as discussed in \cref{sec:methods:framework:vfz-dist}.

	We also developed and tested a barycentric interpolation method, and adapted three other interpolation methods for use in the \gls{vfzo} framework. We provide an easy-to-use, versatile implementation of our methods through an interpolation function \matlab{interp5DOF.m} written in MATLAB (\url{github.com/sgbaird-5dof/interp}, \cite{bairdFiveDegreeofFreedom5DOF2020}) and many companion functions in the \vfzorepo{}.  This approach is general and can be applied to any crystal system (any of the 32 crystallographic point groups can be selected by the parameter \matlab{pgnum}\footnote{While our testing focused on cubic point group symmetry, symmetry operators for other point group symmetries were provided in the TutorialCode/crystal\_symmetry\_ops directory of \url{github.com/ichesser/GB\_octonion\_code} (as of commit: f57f9be). Other point groups (in particular those which are noncentrosymmetric) may give rise to differently shaped/larger VFZs for which a Euclidean distance approximation will have the worst case error of 2 vs. the true value of $\pi$ which represent the furthest Euclidean and arc length distances on a unit hypersphere, respectively. The distance type of \matlab{GBdist4.m} can be changed from \matlab{'norm'} to \matlab{'omega'} to address this issue. We plan to investigate symmetries other than cubic in future work.}). The methods described here may be applicable to other distance metrics (see \citet{morawiecDistancesGrainInterfaces2019} for a comprehensive summary of metrics). We also developed a \gls{gprm} model specifically for better low \gls{gbe} prediction using a non-uniformly distributed, noisy dataset.

	Of the interpolation methods that we present in this work, \Gls{gpr} provided the highest accuracy predictions. It also provided higher accuracy predictions than any of the methods in the literature. The \gls{gpr} interpolation errors (\num{50000} \glspl{vfzo}) for the \gls{brk} validation model are about 2.4 times the intrinsic error that would be expected from reconstruction of noise-free, experimental polycrystalline data via \gls{lobpcg} \cite{shenDeterminingGrainBoundary2019} (\num{180000} \glspl{gb}) with their simpler validation model. Moreover, the interpolation errors for a Fe simulation dataset are on par with the intrinsic errors of the dataset itself (\cref{sec:supp:kim-interp:quality}). While \gls{idw} and \gls{nn} interpolation have the fastest computation times, they also have higher interpolation error. Consequently, we recommend the \gls{gpr} interpolation method for the \gls{vfzo} framework for most applications because it provides the best combination of accuracy and speed and handles \inpt{} noise; however, the other methods can meet niche needs. For example, barycentric interpolation enables rapid and accurate predictions when the function to be evaluated changes, but the \inpt{} and \outpt{} \glspl{gb} remain fixed.

	We anticipate that the \gls{vfzo} framework and corresponding implementation will benefit numerous applications related to \gls{gb} structure and properties, including facilitating \gls{gb} structure-property model development, enabling efficient surrogate modeling of \gls{gb} properties, and larger scale iterative simulations that require repetitive evaluation of computationally expensive structure-property models.

	\section*{Acknowledgement}
	\label{sec:acknowledgement}

	The authors thank Ian Chesser, Toby Francis, Victoria Baird, Brandon Snow, and José Niño for useful discussions. This work was supported by the National Science Foundation under Grant No. 1610077. This work was supported in part through computational resources provided by Brigham Young University's Office of Research Computing.

	\section*{CRediT Statement}
	\textbf{Sterling Baird}: Conceptualization, Methodology, Software, Validation, Formal analysis, Investigation, Data Curation, Writing - Original Draft, Writing - Review \& Editing, Visualization. \textbf{Oliver Johnson}: Supervision, Project administration, Funding acquisition, Conceptualization, Writing - Review \& Editing. \textbf{David Fullwood}: Funding acquisition, Writing - Review \& Editing. \textbf{Eric Homer}: Funding acquisition, Writing - Review \& Editing


	\begin{appendices}

		\crefalias{section}{appsec}
		\crefalias{subsection}{appsec}
		\crefalias{subsubsection}{appsec}

		\section{Active vs. Passive Convention}
		\label{sec:app:convention}
		Misorientation quaternions are represented in the active sense\footnote{The passive convention is used in \cite{francisGeodesicOctonionMetric2019}}:
		\begin{equation}
			q_m = {q_A}^{-1}q_B
		\end{equation}
		where $q_m$, $q_A$, and $q_B$ represent the misorientation quaternion, orientation quaternion of grain A in the sample frame, and orientation quaternion of grain B in the sample frame, respectively. The $^{-1}$ operator denotes a unit quaternion inverse (identical to conjugation of a unit quaternion). Quaternion multiplication is given by equation 23 of \cite{rowenhorstConsistentRepresentationsConversions2015}
		\begin{equation}
			p q \equiv\left(p_{0} q_{0}-\mathbf{p} \cdot \mathbf{q}, q_{0} \mathbf{p}+p_{0} \mathbf{q}+P \mathbf{p} \times \mathbf{q}\right)
		\end{equation}
		where $q_0$ and $p_0$ are scalar components of the quaternions, and $\mathbf{q}$ and $\mathbf{p}$ are the vector components.

		In this work, we use the convention that $P=1$ throughout the various operations in the \vfzorepo{} ($P \equiv $ \matlab{epsijk}) and highly encourage interested readers to refer to \citet{rowenhorstConsistentRepresentationsConversions2015} to understand the redefined versions of quaternion multiplication, quaternion rotation, nuances associated with use of active vs. passive conventions, etc. \Gls{bp} unit normals are expressed pointing away from grain A and in the reference frame of grain A (i.e. the outward-pointing normal convention).

		\section{Detailed Barycentric Interpolation Method}
		\label{sec:app:bary}
		\renewcommand\thefigure{\thesection.\arabic{figure}}
		\setcounter{figure}{0}

		We describe barycentric interpolation applied in the \gls{vfzo} framework in more detail. This includes:
		\begin{itemize}
			\item[1.] triangulation of a \gls{vfz} mesh (\cref{sec:app:bary:tri})
			\item[2.] finding intersections between arbitrary \glspl{vfzo} and the \gls{vfz} mesh (i.e. finding intersecting facets) (\cref{sec:app:bary:int})
			\item[3.] calculating interpolated values of an arbitrary \gls{vfzo} property using the intersecting facet (\cref{sec:app:bary-interp})
		\end{itemize}

		\subsection{Triangulating a \glsentrytitlecase{vfz}{long} Mesh}
		\label{sec:app:bary:tri}

		Creation of a simplicial mesh is necessary to perform barycentric interpolation. Due to the difficulty of visualizing a 7-sphere, we provide visual illustrations of the process as applied to lower-dimensional analogues. After \glspl{gbo} have been symmetrized into a \gls{vfz} (\cref{sec:methods:framework:vfz}), the triangulation process occurs by:
		\begin{enumerate}
			\item[1.1] applying a \gls{svd} transformation to remove the U(1)-symmetry degeneracy inherent in the \gls{vfzo} coordinates (\cref{sec:app:bary:tri:svd1})
			\item[1.2] linearly projecting \glspl{vfzo} onto a hyperplane that is tangent to the vector between the origin and the mean of the \inpt{} \glspl{vfzo} to reduce computational burden of the triangulation
			\item[1.3] performing a second \gls{svd} transformation (\cref{sec:app:bary:tri:svd2})
			\item[1.4] computing the triangulation according to the quickhull algorithm \cite{barberQuickhullAlgorithmConvex1996} using built-in methods
		\end{enumerate}

\begin{figure*}
			\centering
			\includegraphics[scale=1]{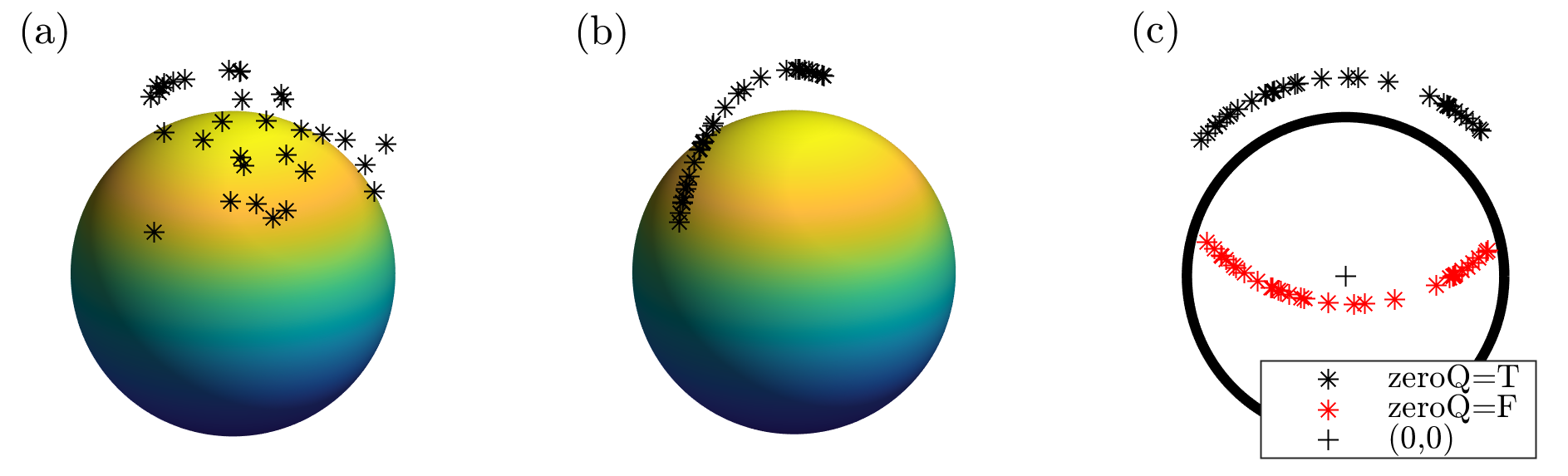}
			\caption{3D Cartesian to 2D Cartesian analogue of 8D Cartesian to 7D Cartesian degeneracy removal via rigid \gls{svd} transformation as used in barycentric interpolation approach. (a) Starting spherical arc points on surface of 2-sphere, (b) rotational symmetrization applied w.r.t. z-axis (analogous to U(1) symmetrization), and (c) degenerate dimension removed via \glsxtrlong{svd} transformation to 2D Cartesian with either the origin (black plus) preserved (black asterisks, \matlab{zeroQ=T}) for triangulation or ignored (red asterisks, \matlab{zeroQ=F}) for mesh intersection. The spheres (a,b) and circle (c) each have a radius of 0.8 and are used as a visualization aid only.}
			\label{fig:bary-remove-deg}
		\end{figure*}

\begin{figure*}
			\centering
			\includegraphics[scale=1]{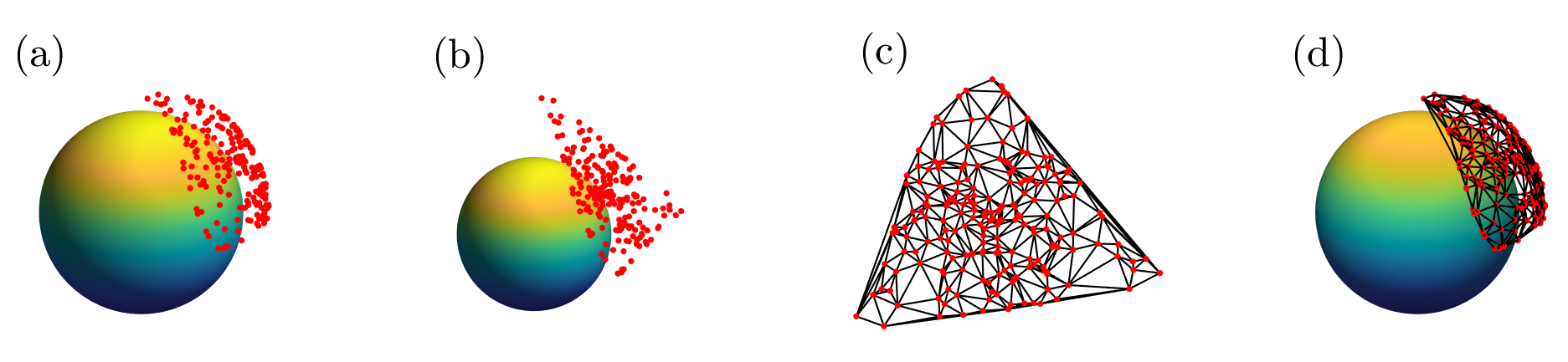}
			\caption{3D Cartesian to 2D Cartesian analogue of 7D Cartesian to 6D Cartesian mesh triangulation used in barycentric interpolation approach. (a) 3D Cartesian \inpt{} points are (b) linearly projected onto hyperplane that is tangent to mean of starting points. (c) The degenerate dimension is removed via a rigid \gls{svd} transformation to 2D Cartesian and the Delaunay triangulation (black lines) is calculated, with \inpt{} vertices (red). Delaunay triangulation superimposed onto normalized \inpt{} points (d). The spheres in (a), (b), and (d) have a radius of 0.8 and are used for visualization only.}
			\label{fig:bary-delaunay}
		\end{figure*}

		In the explanation of each of these steps that follows, we make reference to lower-dimensional visual analogues of the \gls{vfzo} triangulation procedure, which are given in \cref{fig:bary-remove-deg}, \cref{fig:bary-delaunay}, and \cref{fig:bary-interp}. We note that 3D Cartesian coordinates in \cref{fig:bary-remove-deg} correspond to 8D Cartesian coordinates, whereas 3D Cartesian coordinates in \cref{fig:bary-delaunay} and \cref{fig:bary-interp} correspond to 7D Cartesian coordinates. This is intentional for two reasons:
		\begin{itemize}
			\item \cref{fig:bary-remove-deg} illustrates that unsymmetrized 8D Cartesian \glspl{gbo} are analogous to a point cloud on the 2-sphere (\cref{fig:bary-remove-deg}a) and that an 8D Cartesian \gls{vfzo} set, which has already been symmetrized, is analogous to a geodesic arc on the 2-sphere (\cref{fig:bary-remove-deg}b). A \gls{vfzo} set has a degenerate dimension that can then be removed by a rigid \gls{svd} transformation to 7D Cartesian coordinates (analogous to 2D Cartesian coordinates in \cref{fig:bary-remove-deg}c). This sequence would be more difficult to visualize if \cref{fig:bary-remove-deg}a was meant to represent a point cloud on the 3-sphere (4D Cartesian coordinates), etc.
			\item \cref{fig:bary-delaunay} illustrates a second transformation from normalized 7D Cartesian coordinates (\cref{fig:bary-delaunay}a) to a hyperplane (\cref{fig:bary-delaunay}b) which is then transformed into 6D Cartesian coordinates via a second \gls{svd}. In this case, key issues are retained that would otherwise be lost (\cref{sec:supp:bary:artifact}) if an arc on a circle (1-sphere) to 1D Cartesian coordinates were used instead\footnote{Non-intersection issues due to high-aspect ratios and consideration of facets connected up to \matlab{nnMax} \glspl{nn} do not manifest in triangulations on the surface of a 1-sphere because one of the two facets (i.e. line segments) connected to the first \gls{nn} mesh vertex relative to the \outpt{} point is guaranteed to have an intersection.}. Additionally, the use of actual triangles is a more familiar and compelling illustration of \textit{triangulation}.
		\end{itemize}

		While lower dimensional analogues are useful for visualizing and understanding the process of triangulation, a written description is also given in the following sections. As appropriate, we refer back to the teaching figures described in this section.

		\subsubsection{\glsentrytitlecase{svd}{long} Transformation from 8D Cartesian to 7D Cartesian}
		\label{sec:app:bary:tri:svd1}
		To reduce the computational complexity of triangulating a high-dimensional mesh \cite{barberQuickhullAlgorithmConvex1996}, some simplifications are made. First, the degenerate octonion dimension obtained from analytically minimizing $U(1)$ symmetry \cite{francisGeodesicOctonionMetric2019} is removed via a rigid (i.e. distance- and angle-preserving) \gls{svd} transformation,
		analogous to a Cartesian rotation and translation (see 3D to 2D \gls{svd} transformation from \cref{fig:bary-remove-deg}b to \cref{fig:bary-remove-deg}c).

		\subsubsection{Linearly Project onto Hyperplane}
		\label{sec:app:bary:tri:project}
		Next, the resulting 7D Cartesian representation of each \gls{vfzo} is projected onto a hyperplane that is tangent to the centroid (i.e. mean) of the \gls{vfzo} set\footnote{This is \textit{not} a rigid transformation; however, it approximates one with sufficient accuracy to produce a high-quality triangulation in a \gls{vfz}.} (\cref{fig:bary-delaunay}a). By performing this linear projection, one of the dimensions becomes degenerate.

		\subsubsection{\glsentrytitlecase{svd}{long} Transformation from 7D Cartesian to 6D Cartesian}
		\label{sec:app:bary:tri:svd2}
		This additional degeneracy is removed via a second \gls{svd} transformation, this time to 6D Cartesian coordinates (see 3D to 2D projection in \cref{fig:bary-delaunay}a-b). Finally, the resulting points can be triangulated via the quickhull algorithm \cite{barberQuickhullAlgorithmConvex1996} (see \vfzorepo{} function \matlab{sphconvhulln.m} and built-in MATLAB function \matlab{delaunayn()}), which relies on Euclidean distances\footnote{While the triangulation algorithm used in this work relies on Euclidean distances (the use of which is possible via the \gls{vfzo} framework), other distance metrics that are non-Euclidean \cite{morawiecDistancesGrainInterfaces2019} could potentially be incorporated into the barycentric approach such as by doing an edge-length based simplex reconstruction \cite{connorHighdimensionalSimplexesSupermetric2017,boissonnatOnlyDistancesAre2017} using the \gls{vfz} triangulation edge lengths.}. Because the simplicial mesh is defined by a list of edges between vertices for each simplicial facet, this list applies immediately to the \gls{vfzo} set in its 7D Cartesian coordinates (i.e. no reverse transformation is necessary to use the mesh on the 6-sphere in 7D).

		\subsection{Intersections in a \glsentrytitlecase{vfz}{long} Mesh}
		\label{sec:app:bary:int}

		Once the triangulation has been determined, we need to find which facet each \outpt{} point intersects (i.e. find the intersecting facet). There are two sub-steps:
		\begin{itemize}
			\item[2.1] applying the same rigid transformation to the \outpt{} points as was applied to the \inpt{} points (otherwise the \outpt{} points won't line up properly with the mesh) (\cref{sec:app:bary:int:out-svd})
			\item[2.2] identifying facets nearby a \outpt{} point and testing for intersection (\cref{sec:app:bary:int:facets}).
		\end{itemize}
		\subsubsection{Apply Same \glsentrytitlecase{svd}{long} to Input and Prediction Points}
		\label{sec:app:bary:int:out-svd}
		The positions of the \outpt{} points need to be fixed relative to the mesh even after the rigid \gls{svd} transformation. 
		This is accomplished by:
		\begin{itemize}
			\item[2.1a] concatenating both \inpt{} and
			\outpt{} points
			\item[2.1b] using the \matlab{interp5DOF.m} sub-routine \matlab{proj\_down.m} (which depends on MATLAB's built-in \gls{svd} implementation \matlab{svd()}) to perform the transformation
			\item[2.1c] subsequently separating the transformed \inpt{} and \outpt{} points (reverse of concatenation step)
		\end{itemize}

		To map new points onto the mesh, the \matlab{usv} structure output from \matlab{proj\_down.m} needs to be stored and supplied in future calls to \matlab{proj\_down.m}. Likewise, \matlab{usv} need to be supplied to \matlab{proj\_up.m} to perform the reverse \gls{svd} transformation.

		\subsubsection{Testing Nearby Facets for Intersections}
		\label{sec:app:bary:int:facets}
\begin{figure}
			\centering
			\includegraphics[scale=1]{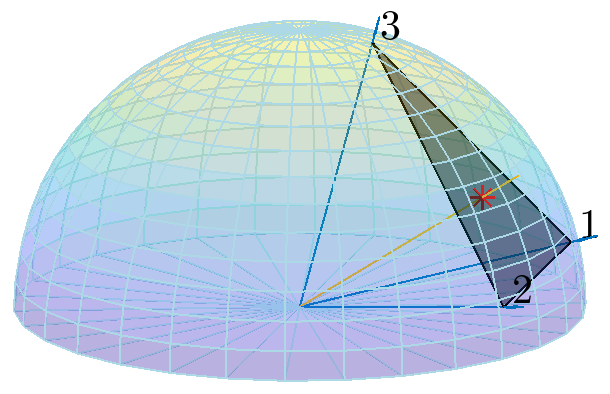}
			\caption{A ray (red line) is linearly projected from the 2-sphere onto the hyperplane of a mesh facet (transparent black), shown as a red asterisk. The barycentric coordinates are computed as $\lambda_{i \in [1,3]} = \frac{1}{3}$. Because all barycentric coordinates are positive, it is determined that the projected point is an intersection with the mesh. Given vertex values of \num{8.183}, \num{3.446}, and \num{3.188} for vertices 1, 2, and 3, respectively, the interpolated value is calculated as \num{4.94} via \cref{eq:bary-interp}.}
			\label{fig:bary-interp}
		\end{figure}

		Once the \outpt{} points are lined up properly with the mesh, the facet containing the \outpt{} point (i.e. intersecting facet) is found. We define the intersecting facet as the one for which a point's barycentric coordinates are positive within a given tolerance. Consequently, we determine facet affiliation by:
		\begin{enumerate}
			\item[2.2a] linearly projecting the \outpt{} point onto the hyperplane defined by a mesh facet's vertices (\cref{fig:bary-interp})
			\item[2.2b] computing the point's barycentric coordinates within the facet \cite{anatoliyCheckIfRay2015,skalaRobustBarycentricCoordinates2013} (see \vfzorepo{} function \matlab{projray2hypersphere.m})
			\item[2.2c] testing that all coordinates are positive \cite{langerSphericalBarycentricCoordinates2006} within a tolerance\footnote{Two tolerances are used: one for the initial computation of barycentric coordinates by projecting onto the hypersphere to determine facet affiliation (\matlab{projtol=1e-4}) and a larger tolerance (\matlab{inttol=1e-2}) for computation of barycentric coordinates to determine interpolated values (\cref{sec:app:bary-interp}). }
			\item[2.2d] repeating steps 2.2a-2.2c until an intersection is found or a stop condition is reached (see \matlab{nnMax} below).
		\end{enumerate}

		For further information on barycentric coordinates and its applications and generalizations, see \cite{anisimovSubdividingBarycentricCoordinates2016,budninskiyPowerCoordinatesGeometric2016,dyerBarycentricCoordinateNeighbourhoods2016,floaterGeneralizedBarycentricCoordinates2015,floaterInjectivityWachspressMean2010,hormannDiscretizingWachspressKernels2017,hormannMaximumEntropyCoordinates2008,langerHigherOrderBarycentric2008,langerSphericalBarycentricCoordinates2006,leiNewCoordinateSystem2020,meyerGeneralizedBarycentricCoordinates2002,peixotoVectorFieldReconstructions2014,pihajokiBarycentricInterpolationRiemannian2019,rustamovBarycentricCoordinatesSurfaces2010,skalaRobustBarycentricCoordinates2013,taoFastNumericalSolver2019,warrenBarycentricCoordinatesConvex2007}.

		Due to the large number of facets per point of a high-dimensional triangulation (approximately \num{2000} facets per vertex for a \num{50000} point \gls{vfz} triangulation, or \num{1e8} total facets), some simplifications are made in order to determine intersections of \outpt{} points with the mesh. If every edge length of every facet were equal, only facets connected to the first \gls{nn} would need to be considered to find a proper intersection. However, since the \glspl{vfzo} are randomly sampled, edge lengths of facets are non-uniform, and non-unity aspect-ratio facets exist (\cref{fig:bary-delaunay}, \cref{fig:high-aspect-non-int}). If the facets have high-aspect ratios, the intersecting facets of \outpt{} points can be far from the \glspl{nn} mesh points relative to the \outpt{} points (see \cref{fig:high-aspect-non-int} inset), especially near the perimeter of a hyperspherical surface mesh. Rather than loop through every facet to find an intersection ($\sim$\num{1e8} facets in a \num{50000} \gls{vfzo} mesh), the \outpt{} point intersections are calculated by considering facets connected to up to some number of \gls{nn} mesh vertices (\matlab{nnMax}) relative to each \outpt{} point (in this work, \matlab{nnMax=10}). The \gls{nn} mesh vertices relative to a \outpt{} point are computed via the MATLAB built-in function \matlab{dsearchn} as in the \gls{nn} approach (\cref{sec:methods:interp:nn}). The facet IDs of facets connected to these \glspl{nn} are computed by calling built-in MATLAB function \matlab{find()}, as in \matlab{find(K==nn)}, where \matlab{K} is the triangulation from \vfzorepo{} function \matlab{sphconvhulln.m} and \matlab{nn} is the ID of one of the \gls{nn} mesh vertices.

		Some \outpt{} points will have no intersecting facet found.
		From our numerical testing, we determine that this non-intersection phenomenon occurs in two situations:
		\begin{itemize}
			\item high-aspect ratio facets (described above)
			\item \outpt{} points that are positioned just outside the bounds of the mesh but within the bounds of the \gls{vfz}, due to the fact that the mesh is a piecewise linear approximation of a surface with a curved perimeter and that randomly sampled points typically do not fall on the true perimeter
		\end{itemize}
		In the first case, barycentric interpolation within high-aspect ratio facets may actually lead to worse interpolation error than a \gls{nn} interpolation strategy due to influence by \glspl{gb} far from the \outpt{} point. In the second case, there is no true intersection between the \outpt{} point and the mesh. Both issues can be addressed with the same strategy: we apply a \gls{nn} approach (\cref{sec:methods:interp:nn}) when an intersecting facet is not found within \matlab{nnMax} \glspl{nn}. In numerical tests, \gls{vfz} meshes composed of \num{388} and \num{50000} vertices produced non-intersection rates of \SI{12.07 \pm 1.02}{\percent} and \SI{0.68 \pm 0.11}{\percent}, respectively, over approximately \num{10} trials and using \num{10000} \outpt{} points for each trial.

		Testing intersections for nearby facets is handled in the \vfzorepo{} function \matlab{intersect\_facet.m} and depends on the barycentric coordinate computations in \matlab{projray2hypersphere.m}.

		\subsection{Interpolation via Barycentric Coordinates}
		\label{sec:app:bary-interp}

		Once a mesh triangulation has been determined (\cref{sec:app:bary:tri}), barycentric coordinates are recomputed for a \outpt{} point within the \inpt{} mesh (\cref{sec:app:bary:int}) using a somewhat larger tolerance; the interpolated value is found by taking the dot product of the \outpt{} point's barycentric coordinates and the properties of the corresponding vertices of the intersecting facet via
		\begin{equation}
			\label{eq:bary-interp}
			v_{m,q}=\underset{i=1}{\overset{N}{\sum }}\lambda _{m,i} v_{m,i}
		\end{equation}
		where $\lambda_{m,i}$, $v_{m,q}$, $v_{m,i}$ and $N$, are the barycentric coordinates of the m-th \outpt{} point, interpolated property at the m-th \outpt{} point, property of the $i$-th vertex of the intersecting facet for the m-th \outpt{} point, and number of vertices in a given facet ($N = 7$ for facets of the simplicial mesh on the degeneracy-free 6-sphere), respectively. Interpolation of many \outpt{} points simultaneously can be accomplished by a simple, vectorized approach via MATLAB built-in function \matlab{dot()} as used in \vfzorepo{} function \matlab{interp\_bary\_fast.m}. This function assumes triangulation and weights have been precomputed. In other words, both \inpt{} and \outpt{} coordinates remain fixed, and only \inpt{} property values change. If this is the case, barycentric interpolation of new points is incredibly fast. By contrast, if \inpt{} coordinates change, the triangulation must be recomputed, and if \outpt{} coordinates change, the intersecting facets must be recomputed. Both triangulation and finding intersecting facets are computationally demanding with respect to memory and runtime (\cref{sec:results:efficiency}).

	\end{appendices}

	\printglossaries

	\bibliographystyle{elsarticle-num-names}
	\bibliography{5dof-gb-energy.bib}

\newcommand{\noopsort}[1]{}
\begin{thebibliography}{95}
\expandafter\ifx\csname natexlab\endcsname\relax\def\natexlab#1{#1}\fi
\providecommand{\url}[1]{\texttt{#1}}
\providecommand{\href}[2]{#2}
\providecommand{\path}[1]{#1}
\providecommand{\DOIprefix}{doi:}
\providecommand{\ArXivprefix}{arXiv:}
\providecommand{\URLprefix}{URL: }
\providecommand{\Pubmedprefix}{pmid:}
\providecommand{\doi}[1]{\href{http://dx.doi.org/#1}{\path{#1}}}
\providecommand{\Pubmed}[1]{\href{pmid:#1}{\path{#1}}}
\providecommand{\bibinfo}[2]{#2}
\ifx\xfnm\relax \def\xfnm[#1]{\unskip,\space#1}\fi
\bibitem[{Jin et~al.(2018)Jin, Huang, Kwon, Zhang, Li, Oh, Dong, Luo, Biswal,
  Cunning, Bakharev, Moon, Yoo, {Camacho-Mojica}, Kim, Lee, Wang, Seong,
  Saxena, Ding, Shin, and Ruoff}]{jinColossalGrainGrowth2018}
\bibinfo{author}{S.~Jin}, \bibinfo{author}{M.~Huang},
  \bibinfo{author}{Y.~Kwon}, \bibinfo{author}{L.~Zhang}, \bibinfo{author}{B.~W.
  Li}, \bibinfo{author}{S.~Oh}, \bibinfo{author}{J.~Dong},
  \bibinfo{author}{D.~Luo}, \bibinfo{author}{M.~Biswal}, \bibinfo{author}{B.~V.
  Cunning}, \bibinfo{author}{P.~V. Bakharev}, \bibinfo{author}{I.~Moon},
  \bibinfo{author}{W.~J. Yoo}, \bibinfo{author}{D.~C. {Camacho-Mojica}},
  \bibinfo{author}{Y.~J. Kim}, \bibinfo{author}{S.~H. Lee},
  \bibinfo{author}{B.~Wang}, \bibinfo{author}{W.~K. Seong},
  \bibinfo{author}{M.~Saxena}, \bibinfo{author}{F.~Ding},
  \bibinfo{author}{H.~J. Shin}, \bibinfo{author}{R.~S. Ruoff},
\newblock \bibinfo{title}{Colossal grain growth yields single-crystal metal
  foils by contact-free annealing},
\newblock \bibinfo{journal}{Science} \bibinfo{volume}{362}
  (\bibinfo{year}{2018}) \bibinfo{pages}{1021--1025}.
  \DOIprefix\doi{10.1126/science.aao3373}.
\bibitem[{Brandenburg et~al.(2014)Brandenburg, {Barrales-Mora}, and
  Molodov}]{brandenburgMigrationFacetingLowangle2014}
\bibinfo{author}{J.~E. Brandenburg}, \bibinfo{author}{L.~A. {Barrales-Mora}},
  \bibinfo{author}{D.~A. Molodov},
\newblock \bibinfo{title}{On migration and faceting of low-angle grain
  boundaries: {{Experimental}} and computational study},
\newblock \bibinfo{journal}{Acta Materialia} \bibinfo{volume}{77}
  (\bibinfo{year}{2014}) \bibinfo{pages}{294--309}.
  \DOIprefix\doi{10.1016/j.actamat.2014.06.006}.
\bibitem[{Huang et~al.(2015)Huang, Bartels, Xu, Osterhoff, Kalbfleisch, Sprung,
  Suzuki, Takahashi, Blanton, Salditt, and
  Miao}]{huangGrainRotationLattice2015}
\bibinfo{author}{Z.~Huang}, \bibinfo{author}{M.~Bartels},
  \bibinfo{author}{R.~Xu}, \bibinfo{author}{M.~Osterhoff},
  \bibinfo{author}{S.~Kalbfleisch}, \bibinfo{author}{M.~Sprung},
  \bibinfo{author}{A.~Suzuki}, \bibinfo{author}{Y.~Takahashi},
  \bibinfo{author}{T.~N. Blanton}, \bibinfo{author}{T.~Salditt},
  \bibinfo{author}{J.~Miao},
\newblock \bibinfo{title}{Grain rotation and lattice deformation during
  photoinduced chemical reactions revealed by in situ {{X}}-ray
  nanodiffraction},
\newblock \bibinfo{journal}{Nature Materials} \bibinfo{volume}{14}
  (\bibinfo{year}{2015}) \bibinfo{pages}{691--695}.
  \DOIprefix\doi{10.1038/nmat4311}.
\bibitem[{Trautt and Mishin(2014)}]{trauttCapillarydrivenGrainBoundary2014}
\bibinfo{author}{Z.~Trautt}, \bibinfo{author}{Y.~Mishin},
\newblock \bibinfo{title}{Capillary-driven grain boundary motion and grain
  rotation in a tricrystal: {{A}} molecular dynamics study},
\newblock \bibinfo{journal}{Acta Materialia} \bibinfo{volume}{65}
  (\bibinfo{year}{2014}) \bibinfo{pages}{19--31}.
  \DOIprefix\doi{10.1016/j.actamat.2013.11.059}.
\bibitem[{Sharma et~al.(2012)Sharma, Huizenga, Bytchkov, Sietsma, and
  Offerman}]{sharmaObservationChangingCrystal2012}
\bibinfo{author}{H.~Sharma}, \bibinfo{author}{R.~M. Huizenga},
  \bibinfo{author}{A.~Bytchkov}, \bibinfo{author}{J.~Sietsma},
  \bibinfo{author}{S.~E. Offerman},
\newblock \bibinfo{title}{Observation of changing crystal orientations during
  grain coarsening},
\newblock \bibinfo{journal}{Acta Materialia} \bibinfo{volume}{60}
  (\bibinfo{year}{2012}) \bibinfo{pages}{229--237}.
  \DOIprefix\doi{10.1016/j.actamat.2011.09.057}.
\bibitem[{Ware et~al.(2018)Ware, Suzuki, Wicker, and
  Cordero}]{wareGrainBoundaryPlane2018}
\bibinfo{author}{L.~G. Ware}, \bibinfo{author}{D.~H. Suzuki},
  \bibinfo{author}{K.~R. Wicker}, \bibinfo{author}{Z.~C. Cordero},
\newblock \bibinfo{title}{Grain boundary plane manipulation in directionally
  solidified bicrystals and tricrystals},
\newblock \bibinfo{journal}{Scripta Materialia} \bibinfo{volume}{152}
  (\bibinfo{year}{2018}) \bibinfo{pages}{98--101}.
  \DOIprefix\doi{10.1016/j.scriptamat.2018.03.047}.
\bibitem[{Li et~al.(2017)Li, Oudriss, Metsue, Bouhattate, and
  Feaugas}]{liAnisotropyHydrogenDiffusion2017}
\bibinfo{author}{J.~Li}, \bibinfo{author}{A.~Oudriss},
  \bibinfo{author}{A.~Metsue}, \bibinfo{author}{J.~Bouhattate},
  \bibinfo{author}{X.~Feaugas},
\newblock \bibinfo{title}{Anisotropy of hydrogen diffusion in nickel single
  crystals: The effects of self-stress and hydrogen concentration on
  diffusion},
\newblock \bibinfo{journal}{Scientific Reports} \bibinfo{volume}{7}
  (\bibinfo{year}{2017}) \bibinfo{pages}{45041}.
  \DOIprefix\doi{10.1038/srep45041}.
\bibitem[{Oudriss et~al.(2012)Oudriss, Creus, Bouhattate, Conforto, Berziou,
  Savall, and Feaugas}]{oudrissGrainSizeGrainboundary2012}
\bibinfo{author}{A.~Oudriss}, \bibinfo{author}{J.~Creus},
  \bibinfo{author}{J.~Bouhattate}, \bibinfo{author}{E.~Conforto},
  \bibinfo{author}{C.~Berziou}, \bibinfo{author}{C.~Savall},
  \bibinfo{author}{X.~Feaugas},
\newblock \bibinfo{title}{Grain size and grain-boundary effects on diffusion
  and trapping of hydrogen in pure nickel},
\newblock \bibinfo{journal}{Acta Materialia} \bibinfo{volume}{60}
  (\bibinfo{year}{2012}) \bibinfo{pages}{6814--6828}.
  \DOIprefix\doi{10.1016/j.actamat.2012.09.004}.
\bibitem[{Metsue et~al.(2016)Metsue, Oudriss, and
  Feaugas}]{metsueHydrogenSolubilityVacancy2016}
\bibinfo{author}{A.~Metsue}, \bibinfo{author}{A.~Oudriss},
  \bibinfo{author}{X.~Feaugas},
\newblock \bibinfo{title}{Hydrogen solubility and vacancy concentration in
  nickel single crystals at thermal equilibrium: {{New}} insights from
  statistical mechanics and ab initio calculations},
\newblock \bibinfo{journal}{Journal of Alloys and Compounds}
  \bibinfo{volume}{656} (\bibinfo{year}{2016}) \bibinfo{pages}{555--567}.
  \DOIprefix\doi{10.1016/j.jallcom.2015.09.252}.
\bibitem[{Huang et~al.(2017)Huang, Chen, Song, McDowell, and
  Zhu}]{huangHydrogenEmbrittlementGrain2017}
\bibinfo{author}{S.~Huang}, \bibinfo{author}{D.~Chen},
  \bibinfo{author}{J.~Song}, \bibinfo{author}{D.~L. McDowell},
  \bibinfo{author}{T.~Zhu},
\newblock \bibinfo{title}{Hydrogen embrittlement of grain boundaries in nickel:
  An atomistic study},
\newblock \bibinfo{journal}{npj Computational Materials} \bibinfo{volume}{3}
  (\bibinfo{year}{2017}) \bibinfo{pages}{1--8}.
  \DOIprefix\doi{10.1038/s41524-017-0031-1}.
\bibitem[{Xia et~al.(2011)Xia, Li, Liu, and Zhou}]{xiaApplingGrainBoundary2011}
\bibinfo{author}{S.~Xia}, \bibinfo{author}{H.~Li}, \bibinfo{author}{T.~G. Liu},
  \bibinfo{author}{B.~X. Zhou},
\newblock \bibinfo{title}{Appling grain boundary engineering to {{Alloy}} 690
  tube for enhancing intergranular corrosion resistance},
\newblock \bibinfo{journal}{Journal of Nuclear Materials} \bibinfo{volume}{416}
  (\bibinfo{year}{2011}) \bibinfo{pages}{303--310}.
  \DOIprefix\doi{10.1016/j.jnucmat.2011.06.017}.
\bibitem[{Demkowicz(2020)}]{demkowiczThresholdDensityHelium2020}
\bibinfo{author}{M.~J. Demkowicz},
\newblock \bibinfo{title}{A threshold density of helium bubbles induces a
  ductile-to-brittle transition at a grain boundary in nickel},
\newblock \bibinfo{journal}{Journal of Nuclear Materials} \bibinfo{volume}{533}
  (\bibinfo{year}{2020}) \bibinfo{pages}{152118}.
  \DOIprefix\doi{10.1016/j.jnucmat.2020.152118}.
\bibitem[{Hanson et~al.(2018)Hanson, Bagri, Lind, Kenesei, Suter, Grade{\v
  c}ak, and Demkowicz}]{hansonCrystallographicCharacterGrain2018}
\bibinfo{author}{J.~P. Hanson}, \bibinfo{author}{A.~Bagri},
  \bibinfo{author}{J.~Lind}, \bibinfo{author}{P.~Kenesei},
  \bibinfo{author}{R.~M. Suter}, \bibinfo{author}{S.~Grade{\v c}ak},
  \bibinfo{author}{M.~J. Demkowicz},
\newblock \bibinfo{title}{Crystallographic character of grain boundaries
  resistant to hydrogen-assisted fracture in {{Ni}}-base alloy 725},
\newblock \bibinfo{journal}{Nature Communications} \bibinfo{volume}{9}
  (\bibinfo{year}{2018}) \bibinfo{pages}{1--11}.
  \DOIprefix\doi{10.1038/s41467-018-05549-y}.
\bibitem[{Jothi et~al.(2016)Jothi, Merzlikin, Croft, Andersson, and
  Brown}]{jothiInvestigationMicromechanismsHydrogen2016}
\bibinfo{author}{S.~Jothi}, \bibinfo{author}{S.~V. Merzlikin},
  \bibinfo{author}{T.~N. Croft}, \bibinfo{author}{J.~Andersson},
  \bibinfo{author}{S.~G. Brown},
\newblock \bibinfo{title}{An investigation of micro-mechanisms in hydrogen
  induced cracking in nickel-based superalloy 718},
\newblock \bibinfo{journal}{Journal of Alloys and Compounds}
  \bibinfo{volume}{664} (\bibinfo{year}{2016}) \bibinfo{pages}{664--681}.
  \DOIprefix\doi{10.1016/j.jallcom.2016.01.033}.
\bibitem[{Zhou et~al.(2016)Zhou, Marchand, McDowell, Zhu, and
  Song}]{zhouChemomechanicalOriginHydrogen2016}
\bibinfo{author}{X.~Zhou}, \bibinfo{author}{D.~Marchand},
  \bibinfo{author}{D.~L. McDowell}, \bibinfo{author}{T.~Zhu},
  \bibinfo{author}{J.~Song},
\newblock \bibinfo{title}{Chemomechanical {{Origin}} of {{Hydrogen Trapping}}
  at {{Grain Boundaries}} in fcc {{Metals}}},
\newblock \bibinfo{journal}{Physical Review Letters} \bibinfo{volume}{116}
  (\bibinfo{year}{2016}) \bibinfo{pages}{1--33}.
  \DOIprefix\doi{10.1103/PhysRevLett.116.075502}.
\bibitem[{Huang et~al.(2020)Huang, Shishehbor, {Guar{\'i}n-Zapata}, Kirchhofer,
  Li, Cruz, Wang, Bhowmick, Stauffer, Manimunda, Bozhilov, Caldwell,
  Zavattieri, and Kisailus}]{huangNaturalImpactresistantBicontinuous2020}
\bibinfo{author}{W.~Huang}, \bibinfo{author}{M.~Shishehbor},
  \bibinfo{author}{N.~{Guar{\'i}n-Zapata}}, \bibinfo{author}{N.~D. Kirchhofer},
  \bibinfo{author}{J.~Li}, \bibinfo{author}{L.~Cruz},
  \bibinfo{author}{T.~Wang}, \bibinfo{author}{S.~Bhowmick},
  \bibinfo{author}{D.~Stauffer}, \bibinfo{author}{P.~Manimunda},
  \bibinfo{author}{K.~N. Bozhilov}, \bibinfo{author}{R.~Caldwell},
  \bibinfo{author}{P.~Zavattieri}, \bibinfo{author}{D.~Kisailus},
\newblock \bibinfo{title}{A natural impact-resistant bicontinuous composite
  nanoparticle coating},
\newblock \bibinfo{journal}{Nature Materials} \bibinfo{volume}{19}
  (\bibinfo{year}{2020}) \bibinfo{pages}{1236--1243}.
  \DOIprefix\doi{10.1038/s41563-020-0768-7}.
\bibitem[{Wang et~al.(2018)Wang, Voisin, McKeown, Ye, Calta, Li, Zeng, Zhang,
  Chen, Roehling, Ott, Santala, Depond, Matthews, Hamza, and
  Zhu}]{wangAdditivelyManufacturedHierarchical2018}
\bibinfo{author}{Y.~M. Wang}, \bibinfo{author}{T.~Voisin},
  \bibinfo{author}{J.~T. McKeown}, \bibinfo{author}{J.~Ye},
  \bibinfo{author}{N.~P. Calta}, \bibinfo{author}{Z.~Li},
  \bibinfo{author}{Z.~Zeng}, \bibinfo{author}{Y.~Zhang},
  \bibinfo{author}{W.~Chen}, \bibinfo{author}{T.~T. Roehling},
  \bibinfo{author}{R.~T. Ott}, \bibinfo{author}{M.~K. Santala},
  \bibinfo{author}{P.~J. Depond}, \bibinfo{author}{M.~J. Matthews},
  \bibinfo{author}{A.~V. Hamza}, \bibinfo{author}{T.~Zhu},
\newblock \bibinfo{title}{Additively manufactured hierarchical stainless steels
  with high strength and ductility},
\newblock \bibinfo{journal}{Nature Materials} \bibinfo{volume}{17}
  (\bibinfo{year}{2018}) \bibinfo{pages}{63--71}.
  \DOIprefix\doi{10.1038/nmat5021}.
\bibitem[{Lin et~al.(2016)Lin, Bierbaum, Schall, Sethna, and
  Cohen}]{linMeasuringNonlinearStresses2016}
\bibinfo{author}{N.~Y.~C. Lin}, \bibinfo{author}{M.~Bierbaum},
  \bibinfo{author}{P.~Schall}, \bibinfo{author}{J.~P. Sethna},
  \bibinfo{author}{I.~Cohen},
\newblock \bibinfo{title}{Measuring nonlinear stresses generated by defects in
  {{3D}} colloidal crystals},
\newblock \bibinfo{journal}{Nature Materials} \bibinfo{volume}{15}
  (\bibinfo{year}{2016}) \bibinfo{pages}{1172--1176}.
  \DOIprefix\doi{10.1038/nmat4715}.
\bibitem[{Yin et~al.(2019)Yin, Chen, Saito, Inoue, and
  Ikuhara}]{yinCeramicPhasesOnedimensional2019}
\bibinfo{author}{D.~Yin}, \bibinfo{author}{C.~Chen},
  \bibinfo{author}{M.~Saito}, \bibinfo{author}{K.~Inoue},
  \bibinfo{author}{Y.~Ikuhara},
\newblock \bibinfo{title}{Ceramic phases with one-dimensional long-range
  order},
\newblock \bibinfo{journal}{Nature Materials} \bibinfo{volume}{18}
  (\bibinfo{year}{2019}) \bibinfo{pages}{19--23}.
  \DOIprefix\doi{10.1038/s41563-018-0240-0}.
\bibitem[{Guan et~al.(2011)Guan, Li, Gong, Liu, Zhang, Chen, Gelb, Yun, Xiong,
  Tian, and Wang}]{guanAnalysisThreedimensionalMicrostructure2011}
\bibinfo{author}{Y.~Guan}, \bibinfo{author}{W.~Li}, \bibinfo{author}{Y.~Gong},
  \bibinfo{author}{G.~Liu}, \bibinfo{author}{X.~Zhang},
  \bibinfo{author}{J.~Chen}, \bibinfo{author}{J.~Gelb},
  \bibinfo{author}{W.~Yun}, \bibinfo{author}{Y.~Xiong},
  \bibinfo{author}{Y.~Tian}, \bibinfo{author}{H.~Wang},
\newblock \bibinfo{title}{Analysis of the three-dimensional microstructure of a
  solid-oxide fuel cell anode using nano {{X}}-ray tomography},
\newblock \bibinfo{journal}{Journal of Power Sources} \bibinfo{volume}{196}
  (\bibinfo{year}{2011}) \bibinfo{pages}{1915--1919}.
  \DOIprefix\doi{10.1016/j.jpowsour.2010.09.059}.
\bibitem[{Vlassiouk et~al.(2018)Vlassiouk, Stehle, Pudasaini, Unocic, Rack,
  Baddorf, Ivanov, Lavrik, List, Gupta, Bets, Yakobson, and
  Smirnov}]{vlassioukEvolutionarySelectionGrowth2018}
\bibinfo{author}{I.~V. Vlassiouk}, \bibinfo{author}{Y.~Stehle},
  \bibinfo{author}{P.~R. Pudasaini}, \bibinfo{author}{R.~R. Unocic},
  \bibinfo{author}{P.~D. Rack}, \bibinfo{author}{A.~P. Baddorf},
  \bibinfo{author}{I.~N. Ivanov}, \bibinfo{author}{N.~V. Lavrik},
  \bibinfo{author}{F.~List}, \bibinfo{author}{N.~Gupta}, \bibinfo{author}{K.~V.
  Bets}, \bibinfo{author}{B.~I. Yakobson}, \bibinfo{author}{S.~N. Smirnov},
\newblock \bibinfo{title}{Evolutionary selection growth of two-dimensional
  materials on polycrystalline substrates},
\newblock \bibinfo{journal}{Nature Materials} \bibinfo{volume}{17}
  (\bibinfo{year}{2018}) \bibinfo{pages}{318--322}.
  \DOIprefix\doi{10.1038/s41563-018-0019-3}.
\bibitem[{Han et~al.(2018)Han, Li, Jung, Marsalis, Qin, Buehler, Li, and
  Muller}]{hanSubnanometreChannelsEmbedded2018}
\bibinfo{author}{Y.~Han}, \bibinfo{author}{M.-Y. Li}, \bibinfo{author}{G.-S.
  Jung}, \bibinfo{author}{M.~A. Marsalis}, \bibinfo{author}{Z.~Qin},
  \bibinfo{author}{M.~J. Buehler}, \bibinfo{author}{L.-J. Li},
  \bibinfo{author}{D.~A. Muller},
\newblock \bibinfo{title}{Sub-nanometre channels embedded in two-dimensional
  materials},
\newblock \bibinfo{journal}{Nature Materials} \bibinfo{volume}{17}
  (\bibinfo{year}{2018}) \bibinfo{pages}{129--133}.
  \DOIprefix\doi{10.1038/nmat5038}.
\bibitem[{Sun et~al.(2020)Sun, Yu, Guo, and Wang}]{sunEnhancingPowerFactor2020}
\bibinfo{author}{J.~Sun}, \bibinfo{author}{J.~Yu}, \bibinfo{author}{Y.~Guo},
  \bibinfo{author}{Q.~Wang},
\newblock \bibinfo{title}{Enhancing power factor of {{SnSe}} sheet with grain
  boundary by doping germanium or silicon},
\newblock \bibinfo{journal}{npj Computational Materials} \bibinfo{volume}{6}
  (\bibinfo{year}{2020}) \bibinfo{pages}{1--6}.
  \DOIprefix\doi{10.1038/s41524-020-00368-6}.
\bibitem[{Johnson et~al.(2015)Johnson, Li, Demkowicz, and
  Schuh}]{johnsonInferringGrainBoundary2015}
\bibinfo{author}{O.~K. Johnson}, \bibinfo{author}{L.~Li},
  \bibinfo{author}{M.~J. Demkowicz}, \bibinfo{author}{C.~A. Schuh},
\newblock \bibinfo{title}{Inferring grain boundary structure\textendash
  property relations from effective property measurements},
\newblock \bibinfo{journal}{Journal of Materials Science} \bibinfo{volume}{50}
  (\bibinfo{year}{2015}) \bibinfo{pages}{6907--6919}.
  \DOIprefix\doi{10.1007/s10853-015-9241-4}.
\bibitem[{Yang et~al.(2001)Yang, Rollett, and
  Mullins}]{yangMeasuringRelativeGrain2001}
\bibinfo{author}{C.-C. Yang}, \bibinfo{author}{A.~Rollett},
  \bibinfo{author}{W.~Mullins},
\newblock \bibinfo{title}{Measuring relative grain boundary energies and
  mobilities in an aluminum foil from triple junction geometry},
\newblock \bibinfo{journal}{Scripta Materialia} \bibinfo{volume}{44}
  (\bibinfo{year}{2001}) \bibinfo{pages}{2735--2740}.
  \DOIprefix\doi{10.1016/S1359-6462(01)00960-5}.
\bibitem[{Zhang et~al.(2020)Zhang, Ludwig, Zhang, S{\o}rensen, Rowenhorst,
  Yamanaka, Voorhees, and Poulsen}]{zhangGrainBoundaryMobilities2020}
\bibinfo{author}{J.~Zhang}, \bibinfo{author}{W.~Ludwig},
  \bibinfo{author}{Y.~Zhang}, \bibinfo{author}{H.~H.~B. S{\o}rensen},
  \bibinfo{author}{D.~J. Rowenhorst}, \bibinfo{author}{A.~Yamanaka},
  \bibinfo{author}{P.~W. Voorhees}, \bibinfo{author}{H.~F. Poulsen},
\newblock \bibinfo{title}{Grain boundary mobilities in polycrystals},
\newblock \bibinfo{journal}{Acta Materialia} \bibinfo{volume}{191}
  (\bibinfo{year}{2020}) \bibinfo{pages}{211--220}.
  \DOIprefix\doi{10.1016/j.actamat.2020.03.044}.
\bibitem[{Han et~al.(2016)Han, Vitek, and
  Srolovitz}]{hanGrainboundaryMetastabilityIts2016}
\bibinfo{author}{J.~Han}, \bibinfo{author}{V.~Vitek}, \bibinfo{author}{D.~J.
  Srolovitz},
\newblock \bibinfo{title}{Grain-boundary metastability and its statistical
  properties},
\newblock \bibinfo{journal}{Acta Materialia} \bibinfo{volume}{104}
  (\bibinfo{year}{2016}) \bibinfo{pages}{259--273}.
  \DOIprefix\doi{10.1016/j.actamat.2015.11.035}.
\bibitem[{Wei et~al.(2021)Wei, Feng, Ishikawa, Yokoi, Matsunaga, Shibata, and
  Ikuhara}]{weiDirectImagingAtomistic2021}
\bibinfo{author}{J.~Wei}, \bibinfo{author}{B.~Feng},
  \bibinfo{author}{R.~Ishikawa}, \bibinfo{author}{T.~Yokoi},
  \bibinfo{author}{K.~Matsunaga}, \bibinfo{author}{N.~Shibata},
  \bibinfo{author}{Y.~Ikuhara},
\newblock \bibinfo{title}{Direct imaging of atomistic grain boundary
  migration},
\newblock \bibinfo{journal}{Nature Materials}  (\bibinfo{year}{2021}).
  \DOIprefix\doi{10.1038/s41563-020-00879-z}.
\bibitem[{Bostanabad et~al.(2016)Bostanabad, Bui, Xie, Apley, and
  Chen}]{bostanabadStochasticMicrostructureCharacterization2016}
\bibinfo{author}{R.~Bostanabad}, \bibinfo{author}{A.~T. Bui},
  \bibinfo{author}{W.~Xie}, \bibinfo{author}{D.~W. Apley},
  \bibinfo{author}{W.~Chen},
\newblock \bibinfo{title}{Stochastic microstructure characterization and
  reconstruction via supervised learning},
\newblock \bibinfo{journal}{Acta Materialia} \bibinfo{volume}{103}
  (\bibinfo{year}{2016}). \DOIprefix\doi{10.1016/j.actamat.2015.09.044}.
\bibitem[{Homer(2019)}]{Homer2019c}
\bibinfo{author}{E.~R. Homer},
\newblock \bibinfo{title}{High-throughput simulations for insight into grain
  boundary structure-property relationships and other complex microstructural
  phenomena},
\newblock \bibinfo{journal}{Computational Materials Science}
  \bibinfo{volume}{161} (\bibinfo{year}{2019}) \bibinfo{pages}{244--254}.
  \DOIprefix\doi{10.1016/j.commatsci.2019.01.041}.
\bibitem[{Jothi et~al.(2015)Jothi, Croft, and Brown}]{Jothi2015h}
\bibinfo{author}{S.~Jothi}, \bibinfo{author}{T.~N. Croft},
  \bibinfo{author}{S.~G. Brown},
\newblock \bibinfo{title}{Multiscale multiphysics model for hydrogen
  embrittlement in polycrystalline nickel},
\newblock \bibinfo{journal}{Journal of Alloys and Compounds}
  \bibinfo{volume}{645} (\bibinfo{year}{2015}) \bibinfo{pages}{S500--S504}.
  \DOIprefix\doi{10.1016/j.jallcom.2014.12.073}.
\bibitem[{Pirgazi(2019)}]{pirgaziAlignment3DEBSD2019}
\bibinfo{author}{H.~Pirgazi},
\newblock \bibinfo{title}{On the alignment of {{3D EBSD}} data collected by
  serial sectioning technique},
\newblock \bibinfo{journal}{Materials Characterization} \bibinfo{volume}{152}
  (\bibinfo{year}{2019}) \bibinfo{pages}{223--229}.
  \DOIprefix\doi{10.1016/j.matchar.2019.04.026}.
\bibitem[{Pirgazi et~al.(2015)Pirgazi, Glowinski, Morawiec, and
  Kestens}]{pirgaziThreedimensionalCharacterizationGrain2015}
\bibinfo{author}{H.~Pirgazi}, \bibinfo{author}{K.~Glowinski},
  \bibinfo{author}{A.~Morawiec}, \bibinfo{author}{L.~A. Kestens},
\newblock \bibinfo{title}{Three-dimensional characterization of grain
  boundaries in pure nickel by serial sectioning via mechanical polishing},
\newblock \bibinfo{journal}{Journal of Applied Crystallography}
  \bibinfo{volume}{48} (\bibinfo{year}{2015}) \bibinfo{pages}{1672--1678}.
  \DOIprefix\doi{10.1107/S1600576715017616}.
\bibitem[{Speidel et~al.(2018)Speidel, Su, {Mitchell-Smith}, Dryburgh,
  Bisterov, Pieris, Li, Patel, Clark, and
  Clare}]{speidelCrystallographicTextureCan2018}
\bibinfo{author}{A.~Speidel}, \bibinfo{author}{R.~Su},
  \bibinfo{author}{J.~{Mitchell-Smith}}, \bibinfo{author}{P.~Dryburgh},
  \bibinfo{author}{I.~Bisterov}, \bibinfo{author}{D.~Pieris},
  \bibinfo{author}{W.~Li}, \bibinfo{author}{R.~Patel},
  \bibinfo{author}{M.~Clark}, \bibinfo{author}{A.~T. Clare},
\newblock \bibinfo{title}{Crystallographic texture can be rapidly determined by
  electrochemical surface analytics},
\newblock \bibinfo{journal}{Acta Materialia} \bibinfo{volume}{159}
  (\bibinfo{year}{2018}) \bibinfo{pages}{89--101}.
  \DOIprefix\doi{10.1016/J.ACTAMAT.2018.07.059}.
\bibitem[{Zheng et~al.(2020)Zheng, Li, Tran, Chen, Horton, Winston, Persson,
  and Ong}]{zhengGrainBoundaryProperties2020}
\bibinfo{author}{H.~Zheng}, \bibinfo{author}{X.~G. Li},
  \bibinfo{author}{R.~Tran}, \bibinfo{author}{C.~Chen},
  \bibinfo{author}{M.~Horton}, \bibinfo{author}{D.~Winston},
  \bibinfo{author}{K.~A. Persson}, \bibinfo{author}{S.~P. Ong},
\newblock \bibinfo{title}{Grain boundary properties of elemental metals},
\newblock \bibinfo{journal}{Acta Materialia} \bibinfo{volume}{186}
  (\bibinfo{year}{2020}) \bibinfo{pages}{40--49}.
  \DOIprefix\doi{10.1016/j.actamat.2019.12.030}.
  \href{http://arxiv.org/abs/1907.08905}{{\tt arXiv:1907.08905}}.
\bibitem[{Keinan et~al.(2018)Keinan, Bale, Gueninchault, Lauridsen, and
  Shahani}]{keinanIntegratedImagingThree2018}
\bibinfo{author}{R.~Keinan}, \bibinfo{author}{H.~Bale},
  \bibinfo{author}{N.~Gueninchault}, \bibinfo{author}{E.~Lauridsen},
  \bibinfo{author}{A.~Shahani},
\newblock \bibinfo{title}{Integrated imaging in three dimensions: {{Providing}}
  a new lens on grain boundaries, particles, and their correlations in
  polycrystalline silicon},
\newblock \bibinfo{journal}{Acta Materialia} \bibinfo{volume}{148}
  (\bibinfo{year}{2018}) \bibinfo{pages}{225--234}.
  \DOIprefix\doi{10.1016/J.ACTAMAT.2018.01.045}.
\bibitem[{Seita et~al.(2016)Seita, Volpi, Patala, McCue, Schuh, Diamanti,
  Erlebacher, and Demkowicz}]{Seita2016}
\bibinfo{author}{M.~Seita}, \bibinfo{author}{M.~Volpi},
  \bibinfo{author}{S.~Patala}, \bibinfo{author}{I.~McCue},
  \bibinfo{author}{C.~A. Schuh}, \bibinfo{author}{M.~V. Diamanti},
  \bibinfo{author}{J.~Erlebacher}, \bibinfo{author}{M.~J. Demkowicz},
\newblock \bibinfo{title}{A high-throughput technique for determining grain
  boundary character non-destructively in microstructures with
  through-thickness grains},
\newblock \bibinfo{journal}{Npj Computational Materials} \bibinfo{volume}{2}
  (\bibinfo{year}{2016}) \bibinfo{pages}{16016}. \URLprefix
  \url{http://dx.doi.org/10.1038/npjcompumats.2016.16}.
\bibitem[{Winiarski et~al.(2017)Winiarski, Gholinia, Mingard, Gee, Thompson,
  and Withers}]{winiarskiBroadIonBeam2017}
\bibinfo{author}{B.~Winiarski}, \bibinfo{author}{A.~Gholinia},
  \bibinfo{author}{K.~Mingard}, \bibinfo{author}{M.~Gee},
  \bibinfo{author}{G.~Thompson}, \bibinfo{author}{P.~Withers},
\newblock \bibinfo{title}{Broad ion beam serial section tomography},
\newblock \bibinfo{journal}{Ultramicroscopy} \bibinfo{volume}{172}
  (\bibinfo{year}{2017}) \bibinfo{pages}{52--64}.
  \DOIprefix\doi{10.1016/j.ultramic.2016.10.014}.
\bibitem[{Kim et~al.(2011)Kim, Ko, Lee, Kim, and
  Lee}]{kimIdentificationSchemeGrain2011}
\bibinfo{author}{H.~K. Kim}, \bibinfo{author}{W.~S. Ko}, \bibinfo{author}{H.~J.
  Lee}, \bibinfo{author}{S.~G. Kim}, \bibinfo{author}{B.~J. Lee},
\newblock \bibinfo{title}{An identification scheme of grain boundaries and
  construction of a grain boundary energy database},
\newblock \bibinfo{journal}{Scripta Materialia} \bibinfo{volume}{64}
  (\bibinfo{year}{2011}) \bibinfo{pages}{1152--1155}.
  \DOIprefix\doi{10.1016/j.scriptamat.2011.03.020}.
\bibitem[{Li et~al.(2019)Li, Yang, and
  Lai}]{liAtomisticSimulationsEnergies2019}
\bibinfo{author}{S.~Li}, \bibinfo{author}{L.~Yang}, \bibinfo{author}{C.~Lai},
\newblock \bibinfo{title}{Atomistic simulations of energies for arbitrary grain
  boundaries. {{Part I}}: {{Model}} and validation},
\newblock \bibinfo{journal}{Computational Materials Science}
  \bibinfo{volume}{161} (\bibinfo{year}{2019}) \bibinfo{pages}{330--338}.
  \DOIprefix\doi{10.1016/j.commatsci.2019.02.003}.
\bibitem[{Li et~al.(2009)Li, Dillon, and Rohrer}]{liRelativeGrainBoundary2009}
\bibinfo{author}{J.~Li}, \bibinfo{author}{S.~J. Dillon}, \bibinfo{author}{G.~S.
  Rohrer},
\newblock \bibinfo{title}{Relative grain boundary area and energy distributions
  in nickel},
\newblock \bibinfo{journal}{Acta Materialia} \bibinfo{volume}{57}
  (\bibinfo{year}{2009}) \bibinfo{pages}{4304--4311}.
  \DOIprefix\doi{10.1016/j.actamat.2009.06.004}.
\bibitem[{Olmsted et~al.(2009{\natexlab{a}})Olmsted, Holm, and
  Foiles}]{olmstedSurveyComputedGrain2009}
\bibinfo{author}{D.~L. Olmsted}, \bibinfo{author}{E.~A. Holm},
  \bibinfo{author}{S.~M. Foiles},
\newblock \bibinfo{title}{Survey of computed grain boundary properties in
  face-centered cubic metals-{{II}}: {{Grain}} boundary mobility},
\newblock \bibinfo{journal}{Acta Materialia} \bibinfo{volume}{57}
  (\bibinfo{year}{2009}{\natexlab{a}}) \bibinfo{pages}{3704--3713}.
  \DOIprefix\doi{10.1016/j.actamat.2009.04.015}.
\bibitem[{Olmsted et~al.(2009{\natexlab{b}})Olmsted, Foiles, and
  Holm}]{olmstedSurveyComputedGrain2009a}
\bibinfo{author}{D.~L. Olmsted}, \bibinfo{author}{S.~M. Foiles},
  \bibinfo{author}{E.~A. Holm},
\newblock \bibinfo{title}{Survey of computed grain boundary properties in
  face-centered cubic metals: {{I}}. {{Grain}} boundary energy},
\newblock \bibinfo{journal}{Acta Materialia} \bibinfo{volume}{57}
  (\bibinfo{year}{2009}{\natexlab{b}}) \bibinfo{pages}{3694--3703}.
  \DOIprefix\doi{10.1016/j.actamat.2009.04.007}.
\bibitem[{Randle et~al.(2008)Randle, Rohrer, Miller, Coleman, and
  Owen}]{randleFiveparameterGrainBoundary2008}
\bibinfo{author}{V.~Randle}, \bibinfo{author}{G.~S. Rohrer},
  \bibinfo{author}{H.~M. Miller}, \bibinfo{author}{M.~Coleman},
  \bibinfo{author}{G.~T. Owen},
\newblock \bibinfo{title}{Five-parameter grain boundary distribution of
  commercially grain boundary engineered nickel and copper},
\newblock \bibinfo{journal}{Acta Materialia} \bibinfo{volume}{56}
  (\bibinfo{year}{2008}) \bibinfo{pages}{2363--2373}.
  \DOIprefix\doi{10.1016/j.actamat.2008.01.039}.
\bibitem[{Saylor et~al.(2000)Saylor, Morawiec, Adams, and
  Rohrer}]{saylorMisorientationDependenceGrain2000}
\bibinfo{author}{D.~M. Saylor}, \bibinfo{author}{A.~Morawiec},
  \bibinfo{author}{B.~L. Adams}, \bibinfo{author}{G.~S. Rohrer},
\newblock \bibinfo{title}{Misorientation dependence of the grain boundary
  energy in magnesia},
\newblock \bibinfo{journal}{Interface Science} \bibinfo{volume}{8}
  (\bibinfo{year}{2000}) \bibinfo{pages}{131--140}.
  \DOIprefix\doi{10.1023/A:1008764219575}.
\bibitem[{Saylor et~al.(2003)Saylor, Morawiec, and
  Rohrer}]{saylorRelativeFreeEnergies2003}
\bibinfo{author}{D.~M. Saylor}, \bibinfo{author}{A.~Morawiec},
  \bibinfo{author}{G.~S. Rohrer},
\newblock \bibinfo{title}{The relative free energies of grain boundaries in
  magnesia as a function of five macroscopic parameters},
\newblock \bibinfo{journal}{Acta Materialia} \bibinfo{volume}{51}
  (\bibinfo{year}{2003}) \bibinfo{pages}{3675--3686}.
  \DOIprefix\doi{10.1016/S1359-6454(03)00182-4}.
\bibitem[{Yang et~al.(2019)Yang, Lai, and
  Li}]{yangAtomisticSimulationsEnergies2019}
\bibinfo{author}{L.~Yang}, \bibinfo{author}{C.~Lai}, \bibinfo{author}{S.~Li},
\newblock \bibinfo{title}{Atomistic simulations of energies for arbitrary grain
  boundaries. {{Part II}}: {{Statistical}} analysis of energies for tilt and
  twist grain boundaries},
\newblock \bibinfo{journal}{Computational Materials Science}
  \bibinfo{volume}{162} (\bibinfo{year}{2019}) \bibinfo{pages}{268--276}.
  \DOIprefix\doi{10.1016/j.commatsci.2019.03.010}.
\bibitem[{Dillon and
  Rohrer(2009)}]{dillonCharacterizationGrainboundaryCharacter2009}
\bibinfo{author}{S.~J. Dillon}, \bibinfo{author}{G.~S. Rohrer},
\newblock \bibinfo{title}{Characterization of the grain-boundary character and
  energy distributions of yttria using automated serial sectioning and ebsd in
  the {{FIB}}},
\newblock \bibinfo{journal}{Journal of the American Ceramic Society}
  \bibinfo{volume}{92} (\bibinfo{year}{2009}) \bibinfo{pages}{1580--1585}.
  \DOIprefix\doi{10.1111/j.1551-2916.2009.03064.x}.
\bibitem[{Restrepo et~al.(2014)Restrepo, Giraldo, and
  Thijsse}]{restrepoUsingArtificialNeural2014}
\bibinfo{author}{S.~E. Restrepo}, \bibinfo{author}{S.~T. Giraldo},
  \bibinfo{author}{B.~J. Thijsse},
\newblock \bibinfo{title}{Using artificial neural networks to predict grain
  boundary energies},
\newblock \bibinfo{journal}{Computational Materials Science}
  \bibinfo{volume}{86} (\bibinfo{year}{2014}) \bibinfo{pages}{170--173}.
  \DOIprefix\doi{10.1016/j.commatsci.2014.01.039}.
\bibitem[{Guziewski et~al.(2021)Guziewski, {Montes de Oca Zapiain},
  Dingreville, and
  Coleman}]{guziewskiMicroscopicMacroscopicCharacterization2021}
\bibinfo{author}{M.~Guziewski}, \bibinfo{author}{D.~{Montes de Oca Zapiain}},
  \bibinfo{author}{R.~Dingreville}, \bibinfo{author}{S.~P. Coleman},
\newblock \bibinfo{title}{Microscopic and {{Macroscopic Characterization}} of
  {{Grain Boundary Energy}} and {{Strength}} in {{Silicon Carbide}} via
  {{Machine}}-{{Learning Techniques}}},
\newblock \bibinfo{journal}{ACS Applied Materials \& Interfaces}
  \bibinfo{volume}{13} (\bibinfo{year}{2021}) \bibinfo{pages}{3311--3324}.
  \DOIprefix\doi{10.1021/acsami.0c15980}.
\bibitem[{Hu et~al.(2020)Hu, Zuo, Chen, Ping~Ong, and
  Luo}]{huGeneticAlgorithmguidedDeep2020}
\bibinfo{author}{C.~Hu}, \bibinfo{author}{Y.~Zuo}, \bibinfo{author}{C.~Chen},
  \bibinfo{author}{S.~Ping~Ong}, \bibinfo{author}{J.~Luo},
\newblock \bibinfo{title}{Genetic algorithm-guided deep learning of grain
  boundary diagrams: {{Addressing}} the challenge of five degrees of freedom},
\newblock \bibinfo{journal}{Materials Today} \bibinfo{volume}{38}
  (\bibinfo{year}{2020}) \bibinfo{pages}{49--57}.
  \DOIprefix\doi{10.1016/j.mattod.2020.03.004}.
  \href{http://arxiv.org/abs/2002.10632}{{\tt arXiv:2002.10632}}.
\bibitem[{Francis et~al.(2019)Francis, Chesser, Singh, Holm, and
  De~Graef}]{francisGeodesicOctonionMetric2019}
\bibinfo{author}{T.~Francis}, \bibinfo{author}{I.~Chesser},
  \bibinfo{author}{S.~Singh}, \bibinfo{author}{E.~A. Holm},
  \bibinfo{author}{M.~De~Graef},
\newblock \bibinfo{title}{A geodesic octonion metric for grain boundaries},
\newblock \bibinfo{journal}{Acta Materialia} \bibinfo{volume}{166}
  (\bibinfo{year}{2019}) \bibinfo{pages}{135--147}.
  \DOIprefix\doi{10.1016/j.actamat.2018.12.034}.
\bibitem[{Chesser et~al.(2020)Chesser, Francis, De~Graef, and
  Holm}]{chesserLearningGrainBoundary2020}
\bibinfo{author}{I.~Chesser}, \bibinfo{author}{T.~Francis},
  \bibinfo{author}{M.~De~Graef}, \bibinfo{author}{E.~Holm},
\newblock \bibinfo{title}{Learning the grain boundary manifold: Tools for
  visualizing and fitting grain boundary properties},
\newblock \bibinfo{journal}{Acta Materialia} \bibinfo{volume}{195}
  (\bibinfo{year}{2020}) \bibinfo{pages}{209--218}.
  \DOIprefix\doi{10.1016/j.actamat.2020.05.024}.
\bibitem[{Morawiec(2019)}]{morawiecDistancesGrainInterfaces2019}
\bibinfo{author}{A.~Morawiec},
\newblock \bibinfo{title}{On distances between grain interfaces in macroscopic
  parameter space},
\newblock \bibinfo{journal}{Acta Materialia} \bibinfo{volume}{181}
  (\bibinfo{year}{2019}) \bibinfo{pages}{399--407}.
  \DOIprefix\doi{10.1016/j.actamat.2019.09.032}.
\bibitem[{Barber et~al.(1996)Barber, Dobkin, and
  Huhdanpaa}]{barberQuickhullAlgorithmConvex1996}
\bibinfo{author}{C.~B. Barber}, \bibinfo{author}{D.~P. Dobkin},
  \bibinfo{author}{H.~Huhdanpaa},
\newblock \bibinfo{title}{The quickhull algorithm for convex hulls},
\newblock \bibinfo{journal}{ACM Transactions on Mathematical Software}
  \bibinfo{volume}{22} (\bibinfo{year}{1996}) \bibinfo{pages}{469--483}.
  \DOIprefix\doi{10.1145/235815.235821}.
\bibitem[{Heinz and
  Neumann(1991)}]{heinzRepresentationOrientationDisorientation1991}
\bibinfo{author}{A.~Heinz}, \bibinfo{author}{P.~Neumann},
\newblock \bibinfo{title}{Representation of orientation and disorientation data
  for cubic, hexagonal, tetragonal and orthorhombic crystals},
\newblock \bibinfo{journal}{Acta Crystallographica Section A}
  \bibinfo{volume}{47} (\bibinfo{year}{1991}) \bibinfo{pages}{780--789}.
  \DOIprefix\doi{10.1107/S0108767391006864}.
\bibitem[{Grimmer(1980)}]{grimmerUniqueDescriptionRelative1980}
\bibinfo{author}{H.~Grimmer},
\newblock \bibinfo{title}{A unique description of the relative orientation of
  neighbouring grains},
\newblock \bibinfo{journal}{Acta Crystallographica Section A}
  \bibinfo{volume}{36} (\bibinfo{year}{1980}) \bibinfo{pages}{382--389}.
  \DOIprefix\doi{10.1107/S0567739480000861}.
\bibitem[{Luong(2020)}]{luongVoronoiSphere2020}
\bibinfo{author}{B.~Luong}, \bibinfo{title}{Voronoi {{Sphere}}},
  \bibinfo{howpublished}{MATLAB Central File Exchange}, \bibinfo{year}{2020}.
  \URLprefix
  \url{https://www.mathworks.com/matlabcentral/fileexchange/40989-voronoi-sphere}.
\bibitem[{Baird and Johnson(2020)}]{bairdFiveDegreeofFreedom5DOF2020}
\bibinfo{author}{S.~Baird}, \bibinfo{author}{O.~Johnson}, \bibinfo{title}{Five
  {{Degree}}-of-{{Freedom}} ({{5DOF}}) {{Interpolation}}},
  \bibinfo{year}{2020}. \URLprefix \url{github.com/sgbaird-5dof/interp}.
\bibitem[{Patala and Schuh(2013)}]{patalaSymmetriesRepresentationGrain2013}
\bibinfo{author}{S.~Patala}, \bibinfo{author}{C.~A. Schuh},
\newblock \bibinfo{title}{Symmetries in the representation of grain
  boundary-plane distributions},
\newblock \bibinfo{journal}{Philosophical Magazine} \bibinfo{volume}{93}
  (\bibinfo{year}{2013}) \bibinfo{pages}{524--573}.
  \DOIprefix\doi{10.1080/14786435.2012.722700}.
\bibitem[{Homer et~al.(2015)Homer, Patala, and
  Priedeman}]{homerGrainBoundaryPlane2015}
\bibinfo{author}{E.~R. Homer}, \bibinfo{author}{S.~Patala},
  \bibinfo{author}{J.~L. Priedeman},
\newblock \bibinfo{title}{Grain {{Boundary Plane Orientation Fundamental
  Zones}} and {{Structure}}-{{Property Relationships}}},
\newblock \bibinfo{journal}{Scientific Reports} \bibinfo{volume}{5}
  (\bibinfo{year}{2015}) \bibinfo{pages}{1--13}.
  \DOIprefix\doi{10.1038/srep15476}.
\bibitem[{Singh and
  De~Graef(2016)}]{singhOrientationSamplingDictionarybased2016}
\bibinfo{author}{S.~Singh}, \bibinfo{author}{M.~De~Graef},
\newblock \bibinfo{title}{Orientation sampling for dictionary-based diffraction
  pattern indexing methods},
\newblock \bibinfo{journal}{Modelling and Simulation in Materials Science and
  Engineering} \bibinfo{volume}{24} (\bibinfo{year}{2016}).
  \DOIprefix\doi{10.1088/0965-0393/24/8/085013}.
\bibitem[{Langer et~al.(2006)Langer, Belyaev, and
  Seidel}]{langerSphericalBarycentricCoordinates2006}
\bibinfo{author}{T.~Langer}, \bibinfo{author}{A.~Belyaev},
  \bibinfo{author}{H.-P. Seidel},
\newblock \bibinfo{title}{Spherical barycentric coordinates},
\newblock \bibinfo{journal}{Proceedings of the fourth Eurographics symposium on
  Geometry processing}  (\bibinfo{year}{2006}) \bibinfo{pages}{81--88}.
  \URLprefix \url{http://portal.acm.org/citation.cfm?id=1281957.1281968}.
\bibitem[{Floater(2015)}]{floaterGeneralizedBarycentricCoordinates2015}
\bibinfo{author}{M.~Floater},
\newblock \bibinfo{title}{Generalized barycentric coordinates and
  applications},
\newblock \bibinfo{journal}{Acta Numerica} \bibinfo{volume}{24}
  (\bibinfo{year}{2015}) \bibinfo{pages}{161--214}.
  \DOIprefix\doi{10.1017/S09624929}.
  \href{http://arxiv.org/abs/1711.05337v1}{{\tt arXiv:1711.05337v1}}.
\bibitem[{Meyer et~al.(2002)Meyer, Barr, Lee, and
  Desbrun}]{meyerGeneralizedBarycentricCoordinates2002}
\bibinfo{author}{M.~Meyer}, \bibinfo{author}{A.~Barr},
  \bibinfo{author}{H.~Lee}, \bibinfo{author}{M.~Desbrun},
\newblock \bibinfo{title}{Generalized {{Barycentric Coordinates}} on
  {{Irregular Polygons}}},
\newblock \bibinfo{journal}{Journal of Graphics Tools} \bibinfo{volume}{7}
  (\bibinfo{year}{2002}) \bibinfo{pages}{13--22}.
  \DOIprefix\doi{10.1080/10867651.2002.10487551}.
\bibitem[{Rasmussen and Williams(2006)}]{rasmussenGaussianProcessesMachine2006}
\bibinfo{author}{C.~E. Rasmussen}, \bibinfo{author}{C.~K.~I. Williams},
  \bibinfo{title}{Gaussian Processes for Machine Learning}, Adaptive
  Computation and Machine Learning, \bibinfo{publisher}{{MIT Press}},
  \bibinfo{address}{{Cambridge, Mass}}, \bibinfo{year}{2006}.
\bibitem[{Tovar(2020)}]{tovarInverseDistanceWeight2020}
\bibinfo{author}{A.~Tovar}, \bibinfo{title}{Inverse distance weight function},
  \bibinfo{howpublished}{MATLAB Central File Exchange}, \bibinfo{year}{2020}.
  \URLprefix
  \url{https://www.mathworks.com/matlabcentral/fileexchange/46350-inverse-distance-weight-function}.
\bibitem[{Kim et~al.(2014)Kim, Kim, Dong, Steinbach, and
  Lee}]{kimPhasefieldModeling3D2014}
\bibinfo{author}{H.-K. Kim}, \bibinfo{author}{S.~G. Kim},
  \bibinfo{author}{W.~Dong}, \bibinfo{author}{I.~Steinbach},
  \bibinfo{author}{B.-J. Lee},
\newblock \bibinfo{title}{Phase-field modeling for {{3D}} grain growth based on
  a grain boundary energy database},
\newblock \bibinfo{journal}{Modelling and Simulation in Materials Science and
  Engineering} \bibinfo{volume}{22} (\bibinfo{year}{2014})
  \bibinfo{pages}{034004}. \DOIprefix\doi{10.1088/0965-0393/22/3/034004}.
\bibitem[{Chesser(2019)}]{chesserGBOctonionCode2019}
\bibinfo{author}{I.~Chesser}, \bibinfo{title}{{{GB Octonion Code}}},
  \bibinfo{year}{2019}. \URLprefix
  \url{https://github.com/ichesser/GB_octonion_code}.
\bibitem[{Bulatov et~al.(2014)Bulatov, Reed, and
  Kumar}]{bulatovGrainBoundaryEnergy2014}
\bibinfo{author}{V.~V. Bulatov}, \bibinfo{author}{B.~W. Reed},
  \bibinfo{author}{M.~Kumar},
\newblock \bibinfo{title}{Grain boundary energy function for fcc metals},
\newblock \bibinfo{journal}{Acta Materialia} \bibinfo{volume}{65}
  (\bibinfo{year}{2014}) \bibinfo{pages}{161--175}.
  \DOIprefix\doi{10.1016/j.actamat.2013.10.057}.
\bibitem[{Bean(2020)}]{beanHexscatter2020}
\bibinfo{author}{G.~Bean}, \bibinfo{title}{Hexscatter},
  \bibinfo{howpublished}{MATLAB Central File Exchange}, \bibinfo{year}{2020}.
  \URLprefix
  \url{https://www.mathworks.com/matlabcentral/fileexchange/45639-hexscatter-m}.
\bibitem[{Dolan et~al.(2004)Dolan, More, and
  Munson}]{dolanBenchmarkingOptimizationSoftware2004}
\bibinfo{author}{E.~D. Dolan}, \bibinfo{author}{J.~J. More},
  \bibinfo{author}{T.~S. Munson}, \bibinfo{title}{Benchmarking Optimization
  Software with {{COPS}} 3.0}, \bibinfo{type}{Technical Report}, {Argonne
  National Laboratory (ANL)}, \bibinfo{address}{{United States}},
  \bibinfo{year}{2004}. \DOIprefix\doi{10.2172/834714}.
\bibitem[{MATLAB Optimization
  Toolbox(2020)}]{ConstrainedElectrostaticNonlinear2020}
MATLAB Optimization Toolbox, \bibinfo{title}{Constrained {{Electrostatic
  Nonlinear Optimization}}, {{Problem}}-{{Based}}}, \bibinfo{year}{2020}.
  \URLprefix
  \url{https://www.mathworks.com/help/optim/ug/constrained-electrostatic-problem-based-optimization.html}.
\bibitem[{Shen et~al.(2019)Shen, Zhong, Liu, Suter, Morawiec, and
  Rohrer}]{shenDeterminingGrainBoundary2019}
\bibinfo{author}{Y.~F. Shen}, \bibinfo{author}{X.~Zhong},
  \bibinfo{author}{H.~Liu}, \bibinfo{author}{R.~M. Suter},
  \bibinfo{author}{A.~Morawiec}, \bibinfo{author}{G.~S. Rohrer},
\newblock \bibinfo{title}{Determining grain boundary energies from triple
  junction geometries without discretizing the five-parameter space},
\newblock \bibinfo{journal}{Acta Materialia} \bibinfo{volume}{166}
  (\bibinfo{year}{2019}) \bibinfo{pages}{126--134}.
  \DOIprefix\doi{10.1016/j.actamat.2018.12.022}.
\bibitem[{De~Graef(2020)}]{degraefEMSoft2020}
\bibinfo{author}{M.~De~Graef}, \bibinfo{title}{{{EMSoft}}},
  \bibinfo{year}{2020}. \DOIprefix\doi{10.5281/zenodo.3489720}.
\bibitem[{Dimokrati et~al.(2020)Dimokrati, Le~Bouar, Benyoucef, and
  Finel}]{dimokratiSPFMModelIdeal2020}
\bibinfo{author}{A.~Dimokrati}, \bibinfo{author}{Y.~Le~Bouar},
  \bibinfo{author}{M.~Benyoucef}, \bibinfo{author}{A.~Finel},
\newblock \bibinfo{title}{S-{{PFM}} model for ideal grain growth},
\newblock \bibinfo{journal}{Acta Materialia} \bibinfo{volume}{201}
  (\bibinfo{year}{2020}) \bibinfo{pages}{147--157}.
  \DOIprefix\doi{10.1016/j.actamat.2020.09.073}.
\bibitem[{Miyoshi et~al.(2021)Miyoshi, Takaki, Sakane, Ohno, Shibuta, and
  Aoki}]{miyoshiLargescalePhasefieldStudy2021}
\bibinfo{author}{E.~Miyoshi}, \bibinfo{author}{T.~Takaki},
  \bibinfo{author}{S.~Sakane}, \bibinfo{author}{M.~Ohno},
  \bibinfo{author}{Y.~Shibuta}, \bibinfo{author}{T.~Aoki},
\newblock \bibinfo{title}{Large-scale phase-field study of anisotropic grain
  growth: {{Effects}} of misorientation-dependent grain boundary energy and
  mobility},
\newblock \bibinfo{journal}{Computational Materials Science}
  \bibinfo{volume}{186} (\bibinfo{year}{2021}) \bibinfo{pages}{109992}.
  \DOIprefix\doi{10.1016/j.commatsci.2020.109992}.
\bibitem[{Rowenhorst et~al.(2015)Rowenhorst, Rollett, Rohrer, Groeber, Jackson,
  Konijnenberg, and
  De~Graef}]{rowenhorstConsistentRepresentationsConversions2015}
\bibinfo{author}{D.~Rowenhorst}, \bibinfo{author}{A.~D. Rollett},
  \bibinfo{author}{G.~S. Rohrer}, \bibinfo{author}{M.~Groeber},
  \bibinfo{author}{M.~Jackson}, \bibinfo{author}{P.~J. Konijnenberg},
  \bibinfo{author}{M.~De~Graef},
\newblock \bibinfo{title}{Consistent representations of and conversions between
  {{3D}} rotations},
\newblock \bibinfo{journal}{Modelling and Simulation in Materials Science and
  Engineering} \bibinfo{volume}{23} (\bibinfo{year}{2015})
  \bibinfo{pages}{083501}. \DOIprefix\doi{10.1088/0965-0393/23/8/083501}.
\bibitem[{Connor et~al.(2017)Connor, Vadicamo, and
  Rabitti}]{connorHighdimensionalSimplexesSupermetric2017}
\bibinfo{author}{R.~Connor}, \bibinfo{author}{L.~Vadicamo},
  \bibinfo{author}{F.~Rabitti},
\newblock \bibinfo{title}{High-dimensional simplexes for supermetric search},
\newblock \bibinfo{journal}{Lecture Notes in Computer Science (including
  subseries Lecture Notes in Artificial Intelligence and Lecture Notes in
  Bioinformatics)} \bibinfo{volume}{10609 LNCS} (\bibinfo{year}{2017})
  \bibinfo{pages}{96--109}. \DOIprefix\doi{10.1007/978-3-319-68474-1_7}.
  \href{http://arxiv.org/abs/1707.08370}{{\tt arXiv:1707.08370}}.
\bibitem[{Boissonnat et~al.(2017)Boissonnat, Dyer, Ghosh, and
  Oudot}]{boissonnatOnlyDistancesAre2017}
\bibinfo{author}{J.~D. Boissonnat}, \bibinfo{author}{R.~Dyer},
  \bibinfo{author}{A.~Ghosh}, \bibinfo{author}{S.~Y. Oudot},
\newblock \bibinfo{title}{Only distances are required to reconstruct
  submanifolds},
\newblock \bibinfo{journal}{Computational Geometry: Theory and Applications}
  \bibinfo{volume}{66} (\bibinfo{year}{2017}) \bibinfo{pages}{32--67}.
  \DOIprefix\doi{10.1016/j.comgeo.2017.08.001}.
  \href{http://arxiv.org/abs/1410.7012}{{\tt arXiv:1410.7012}}.
\bibitem[{Anatoliy(2015)}]{anatoliyCheckIfRay2015}
\bibinfo{author}{T.~Anatoliy}, \bibinfo{title}{Check if ray intersects
  internals of {{D}}-facet}, \bibinfo{howpublished}{Mathematics Stack
  Exchange}, \bibinfo{year}{2015}. \URLprefix
  \url{https://math.stackexchange.com/q/1256236}.
\bibitem[{Skala(2013)}]{skalaRobustBarycentricCoordinates2013}
\bibinfo{author}{V.~Skala},
\newblock \bibinfo{title}{Robust {{Barycentric Coordinates Computation}} of the
  {{Closest Point}} to a {{Hyperplane}} in {{E}}\^n},
\newblock \bibinfo{journal}{Proceedings of the 2013 Internation Conference on
  Applies Mathematics and Computational Methods in Engineering}
  (\bibinfo{year}{2013}) \bibinfo{pages}{239--244}.
\bibitem[{Anisimov et~al.(2016)Anisimov, Deng, and
  Hormann}]{anisimovSubdividingBarycentricCoordinates2016}
\bibinfo{author}{D.~Anisimov}, \bibinfo{author}{C.~Deng},
  \bibinfo{author}{K.~Hormann},
\newblock \bibinfo{title}{Subdividing barycentric coordinates},
\newblock \bibinfo{journal}{Computer Aided Geometric Design}
  \bibinfo{volume}{43} (\bibinfo{year}{2016}) \bibinfo{pages}{172--185}.
  \DOIprefix\doi{10.1016/j.cagd.2016.02.005}.
\bibitem[{Budninskiy et~al.(2016)Budninskiy, Liu, Tong, and
  Desbrun}]{budninskiyPowerCoordinatesGeometric2016}
\bibinfo{author}{M.~Budninskiy}, \bibinfo{author}{B.~Liu},
  \bibinfo{author}{Y.~Tong}, \bibinfo{author}{M.~Desbrun},
\newblock \bibinfo{title}{Power coordinates: {{A}} geometric construction of
  barycentric coordinates on convex polytopes},
\newblock \bibinfo{journal}{ACM Transactions on Graphics} \bibinfo{volume}{35}
  (\bibinfo{year}{2016}). \DOIprefix\doi{10.1145/2980179.2982441}.
\bibitem[{Dyer et~al.(2016)Dyer, Vegter, and
  Wintraecken}]{dyerBarycentricCoordinateNeighbourhoods2016}
\bibinfo{author}{R.~Dyer}, \bibinfo{author}{G.~Vegter},
  \bibinfo{author}{M.~Wintraecken},
\newblock \bibinfo{title}{Barycentric coordinate neighbourhoods in
  {{Riemannian}} manifolds},
\newblock \bibinfo{journal}{arXiv}  (\bibinfo{year}{2016}). \URLprefix
  \url{http://arxiv.org/abs/1606.01585}.
  \href{http://arxiv.org/abs/1606.01585}{{\tt arXiv:1606.01585}}.
\bibitem[{Floater and Kosinka(2010)}]{floaterInjectivityWachspressMean2010}
\bibinfo{author}{M.~S. Floater}, \bibinfo{author}{J.~Kosinka},
\newblock \bibinfo{title}{On the injectivity of {{Wachspress}} and mean value
  mappings between convex polygons},
\newblock \bibinfo{journal}{Advances in Computational Mathematics}
  \bibinfo{volume}{32} (\bibinfo{year}{2010}) \bibinfo{pages}{163--174}.
  \DOIprefix\doi{10.1007/s10444-008-9098-z}.
\bibitem[{Hormann and Kosinka(2017)}]{hormannDiscretizingWachspressKernels2017}
\bibinfo{author}{K.~Hormann}, \bibinfo{author}{J.~Kosinka},
\newblock \bibinfo{title}{Discretizing {{Wachspress}} kernels is safe},
\newblock \bibinfo{journal}{Computer Aided Geometric Design}
  \bibinfo{volume}{52-53} (\bibinfo{year}{2017}) \bibinfo{pages}{126--134}.
  \DOIprefix\doi{10.1016/j.cagd.2017.02.015}.
\bibitem[{Hormann and Sukumar(2008)}]{hormannMaximumEntropyCoordinates2008}
\bibinfo{author}{K.~Hormann}, \bibinfo{author}{N.~Sukumar},
\newblock \bibinfo{title}{Maximum entropy coordinates for arbitrary polytopes},
\newblock \bibinfo{journal}{Eurographics Symposium on Geometry Processing}
  \bibinfo{volume}{27} (\bibinfo{year}{2008}) \bibinfo{pages}{1513--1520}.
\bibitem[{Langer and Seidel(2008)}]{langerHigherOrderBarycentric2008}
\bibinfo{author}{T.~Langer}, \bibinfo{author}{H.~P. Seidel},
\newblock \bibinfo{title}{Higher order barycentric coordinates},
\newblock \bibinfo{journal}{Computer Graphics Forum} \bibinfo{volume}{27}
  (\bibinfo{year}{2008}) \bibinfo{pages}{459--466}.
  \DOIprefix\doi{10.1111/j.1467-8659.2008.01143.x}.
\bibitem[{Lei et~al.(2020)Lei, Qi, and Tian}]{leiNewCoordinateSystem2020}
\bibinfo{author}{K.~Lei}, \bibinfo{author}{D.~Qi}, \bibinfo{author}{X.~Tian},
\newblock \bibinfo{title}{A {{New Coordinate System}} for {{Constructing
  Spherical Grid Systems}}},
\newblock \bibinfo{journal}{Applied Sciences} \bibinfo{volume}{10}
  (\bibinfo{year}{2020}) \bibinfo{pages}{655}.
  \DOIprefix\doi{10.3390/app10020655}.
\bibitem[{Peixoto and Barros(2014)}]{peixotoVectorFieldReconstructions2014}
\bibinfo{author}{P.~S. Peixoto}, \bibinfo{author}{S.~R. Barros},
\newblock \bibinfo{title}{On vector field reconstructions for
  semi-{{Lagrangian}} transport methods on geodesic staggered grids},
\newblock \bibinfo{journal}{Journal of Computational Physics}
  \bibinfo{volume}{273} (\bibinfo{year}{2014}) \bibinfo{pages}{185--211}.
  \DOIprefix\doi{10.1016/j.jcp.2014.04.043}.
\bibitem[{Pihajoki et~al.(2019)Pihajoki, Mannerkoski, and
  Johansson}]{pihajokiBarycentricInterpolationRiemannian2019}
\bibinfo{author}{P.~Pihajoki}, \bibinfo{author}{M.~Mannerkoski},
  \bibinfo{author}{P.~H. Johansson},
\newblock \bibinfo{title}{Barycentric interpolation on {{Riemannian}} and
  semi-{{Riemannian}} spaces},
\newblock \bibinfo{journal}{Monthly Notices of the Royal Astronomical Society}
  \bibinfo{volume}{489} (\bibinfo{year}{2019}) \bibinfo{pages}{4161--4169}.
  \DOIprefix\doi{10.1093/mnras/stz2447}.
  \href{http://arxiv.org/abs/1907.09487}{{\tt arXiv:1907.09487}}.
\bibitem[{Rustamov(2010)}]{rustamovBarycentricCoordinatesSurfaces2010}
\bibinfo{author}{R.~M. Rustamov},
\newblock \bibinfo{title}{Barycentric coordinates on surfaces},
\newblock \bibinfo{journal}{Computer Graphics Forum} \bibinfo{volume}{29}
  (\bibinfo{year}{2010}) \bibinfo{pages}{1507--1516}.
  \DOIprefix\doi{10.1111/j.1467-8659.2010.01759.x}.
\bibitem[{Tao et~al.(2019)Tao, Deng, and Zhang}]{taoFastNumericalSolver2019}
\bibinfo{author}{J.~Tao}, \bibinfo{author}{B.~Deng},
  \bibinfo{author}{J.~Zhang},
\newblock \bibinfo{title}{A fast numerical solver for local barycentric
  coordinates},
\newblock \bibinfo{journal}{Computer Aided Geometric Design}
  \bibinfo{volume}{70} (\bibinfo{year}{2019}) \bibinfo{pages}{46--58}.
  \DOIprefix\doi{10.1016/j.cagd.2019.04.006}.
\bibitem[{Warren et~al.(2007)Warren, Schaefer, Hirani, and
  Desbrun}]{warrenBarycentricCoordinatesConvex2007}
\bibinfo{author}{J.~Warren}, \bibinfo{author}{S.~Schaefer},
  \bibinfo{author}{A.~N. Hirani}, \bibinfo{author}{M.~Desbrun},
\newblock \bibinfo{title}{Barycentric coordinates for convex sets},
\newblock \bibinfo{journal}{Advances in Computational Mathematics}
  \bibinfo{volume}{27} (\bibinfo{year}{2007}) \bibinfo{pages}{319--338}.
  \DOIprefix\doi{10.1007/s10444-005-9008-6}.

\end{thebibliography}


\newcommand{\noopsort}[1]{}
\begin{thebibliography}{7}
\expandafter\ifx\csname natexlab\endcsname\relax\def\natexlab#1{#1}\fi
\providecommand{\url}[1]{\texttt{#1}}
\providecommand{\href}[2]{#2}
\providecommand{\path}[1]{#1}
\providecommand{\DOIprefix}{doi:}
\providecommand{\ArXivprefix}{arXiv:}
\providecommand{\URLprefix}{URL: }
\providecommand{\Pubmedprefix}{pmid:}
\providecommand{\doi}[1]{\href{http://dx.doi.org/#1}{\path{#1}}}
\providecommand{\Pubmed}[1]{\href{pmid:#1}{\path{#1}}}
\providecommand{\bibinfo}[2]{#2}
\ifx\xfnm\relax \def\xfnm[#1]{\unskip,\space#1}\fi
\bibitem[{Baird and Johnson(2020)}]{bairdFiveDegreeofFreedom5DOF2020}
\bibinfo{author}{S.~Baird}, \bibinfo{author}{O.~Johnson}, \bibinfo{title}{Five
  {{Degree}}-of-{{Freedom}} ({{5DOF}}) {{Interpolation}}},
  \bibinfo{year}{2020}. \URLprefix \url{github.com/sgbaird-5dof/interp}.
\bibitem[{Francis et~al.(2019)Francis, Chesser, Singh, Holm, and
  De~Graef}]{francisGeodesicOctonionMetric2019}
\bibinfo{author}{T.~Francis}, \bibinfo{author}{I.~Chesser},
  \bibinfo{author}{S.~Singh}, \bibinfo{author}{E.~A. Holm},
  \bibinfo{author}{M.~De~Graef},
\newblock \bibinfo{title}{A geodesic octonion metric for grain boundaries},
\newblock \bibinfo{journal}{Acta Materialia} \bibinfo{volume}{166}
  (\bibinfo{year}{2019}) \bibinfo{pages}{135--147}.
  \DOIprefix\doi{10.1016/j.actamat.2018.12.034}.
\bibitem[{Chesser et~al.(2020)Chesser, Francis, De~Graef, and
  Holm}]{chesserLearningGrainBoundary2020}
\bibinfo{author}{I.~Chesser}, \bibinfo{author}{T.~Francis},
  \bibinfo{author}{M.~De~Graef}, \bibinfo{author}{E.~Holm},
\newblock \bibinfo{title}{Learning the grain boundary manifold: Tools for
  visualizing and fitting grain boundary properties},
\newblock \bibinfo{journal}{Acta Materialia} \bibinfo{volume}{195}
  (\bibinfo{year}{2020}) \bibinfo{pages}{209--218}.
  \DOIprefix\doi{10.1016/j.actamat.2020.05.024}.
\bibitem[{Morawiec(2019)}]{morawiecDistancesGrainInterfaces2019}
\bibinfo{author}{A.~Morawiec},
\newblock \bibinfo{title}{On distances between grain interfaces in macroscopic
  parameter space},
\newblock \bibinfo{journal}{Acta Materialia} \bibinfo{volume}{181}
  (\bibinfo{year}{2019}) \bibinfo{pages}{399--407}.
  \DOIprefix\doi{10.1016/j.actamat.2019.09.032}.
\bibitem[{Bulatov et~al.(2014)Bulatov, Reed, and
  Kumar}]{bulatovGrainBoundaryEnergy2014}
\bibinfo{author}{V.~V. Bulatov}, \bibinfo{author}{B.~W. Reed},
  \bibinfo{author}{M.~Kumar},
\newblock \bibinfo{title}{Grain boundary energy function for fcc metals},
\newblock \bibinfo{journal}{Acta Materialia} \bibinfo{volume}{65}
  (\bibinfo{year}{2014}) \bibinfo{pages}{161--175}.
  \DOIprefix\doi{10.1016/j.actamat.2013.10.057}.
\bibitem[{Kim et~al.(2014)Kim, Kim, Dong, Steinbach, and
  Lee}]{kimPhasefieldModeling3D2014}
\bibinfo{author}{H.-K. Kim}, \bibinfo{author}{S.~G. Kim},
  \bibinfo{author}{W.~Dong}, \bibinfo{author}{I.~Steinbach},
  \bibinfo{author}{B.-J. Lee},
\newblock \bibinfo{title}{Phase-field modeling for {{3D}} grain growth based on
  a grain boundary energy database},
\newblock \bibinfo{journal}{Modelling and Simulation in Materials Science and
  Engineering} \bibinfo{volume}{22} (\bibinfo{year}{2014})
  \bibinfo{pages}{034004}. \DOIprefix\doi{10.1088/0965-0393/22/3/034004}.
\bibitem[{Kim et~al.(2011)Kim, Ko, Lee, Kim, and
  Lee}]{kimIdentificationSchemeGrain2011}
\bibinfo{author}{H.~K. Kim}, \bibinfo{author}{W.~S. Ko}, \bibinfo{author}{H.~J.
  Lee}, \bibinfo{author}{S.~G. Kim}, \bibinfo{author}{B.~J. Lee},
\newblock \bibinfo{title}{An identification scheme of grain boundaries and
  construction of a grain boundary energy database},
\newblock \bibinfo{journal}{Scripta Materialia} \bibinfo{volume}{64}
  (\bibinfo{year}{2011}) \bibinfo{pages}{1152--1155}.
  \DOIprefix\doi{10.1016/j.scriptamat.2011.03.020}.

\end{thebibliography}

\end{document}


\sloppy 
	
	\begin{frontmatter}
		
		\title{\mytitle{}: Supplementary Information}
		
		\author[myu]{Sterling G. Baird\corref{cor1}}
\ead{ster.g.baird@gmail.com}
\author[myu]{Eric R. Homer}
\author[myu]{David T. Fullwood}
\author[myu]{Oliver K. Johnson}

\address[myu]{Department of Mechanical Engineering, Brigham Young University, Provo, UT 84602, USA}

\cortext[cor1]{Corresponding author.}

\date{February 2020}
		
	\end{frontmatter}
	
	\tableofcontents
	
	\subsection{Use of Interpolation Function}
	\label{sec:methods:repofn}
	To facilitate easy application of the presented methods, a vectorized, parallelized, MATLAB implementation, \matlab{interp5DOF.m}, is made available in the \vfzorepo{} \cite{bairdFiveDegreeofFreedom5DOF2020} with similar input/output structure to that of built-in MATLAB interpolation functions (e.g. \matlab{scatteredInterpolant()}, \matlab{griddatan()}). A typical function call is as follows: \matlab{ypred = interp5DOF(qm,nA,y,qm2,nA2,method)}. The argument \matlab{y} is a vector of known property values corresponding to the GBs defined by (\matlab{qm},\matlab{nA}), which respectively denote pairs of GB misorientation quaternions and \gls{bp} normals. The result, \matlab{ypred}, is a vector of predicted/interpolated property values corresponding to the \outpt{} \glspl{gb} defined by (\matlab{qm2},\matlab{nA2}). 
	
	Internally, these are converted to octonions and interpolation is performed using the selected \matlab{method}. For the validation function, these can be compared to the true \glspl{gbe} \matlab{ytrue}. The methods used in this work are \matlab{'pbary'}, \matlab{'gpr'}, \matlab{'idw'}, and \matlab{'nn'}, corresponding to planar barycentric, \gls{gpr}, \gls{idw}, and \gls{nn} interpolation, respectively. A placeholder template with instructions for implementing additional interpolation schemes is also provided in \matlab{interp5DOF.m}. See \citet{francisGeodesicOctonionMetric2019} and \matlab{five2oct.m} \cite{bairdFiveDegreeofFreedom5DOF2020} treatments of conversions to octonion coordinates in the passive and active senses, respectively (\cref{sec:app:convention}).
	

	\section{Euclidean and Arc Length Distances}
	\label{sec:supp:dist-parity}
\begin{figure}
		\centering
		\includegraphics[scale=1]{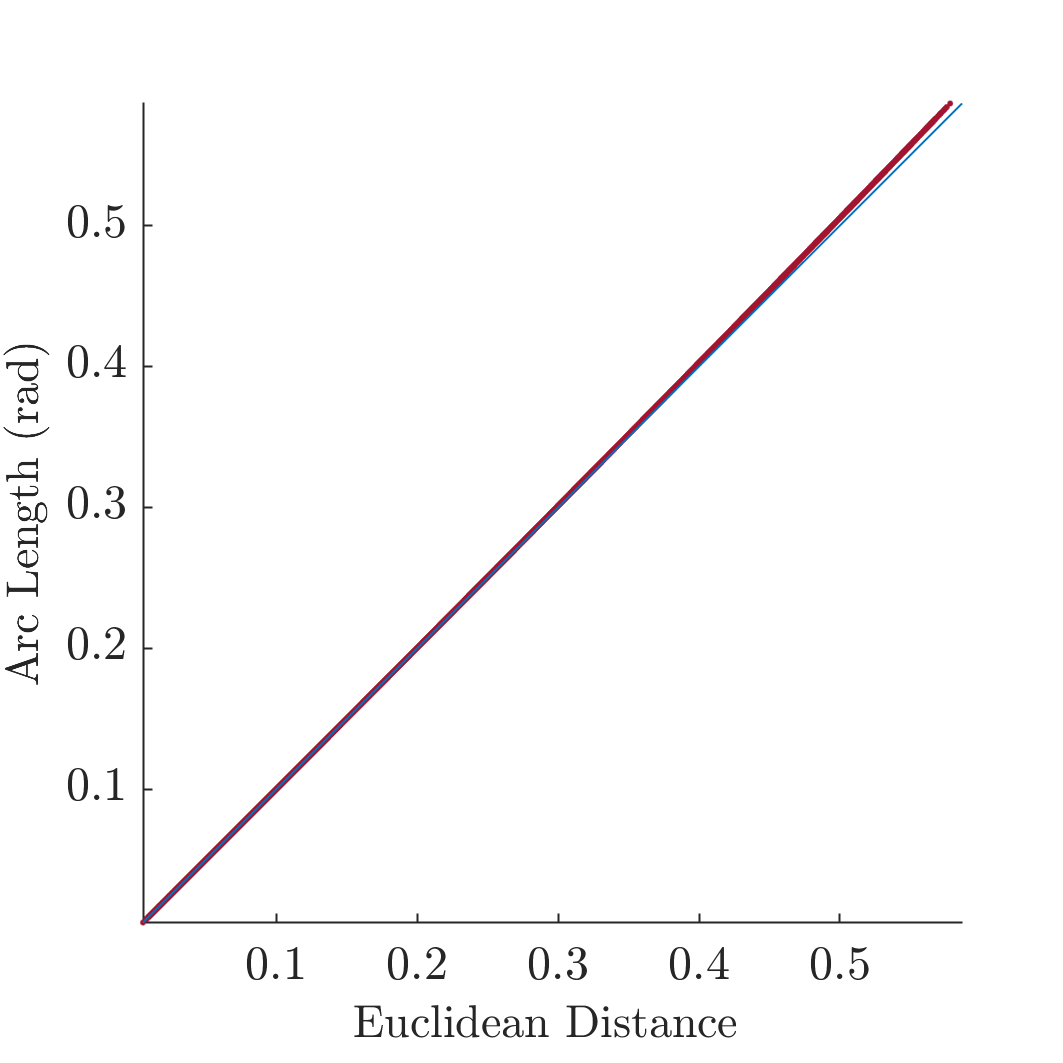}
		\caption{Parity plot of 8D Cartesian hyperspherical arc length vs. 8D Cartesian Euclidean distance for pairwise distances in a ($m\bar{3}m$) symmetrized set of \num{10000} randomly sampled \glspl{vfzo}. The max arc length is approximately \SI{0.58}{\radian}, indicating a max octonion distance of approximately \SI{1.16}{\radian} or \SI{66.5}{\degree} between any two points in the \gls{vfz}. The close correlation between arc length and Euclidean distance supports the validity of using Euclidean distance instead of arc length in the interpolation methods. This is \textit{separate} from the correlation between \gls{vfzo} Euclidean or arc length distances with the traditional octonion distance \cite{chesserLearningGrainBoundary2020}.}
		\label{fig:dist-parity}
	\end{figure}

	The close correlation between Euclidean and arc length distances in the \gls{vfzo} sense is shown in \cref{fig:dist-parity} using pairwise distances of \num{10000} \glspl{vfzo}. This justifies our use of Euclidean distance as an approximation of hyperspherical arc length (and by extension, that a scaled Euclidean distance approximates a non-symmetrized octonion distance, see \cref{eq:8Deuclidean_dist,eq:7sphere_arc_length,eq:omega} of the main paper). However, comparison with the original octonion metric \cite{francisGeodesicOctonionMetric2019} gives overestimation for some boundaries. This is an inherent feature of the \gls{vfzo} framework that can be addressed via use of the ensemble methods described in \cref{sec:methods:framework:vfz-dist} (see also \cref{fig:dist-ensemble-k1-2-10-20,fig:dist-ensemble-rmse-mae}).

\begin{figure}
		\centering
		\includegraphics{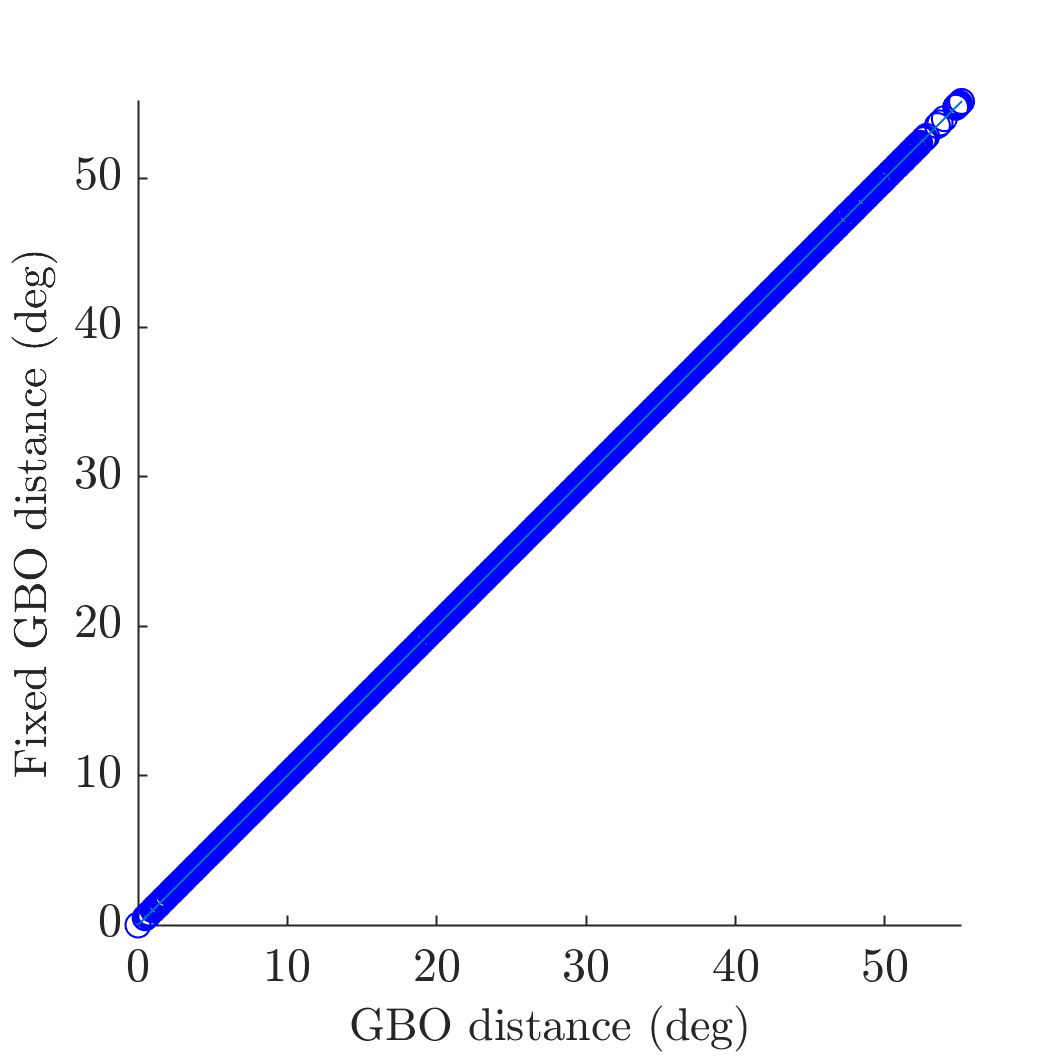}
		\caption{\Gls{gb} distances calculated with one \gls{gbo} fixed vs. the traditional calculations in \citet{chesserLearningGrainBoundary2020} show that the isometry equivalence discussed in \citet{morawiecDistancesGrainInterfaces2019} applies to \glspl{gbo}. }
		\label{fig:pd-fix}
	\end{figure}

	Additionally, the use of an isometry equivalence relationship in \citet{morawiecDistancesGrainInterfaces2019} in a non-\gls{vfz} sense results in identical distance results within numerical tolerance (\cref{fig:pd-fix}).
	
	\section{Additional Interpolation Results}
	
	\subsection{\num{388} and \num{10000} \inpt{} GBs}
\begin{figure}[!ht]
		\centering
		\includegraphics[scale=1]{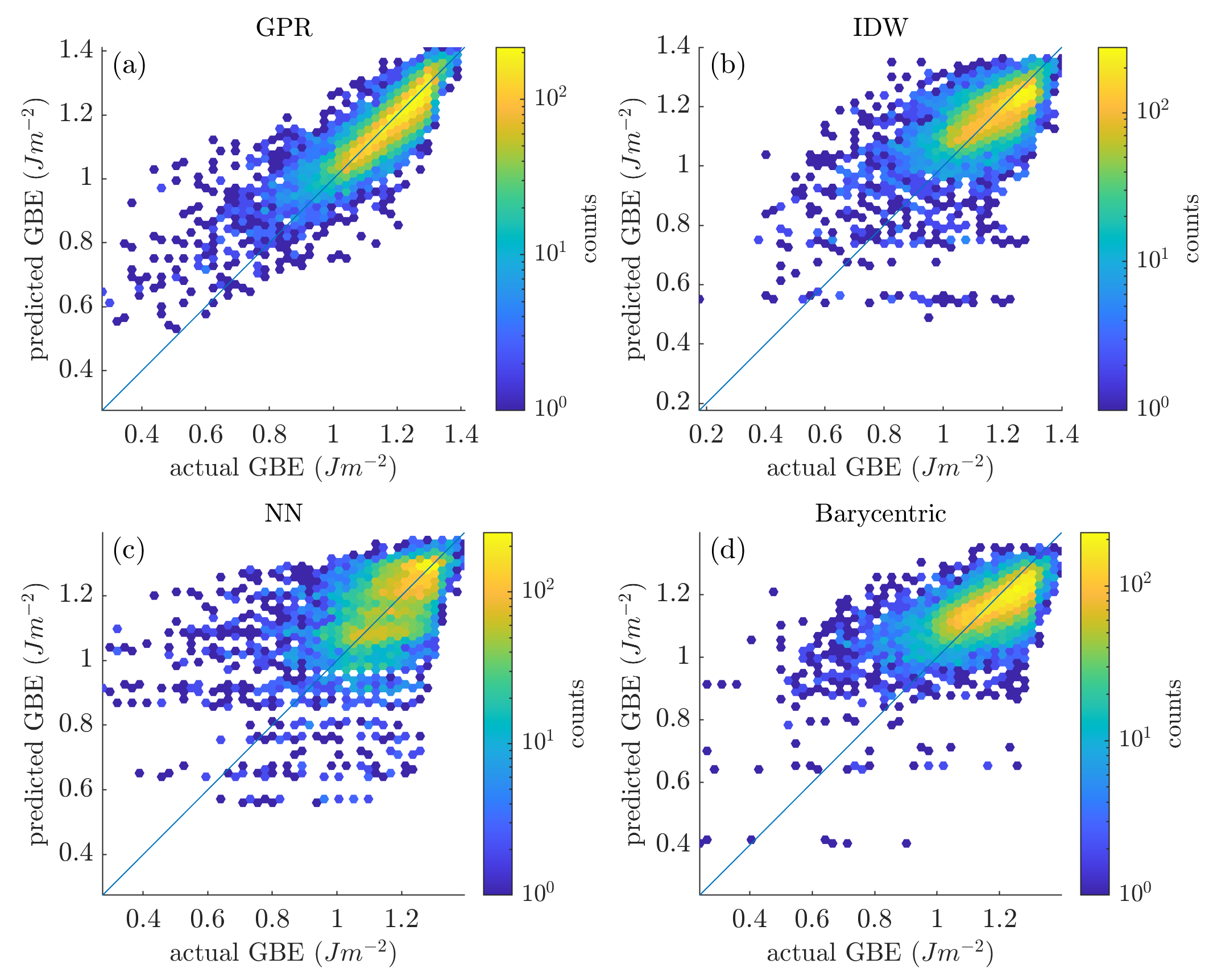}
		\caption{Hexagonally binned parity plots for \num{388} \inpt{} and \num{10000} \outpt{} octonions formed via pairs of a random cubochorically sampled quaternion and a spherically sampled random boundary plane normal. Interpolation via (a) \gls{gpr}, (b) \gls{idw}, (c) \gls{nn}, and (d) barycentric coordinates.  \gls{brk} \gls{gbe} function for \gls{fcc} Ni \cite{bulatovGrainBoundaryEnergy2014} was used as the test function.}
		\label{fig:brkparity388}
	\end{figure}

\begin{figure}[!ht]
		\centering
		\includegraphics[scale=1]{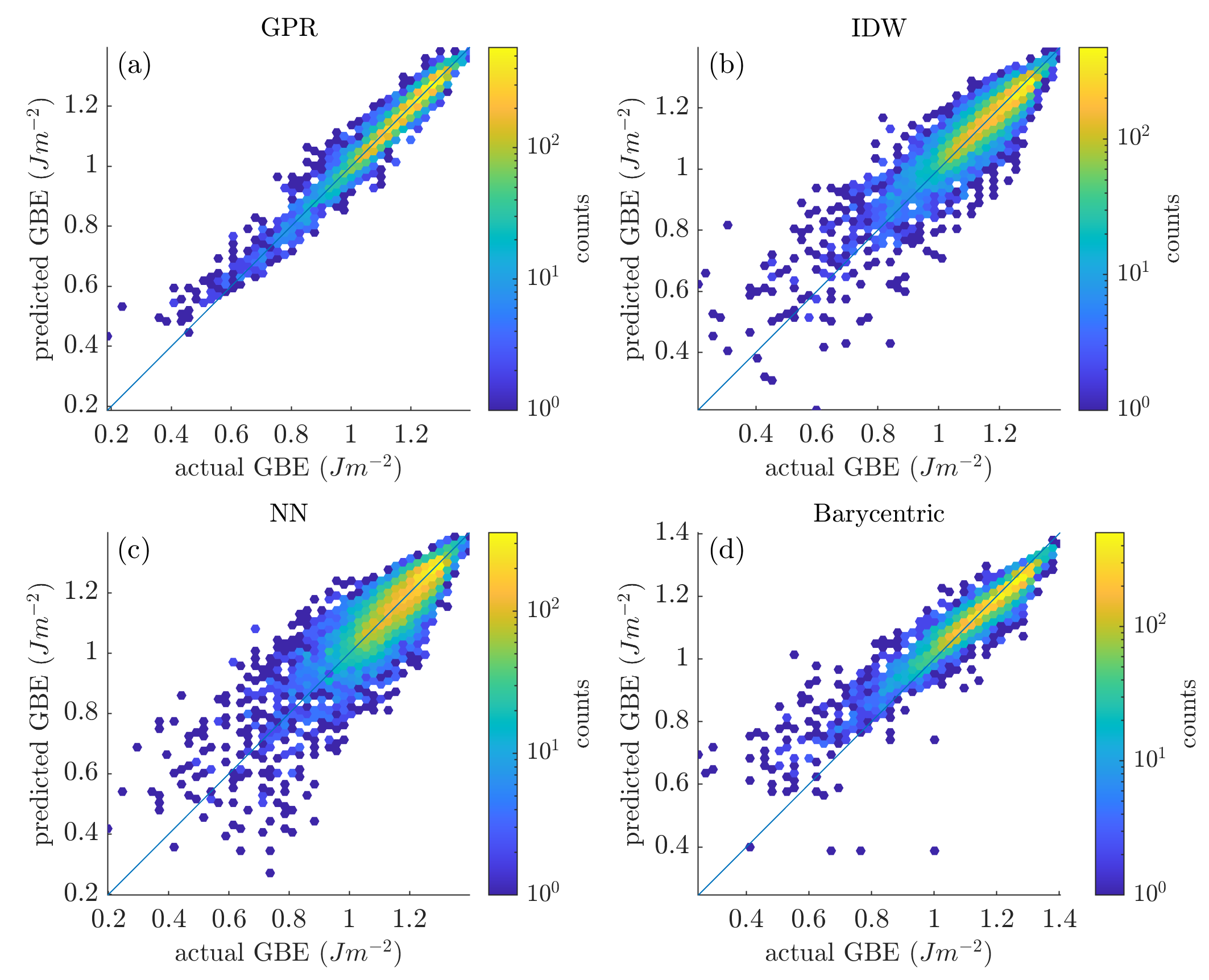}
		\caption{Hexagonally binned parity plots for \num{10000} \inpt{} and \num{10000} \outpt{} octonions formed via pairs of a random cubochorically sampled quaternion and a spherically sampled random boundary plane normal. Interpolation via (a) \gls{gpr}, (b) \gls{idw}, (c) \gls{nn}, and (d) barycentric coordinates.  \gls{brk} \gls{gbe} function for \gls{fcc} Ni \cite{bulatovGrainBoundaryEnergy2014} was used as the test function.}
		\label{fig:brkparity10000}
	\end{figure}

	Interpolation results for \num{388} and \num{10000} \glspl{gb} are given in \cref{fig:brkparity388} and \cref{fig:brkparity10000}, respectively.
	
	\section{Ensemble Interpolation Results}
	\label{sec:ensemble-interp}
	Ensemble interpolation is a classic technique that can be used to enhance predictive performance of models. Here we describe our methods (\cref{sec:ensemble-interp:methods}), results (\cref{sec:ensemble-interp:results}), and the potential of integrating ensemble interpolation with a \gls{gprm} scheme (\cref{sec:ensemble-interp:egprm}).
	
	\subsection{Methods}
	\label{sec:ensemble-interp:methods}
	\Gls{vfzo} ensemble\footnote{Ours is a "bagging"-esque ensemble scheme because the same interpolation method (\glsxtrshort{gpr}) is used but with different representations for the \inpt{} data. } interpolation occurs by:
	\begin{enumerate}
		\item generating multiple reference octonions to define multiple \glspl{vfz}
		\item obtaining multiple \gls{vfzo} representations for a set of \glspl{gb} based on the various reference octonions
		\item performing an interpolation (e.g. \gls{gpr}) for each of the representations
		\item homogenizing the ensemble of models (e.g. by taking the mean or median of the various models)
	\end{enumerate}
	
	\subsection{Results}
	\label{sec:ensemble-interp:results}
	
\begin{figure}
		\centering
		\includegraphics[scale=1]{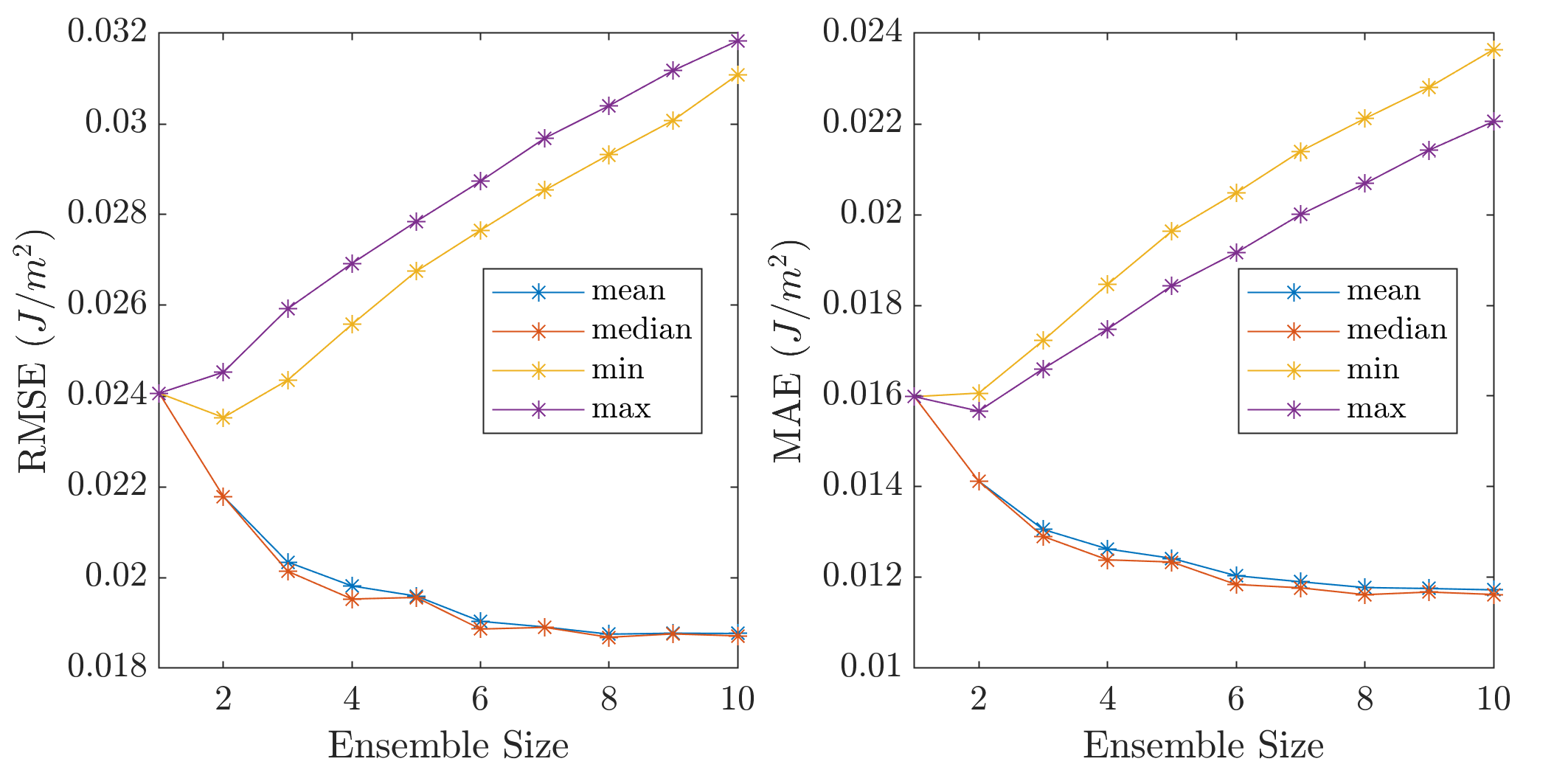}
		\caption{(a) \Gls{rmse} and (b) \gls{mae} vs. ensemble size for mean, median, minimum, and maximum homogenization functions. A \gls{gpr} model with \num{50000} \inpt{} and \num{10000} \outpt{} \glspl{vfzo} was used. }
		\label{fig:ensemble-interp-rmse-mae}
	\end{figure}

	Use of an ensemble interpolation scheme decreases interpolation error for a \gls{gpr} model with \num{50000} \inpt{} and \num{10000} \outpt{} \glspl{vfzo}. By using an ensemble size of 10 (i.e. 10 \gls{gpr} models each with different reference octonions and therefore different \glspl{vfz}), \gls{rmse} and \gls{mae} decreased from \SIlist{0.0241;0.0160}{\J\per\square\m} to \SIlist{0.0187;0.0116}{\J\per\square\m}, respectively, using the median homogenization function (\cref{fig:ensemble-interp-rmse-mae}). 
	
\begin{figure}[h!]
		\centering
		\includegraphics[scale=1]{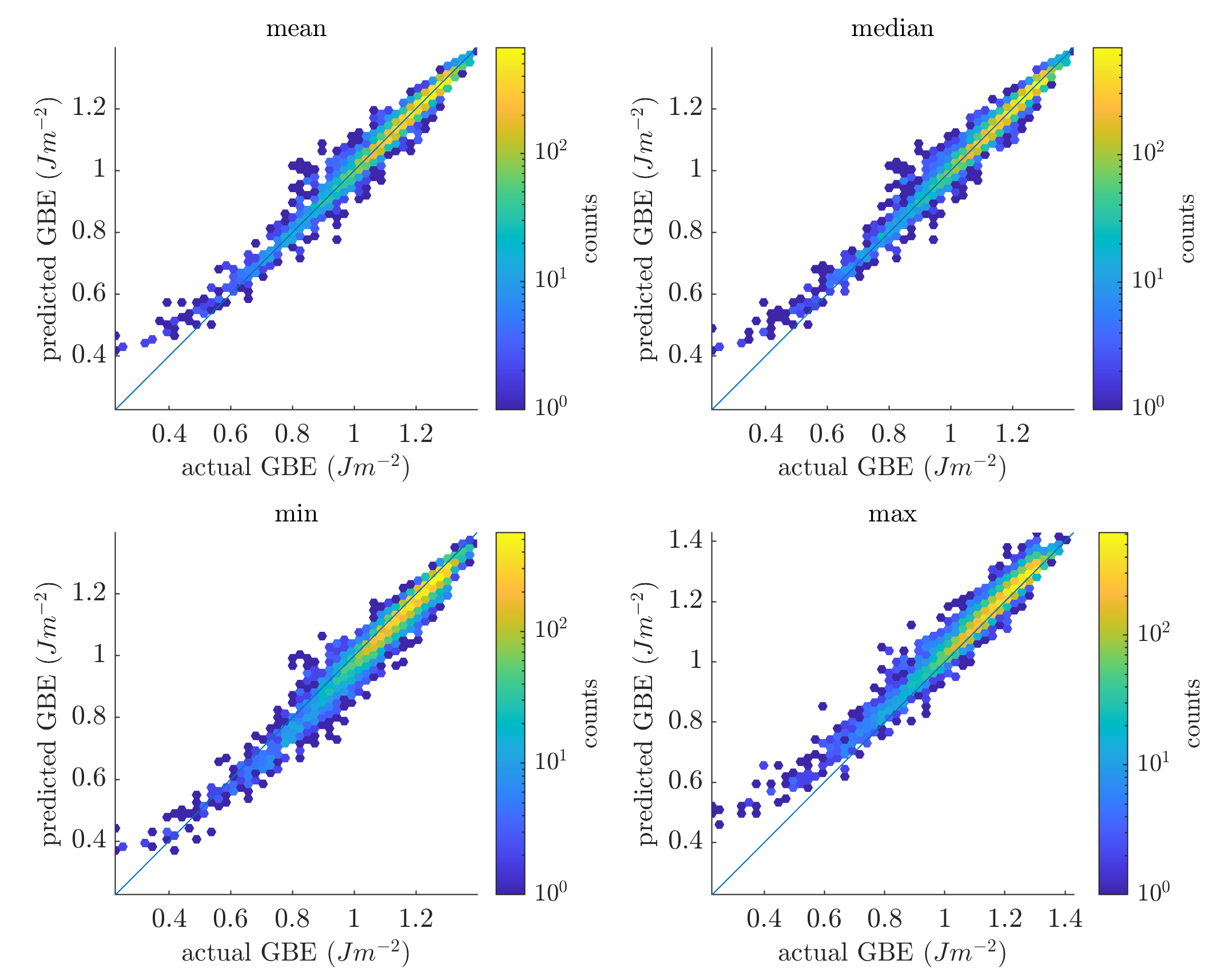}
		\caption{Hexagonally binned parity plots for (a) mean, (b) median, (c) minimum, and (d) maximum ensemble homogenization functions. A \gls{gpr} model with \num{50000} \inpt{} and \num{10000} \outpt{} \glspl{vfzo} was used. }
		\label{fig:ensemble-interp}
	\end{figure}

	\Cref{fig:ensemble-interp} shows the hexagonally binned parity plots for predictions made using the mean, median, minimum, and maximum predicted values over an ensemble of 10 \glspl{vfz}. Qualitatively, the ensemble mean and ensemble median parity plots look similar to those from the main text (\cref{fig:brkparity50000}), though the distributions of the ensemble scheme are somewhat tighter. The ensemble minimum produces better predictions of low \gls{gbe} than any of the other models, but underestimates high \gls{gbe} as expected. Naturally, the ensemble maximum overestimates in general. Diminishing returns manifest in \cref{fig:ensemble-interp-rmse-mae} for mean and median homogenizations. This is to be expected because the original octonion distances \cite{francisGeodesicOctonionMetric2019} are well-approximated using an ensemble size of 10 (\cref{fig:dist-ensemble-k1-2-10-20}c and \cref{fig:dist-ensemble-rmse-mae}).
	
	\subsection{Possibility: Combining Ensemble with \glsentrytitlecase{gpr}{long} Mixture}
	\label{sec:ensemble-interp:egprm}
	
	A scheme which preferentially favors the ensemble minimum for low \gls{gbe} predictions and defaults to ensemble mean or median for all other \glspl{gbe} may produce even better results across the full range of \glspl{gbe}. For example, this could be accomplished by combining the ensemble scheme described here with the \gls{gpr} mixture model described in \cref{sec:supp:kim-interp:method}.
	
	\section{Barycentric Interpolation}
	\label{sec:supp:bary}
	
	\subsection{High-Aspect Ratios}
	\label{sec:supp:bary:artifact}
\begin{figure*}
		\centering
		\includegraphics[scale=1]{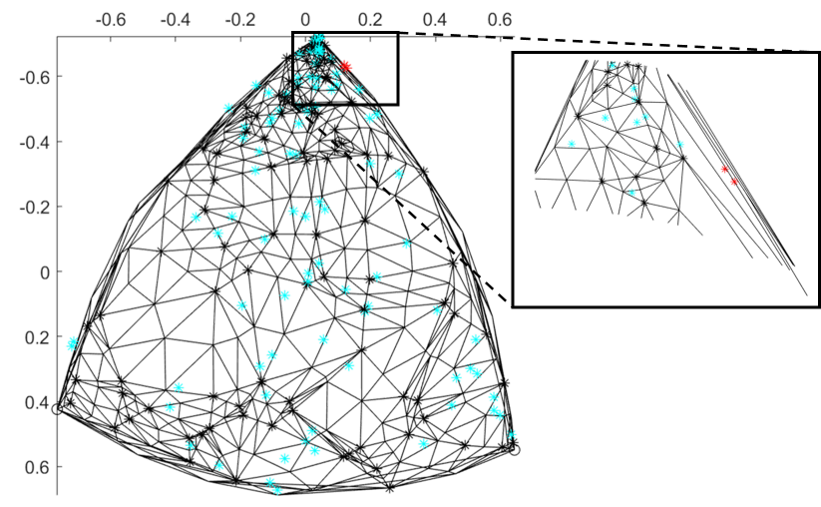}
		\caption{Illustration of two \outpt{} points (red) for which no intersecting facet is found due to being positioned within a high-aspect ratio facet. The inset shows that facets connected to the \gls{nn} do not contain the \outpt{} point. Many \glspl{nn} would need to be considered before an intersection is found. Additionally, it is expected that if found, the interpolation will suffer from higher error due to use of facet vertices far from the interpolation point. Proper intersections of \outpt{} points with the mesh are shown in blue.}
		\label{fig:high-aspect-non-int}
	\end{figure*}

	An artifact of the barycentric interpolation method which occurs due to the presence of high-aspect ratio facets is shown in \cref{fig:high-aspect-non-int}. As the dimensionality increases for a constant number of points and from our numerical tests, the rate of missed facet intersections increases. This artifact and our method for addressing it are discussed in \cref{sec:app:bary:int} of the main text.
	
	\section{Kim Interpolation}
	\label{sec:supp:kim-interp}
	
	A \gls{gpr} mixing model is developed to accommodate the non-uniformly distributed, noisy Fe simulation data \cite{kimPhasefieldModeling3D2014} and better predict low \gls{gbe}. Details of the method (\cref{sec:supp:kim-interp:method}) and an analysis of the input data quality (\cref{sec:supp:kim-interp:quality}) are given. The code implementation is given in \texttt{gprmix.m} and \texttt{gprmix\_test.m} of the \vfzorepo{} \cite{bairdFiveDegreeofFreedom5DOF2020}.
	
	\subsection{Details of \glsentrytitlecase{gpr}{long} Mixture}
	\label{sec:supp:kim-interp:method}
\begin{figure}
		\centering
		\includegraphics[scale=1]{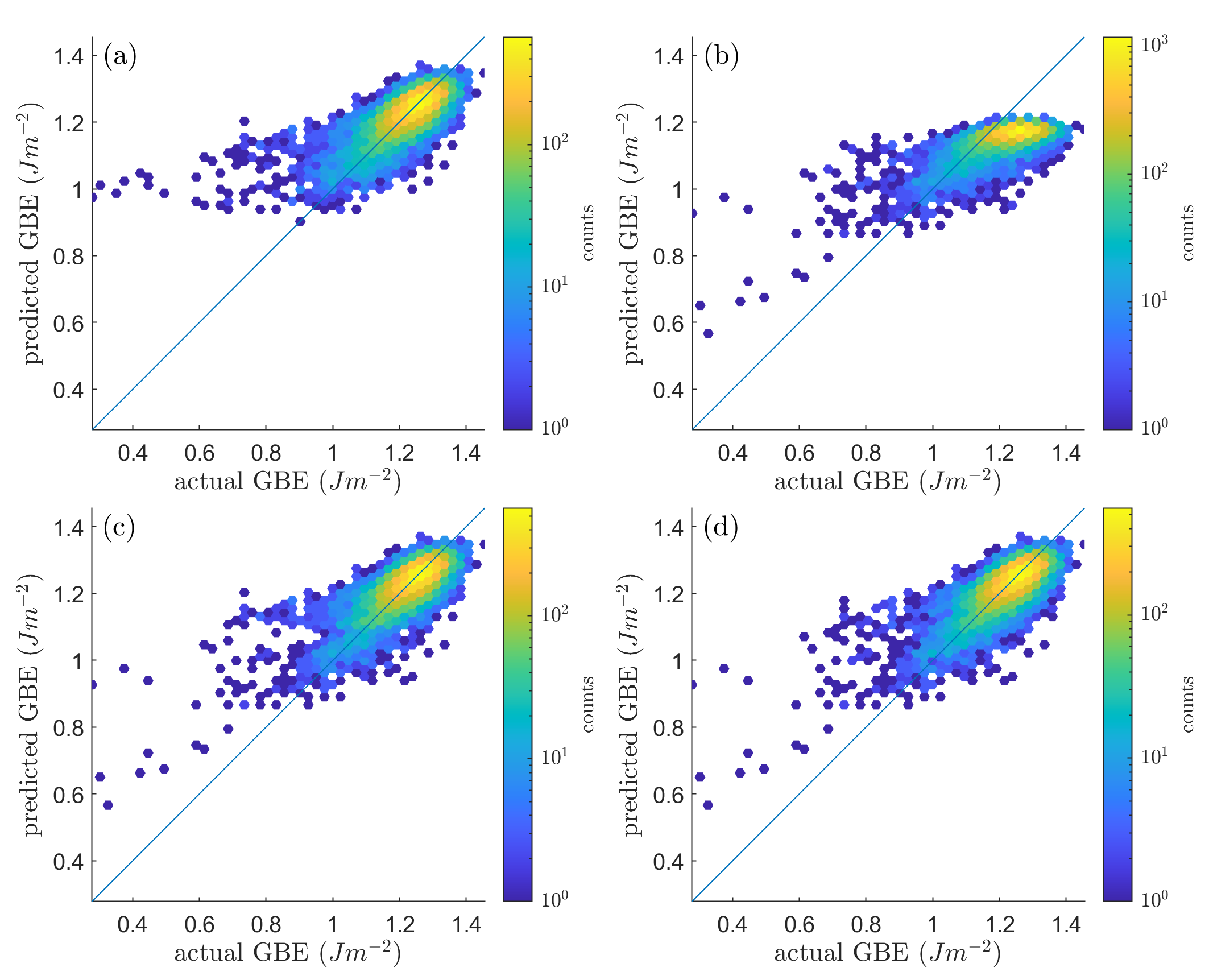}
		\caption{(a) Hexagonally binned parity plot of the standard \gls{gpr} model. (b) All prediction \glspl{gb} based on the model using only training \glspl{gb} with a \gls{gbe} less than \thrtwo{}. (c) Combined disjoint model as explained in the text. (d) Hexagonally binned parity plots of the final \gls{gpr} mixing model. Points in (c) are produced by splitting the prediction data into less than and greater than \thr{}. A sigmoid mixing function (\cref{fig:gprmix-sigmoid}) is then applied where the predicted \glspl{gbe} shown in (c) determines the mixing fraction ($f$) to produce a weighted average of models (a) and (b). A large Fe simulation database \cite{kimPhasefieldModeling3D2014} using \num{46883} training datapoints and \num{11721} validation datapoints in an 80\%/20\% split. The \gls{gpr} mixture model decreases error for low \gls{gbe} and changes overall \gls{rmse} and \gls{mae} from \SI{0.057932}{\J\per\square\meter} and \SI{0.039857}{\J\per\square\meter} in the original model (shown in (a)) to \SI{0.055035}{\J\per\square\meter} and \SI{0.039185}{\J\per\square\meter} (shown in (d)), respectively.}
		\label{fig:kim-interp-teach}
	\end{figure}

	As shown in \cref{fig:kim-interp-teach}a, prediction using the standard approach of the main document (termed the $\epsilon_1$ model) overestimates low \glspl{gbe} for this dataset. By training the model on only \glspl{gb} with a \gls{gbe} less than \thrtwo{} (termed the $\epsilon_2$ model) and by using an exponential (\matlab{KernelFunction='exponential'}) rather than a squared exponential kernel, prediction of low \glspl{gbe} improves, but naturally underestimation occurs for higher \glspl{gbe} (\cref{fig:kim-interp-teach}b).
	
\begin{figure}
		\centering
		\includegraphics[scale=1]{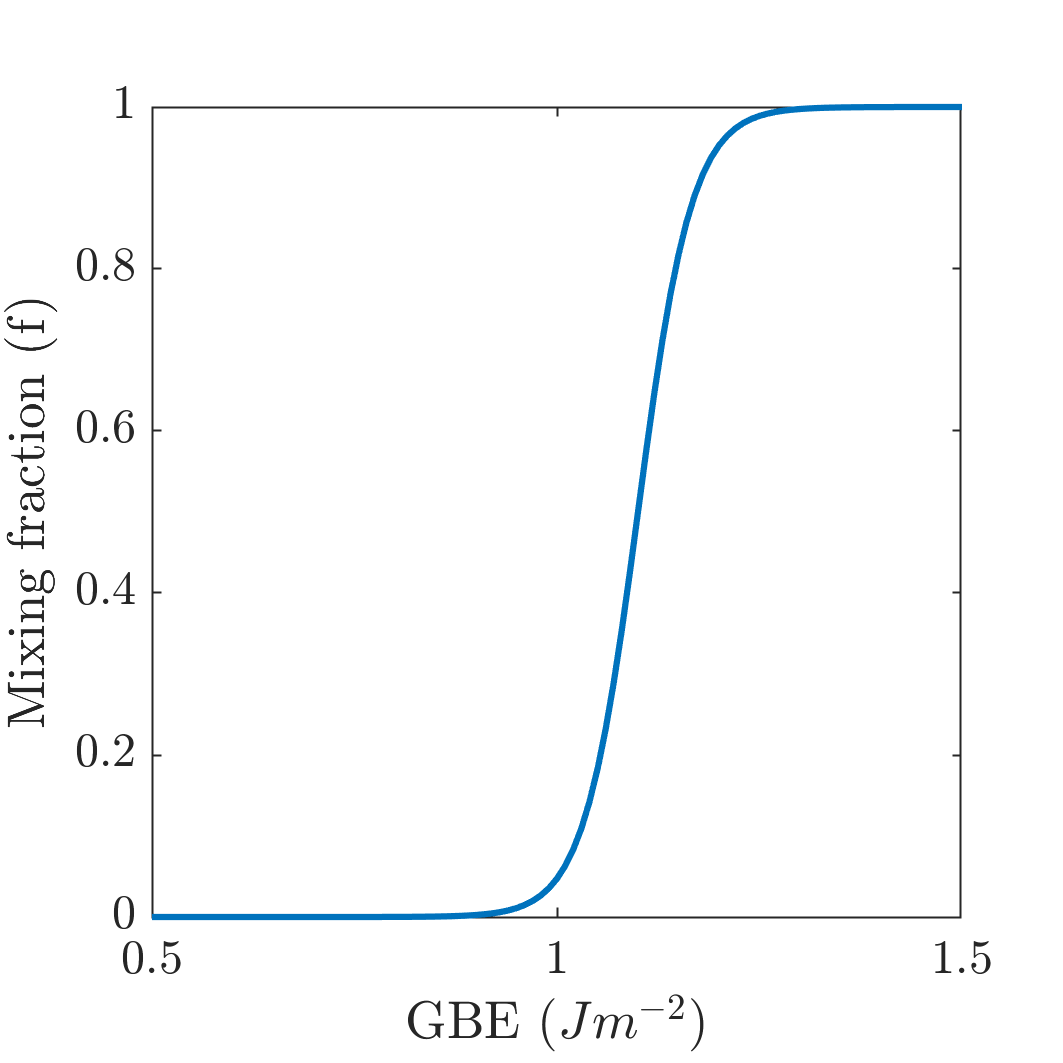}
		\caption{Sigmoid mixing function used in the \gls{gpr} mixing model with $m=30$ and $b=\sigthr{}$ (\cref{eq:sigmoid}).}
		\label{fig:gprmix-sigmoid}
	\end{figure}

	A combined, disjoint model (\cref{fig:kim-interp-teach}c) is taken ($\epsilon_3$) by replacing $\epsilon_1$ \gls{gbe} predictions for \glspl{gb} with \gls{gbe} less than \thr{} with the corresponding $\epsilon_2$ predictions. Finally, a weighted average (\cref{eq:gprmix}) is taken according to:
	
	\begin{equation}
		\epsilon_{mix} = f \epsilon_1+(f-1) \epsilon_2
		\label{eq:gprmix}
	\end{equation}
	where $\epsilon_1$ and $\epsilon_2$ represent the standard \gls{gpr} model and the \gls{gpr} model trained on the subset of \glspl{gb} with a \gls{gbe} less than \thrtwo{}, respectively, and $f$ is the sigmoid mixing fraction given by:
	
	\begin{equation}
		f=\frac{1}{e^{-m \left(\epsilon_3-b\right)}+1}
		\label{eq:sigmoid}
	\end{equation}
	and shown in \cref{fig:gprmix-sigmoid} with $m=30$ and $b=\sigthr{}$, as used in this work. Larger values of $m$ yield a steeper sigmoid function and larger values of $b$ shift the sigmoid function further to the right. Specific values for $m$ and $b$ were chosen by visual inspection and trial and error. This results in a \gls{gpr} mixing model which better predicts low \glspl{gbe} while retaining overall predictive accuracy (\cref{fig:kim-interp-teach}d).
	
	Uncertainty of the \gls{gpr} mixing model is similarly obtained by taking a weighted average of the uncertainties of each model according to:
	
	\begin{equation}
		\sigma_{mix} = f \sigma_1+(f-1) \sigma_2
		\label{eq:gprmix-sigma}
	\end{equation}
	where $\sigma_1$ and $\sigma_2$ are the corresponding uncertainties of $\epsilon_1$ and $\epsilon_2$, respectively, and $f$ is given by \cref{eq:sigmoid}. 
	
	\subsection{Input Data Quality}
	\label{sec:supp:kim-interp:quality}
\begin{figure}
		\centering
		\includegraphics[scale=1]{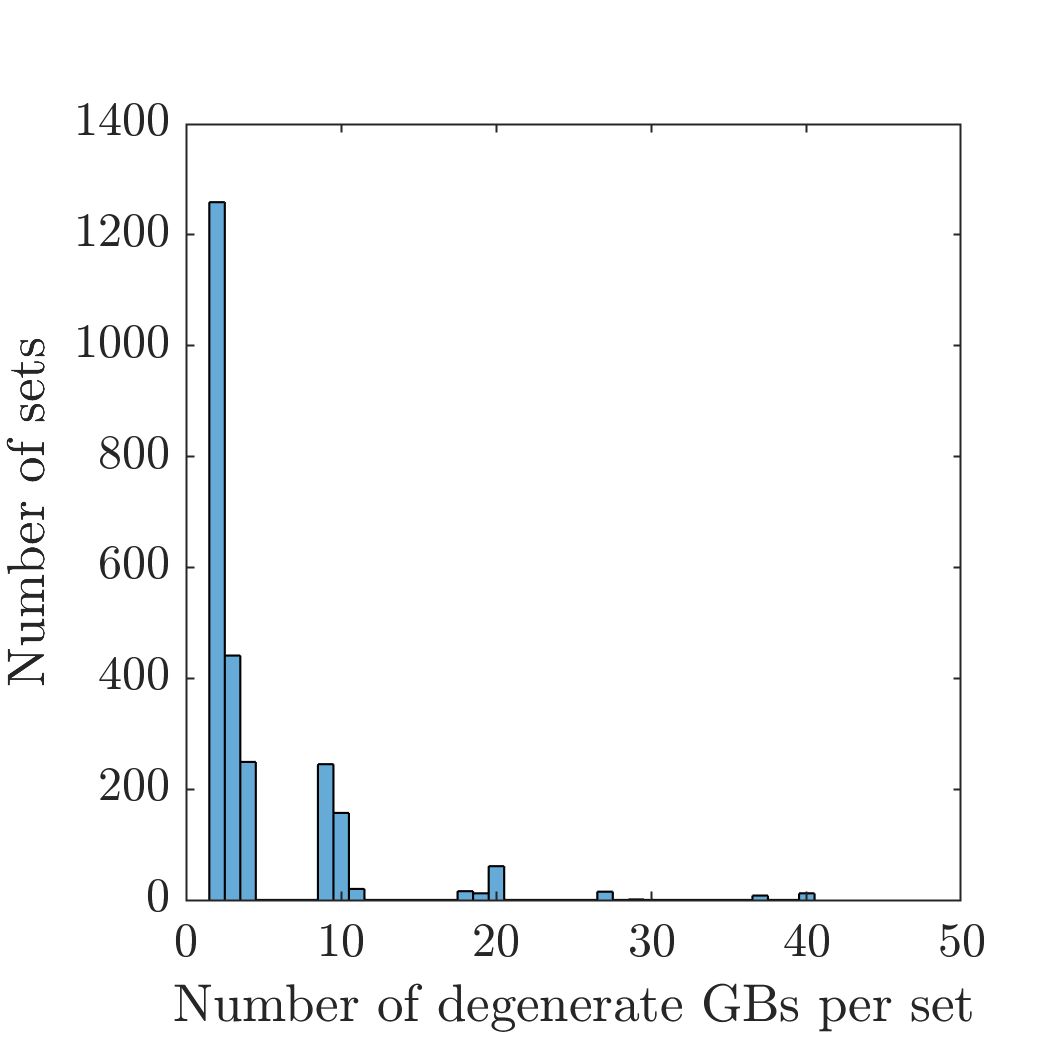}
		\caption{Histogram of number of sets vs. number of degenerate \glspl{gb} per set for the Fe simulation dataset \cite{kimPhasefieldModeling3D2014}. Most sets have a degeneracy of fewer than 5.}
		\label{fig:kim-interp-degeneracy-sets}
	\end{figure}

	Of the $\sim$\num{60000}\footnote{The "no-boundary" \glspl{gb} (i.e. \glspl{gb} with close to \SI{0}{\joule\per\square\meter} \gls{gbe}) were removed before testing for degeneracy.} \glspl{gb} in \cite{kimPhasefieldModeling3D2014}, $\sim$\num{10000} \glspl{gb} were repeats that were identified by converting to \glspl{vfzo} and applying \vfzorepo{} function \texttt{avg\_repeats.m}. In \cite{kimPhasefieldModeling3D2014}, mechanically selected \glspl{gb} were those which involved sampling in equally spaced increments\footnote{In some cases, this was equally spaced increments of the argument of a trigonometric function.} for each \gls{5dof} parameter, and a few thousand intentionally selected \glspl{gb} (i.e. special \glspl{gb}) were also considered. Of mechanically and intentionally selected \glspl{gb}, \numlist{9170;112} are repeats, respectively, with a total of \num{2496} degenerate sets\footnote{A degenerate "set" is distinct from a \gls{vfzo} "set", the latter of which is often used in the main text.} (see \cref{fig:kim-interp-degeneracy-sets} for a degeneracy histogram). Thus, on average there is a degeneracy of approximately four per set of degenerate \glspl{gb}.
	
\begin{figure}
		\centering
		\includegraphics[scale=1]{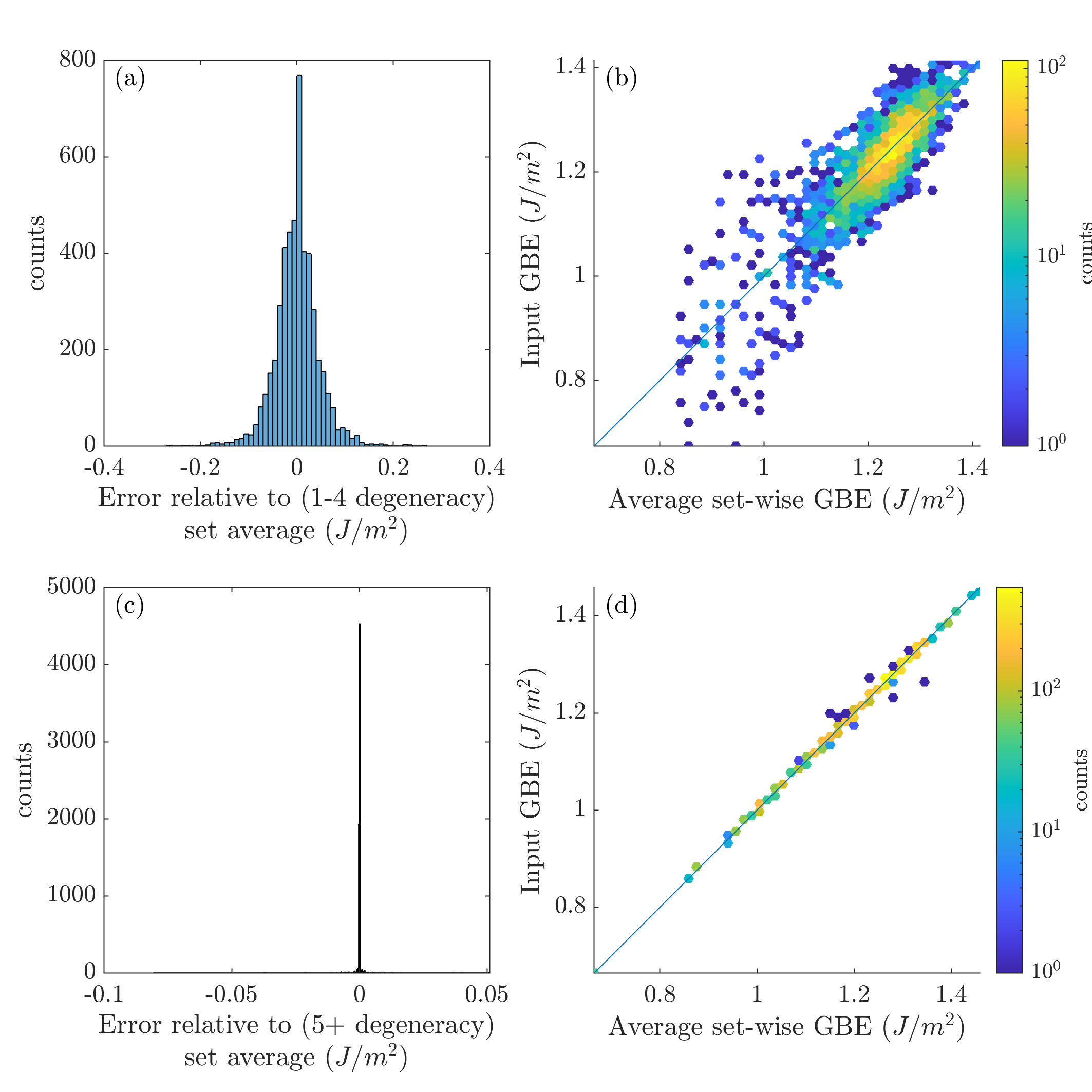}
		\caption{Degenerate \glspl{gb} sets are split into those with a degeneracy of fewer than 5 and greater than or equal to 5 and plotted as ( (a) and (c), respectively) error histograms and ( (b) and (d), respectively) hexagonally-binned parity plots. Large degenerate sets tend to have very low error, whereas small degenerate sets tend to have higher error. In other words, \glspl{gb} that are more likely to be repeated many times based on the sampling scheme in \cite{kimPhasefieldModeling3D2014} tend to give similar results, whereas \glspl{gb} that are less likely to be repeated often have larger variability in the simulation output. We do not know if this has physical meaning or is an artifact of the simulation setup.}
		\label{fig:kim-interp-degeneracy-results}
	\end{figure}

	By comparing \gls{gbe} values of (unintentionally\footnote{To our knowledge, the presence of repeat \glspl{gb} were not mentioned in \cite{kimPhasefieldModeling3D2014} or \cite{kimIdentificationSchemeGrain2011}}) repeated \glspl{gb} in the Fe simulation dataset \cite{kimPhasefieldModeling3D2014}, we can estimate the intrinsic error of the \inpt{} data. For example, minimum and maximum deviations from the average value of a degenerate set are \SIlist{-0.2625;0.2625}{\joule\per\square\meter}, respectively, indicating that a repeated Fe \gls{gb} simulation from \cite{kimPhasefieldModeling3D2014} can vary by as much as \SI{0.525}{\joule\per\square\meter}, though rare. Additionally, \Gls{rmse} and \gls{mae} values can be obtained within each degenerate set by comparing against the set mean. Overall \gls{rmse} and \gls{mae} are then obtained by averaging and weighting by the number of \glspl{gb} in each degenerate set. Following this procedure, we obtain an average set-wise \gls{rmse} and \gls{mae} of \SIlist{0.06529;0.06190}{\joule\per\square\meter}, respectively, which is an approximate measure of the intrinsic error of the data. \cref{fig:kim-interp-degeneracy-results} shows histograms and parity plots of the intrinsic error. The overestimation of intrinsic error mentioned in the main text (\cref{sec:results:simulation}) could stem from bias as to what type of \glspl{gb} exhibit repeats based on the sampling scheme used in \cite{kimPhasefieldModeling3D2014} and/or that many of the degenerate sets contain a low number of repeats (\cref{fig:kim-interp-degeneracy-sets}).
	
	Next, we see that by binning \glspl{gb} into degenerate sets, most degenerate sets have a degeneracy of fewer than 5 \cref{fig:kim-interp-degeneracy-sets}. We split the repeated data into sets with a degeneracy of fewer than 5 and greater than or equal to 5 and plot the errors (relative to the respective set mean) in both histogram form (\cref{fig:kim-interp-degeneracy-results}a and \cref{fig:kim-interp-degeneracy-results}c, respectively) and as hexagonally-binned parity plots (\cref{fig:kim-interp-degeneracy-results}b and \cref{fig:kim-interp-degeneracy-results}d, respectively). While heavily repeated \glspl{gb} tend to give similar results, occasionally repeated \glspl{gb} often have larger \gls{gbe} variability. This could have physical meaning: Certain types of (e.g. high-symmetry) \glspl{gb} tend to have less variation (i.e. fewer and/or more tightly distributed metastable states). However, it could also be an artifact of the simulation setup that produced this data (e.g. deterministic simulation output for certain types of \glspl{gb}).
	
	\section{Olmsted Interpolation}
	
\begin{figure}
		\centering
		\begin{subfigure}[b]{0.5\textwidth}
			\centering
			\includegraphics[width=\textwidth]{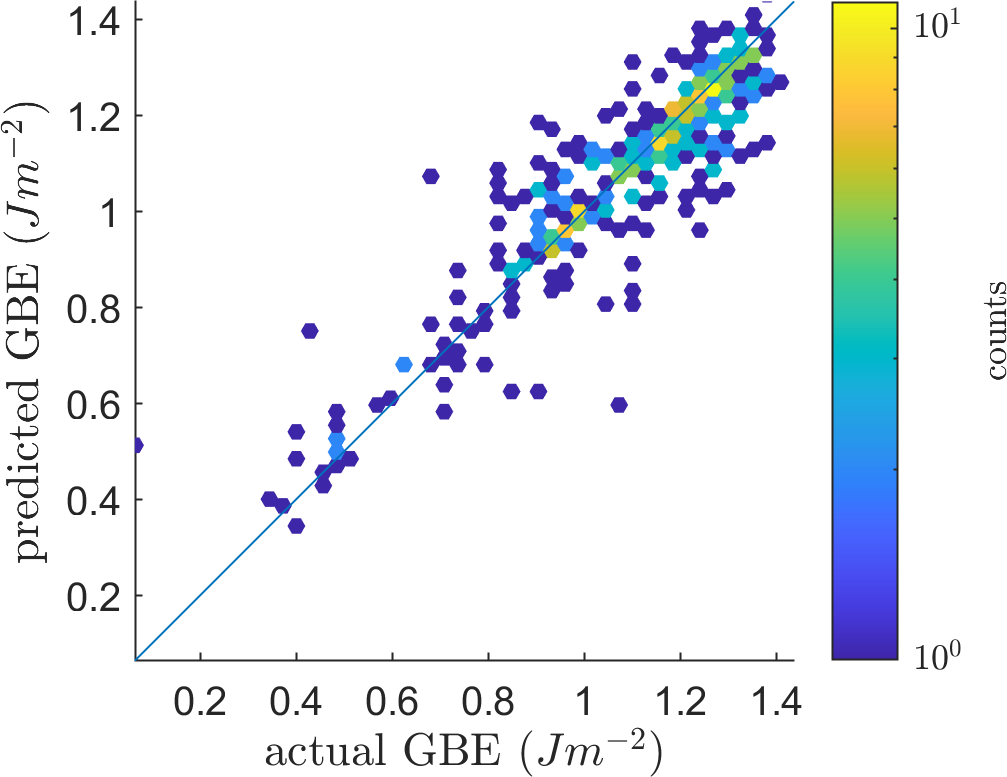}
			\label{fig:our-loocv}
			\hfill
		\end{subfigure}
		\begin{subfigure}[b]{0.4\textwidth}
			\centering
			\includegraphics[width=\textwidth]{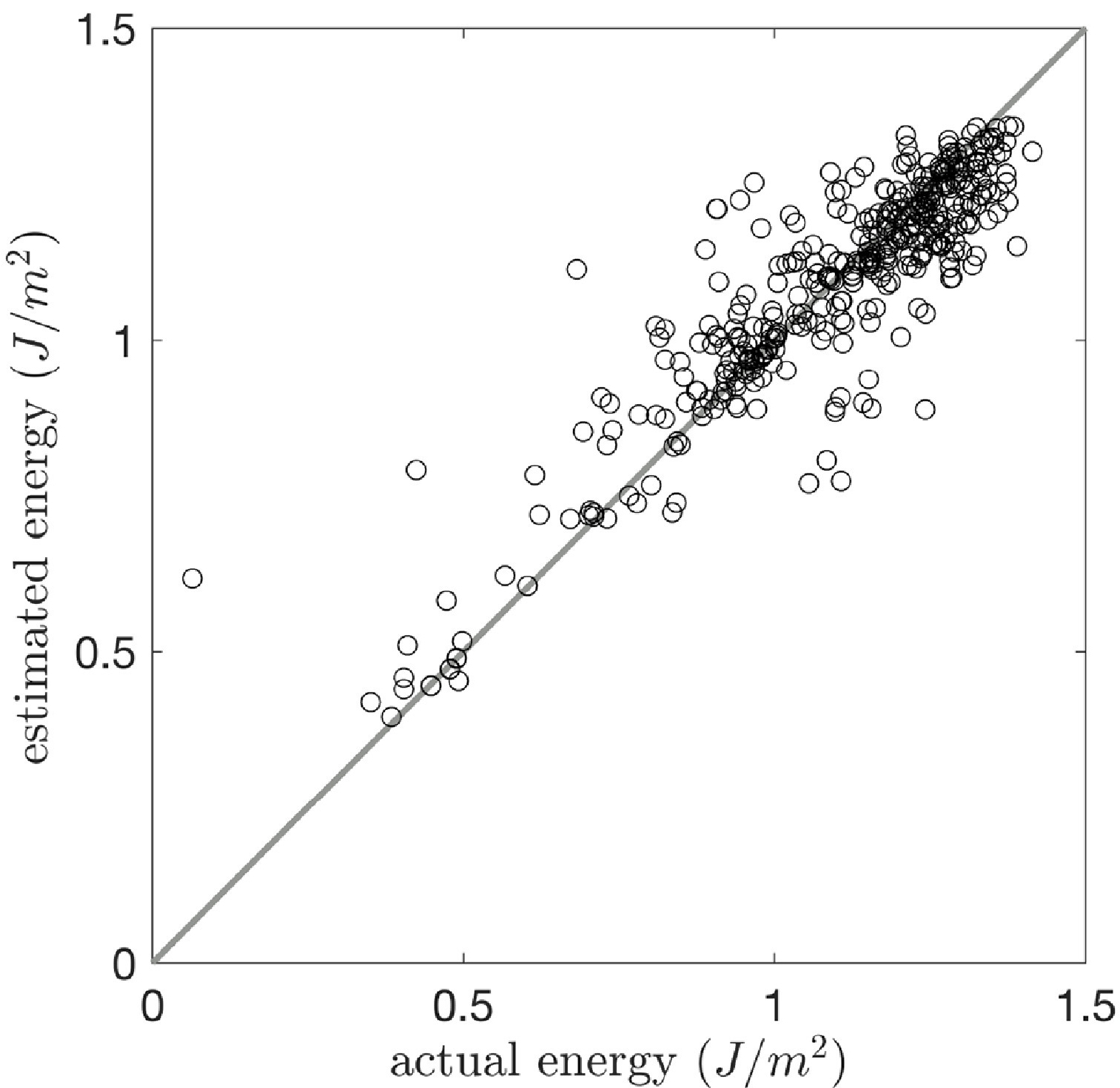}
			\label{fig:chesser-loocv}
		\end{subfigure}
		\caption{(left) Hexagonally binned parity plot for Ni simulation \glsxtrfull{gbe} interpolation using \glsxtrshort{loocv}. (right) Parity plot for \glsxtrfull{loocv} interpolation results reproduced from Figure 6a of \citet{chesserLearningGrainBoundary2020} under CC-BY Creative Commons license. }
		\label{fig:olmsted-Ni-loocv}
	\end{figure}

	As illustrated in \cref{fig:olmsted-Ni-loocv}, \gls{loocv} interpolation results for \SI{0}{\kelvin} \gls{ms} low-noise Ni simulations using the \gls{gpr} method are similar to \gls{lkr} results reported in Figure 6a of \citet{chesserLearningGrainBoundary2020} (reproduced on the right of \cref{fig:olmsted-Ni-loocv} for convenience).
	
	\newpage
	\clearpage 
	\printglossaries
	
	\newpage
	\bibliographystyle{elsarticle-num-names}
	\bibliography{5dof-gb-energy.bib}